\documentstyle[11pt,aaspp4,tighten]{article}

\tightenlines

\def\gsimeq
{\hbox{\raise0.5ex\hbox{$>\lower1.06ex\hbox{$\kern-1.07em{\sim}$}$}}}
\def\lsimeq
{\hbox{\raise0.5ex\hbox{$<\lower1.06ex\hbox{$\kern-1.07em{\sim}$}$}}}
\def\ion#1#2{#1$\;${\small\rm{#2}}\relax}

\begin{document}

\title{A Comprehensive Spectral and Variability Study of Narrow-Line
Seyfert 1 Galaxies Observed by ASCA: II. Spectral Analysis and
Correlations}

\author{Karen M. Leighly}
\affil{Columbia Astrophysics Laboratory, 550 West 120th Street, New
York, NY 10027, USA, leighly@ulisse.phys.columbia.edu}

\slugcomment{Submitted to{\it The Astrophysical Journal}}


\begin{abstract}

I present a comprehensive and uniform analysis of 25 {\it ASCA}
observations from 23 Narrow-line Seyfert 1 galaxies.  The spectral
analysis and correlations are presented in this paper, Part 2; the
reduction and time series analysis is presented in the companion
paper, Part 1.

A maximum likelihood analysis confirms that the hard X-ray photon
index is significantly steeper at $>$90\% confidence in this sample of
NLS1s compared with a random sample of Seyfert galaxies with broad
optical lines.  Soft excess emission was detected in 17 of the 19
objects which had no significant absorption, a result that
demonstrates that soft excesses appear considerably more frequently in
NLS1s than in Seyfert 1 galaxies with broad optical lines.  The
strength of the soft excess, parameterized using a flux ratio obtained
from the black body plus power law model, has a wide range of values
in these objects (a factor of 50). The photon index was found to be
correlated with the H$\beta$ FWHM, despite the small range of the
latter parameter; however, neither parameter is correlated with the
strength of the soft excess or {\it ROSAT} slope.  Therefore, assuming
that an excess of soft photons results in the steep photon index and
narrow H$\beta$ FWHM, that excess may lie primarily in the
unobservable EUV.  The strength of the soft excess is correlated with
the variability parameters, so that objects with strong soft excesses
show higher amplitude variability; this potentially important result
is not easily explained.  While a range of two orders of magnitude in
luminosity is represented, the temperature of the soft excess is
approximately consistent throughout the sample, in contrast with
expectations of simple accretion disk models.  The presence of ionized
absorption was sought using a two-edge model. It was found that this
component appears to be typically less common in NLS1s and some
evidence was found that the typical ionization state is lower compared
with broad-line Seyfert galaxies.  This fact, plus evidence for a
correlation between the presence of the warm absorber and significant
optical polarization, may imply that the inner warm absorber is
missing or is too highly ionized to be detected, and only the outer,
dusty, less ionized warm absorber is present in many cases.  The iron
line equivalent width appears to be similar among narrow and
broad-line Seyfert galaxies.  This could mean that reprocessing occurs
with similar geometry in both classes of objects; however, the
detection of ionized iron lines in a few objects, implying possibly
altered fluorescence yields, and poor statistics, makes this
conclusion tentative.  Constraints on physical processes and models
based on extreme values of orientation and accretion rate for NLS1s
are examined in light of the observational results.

\end{abstract}

\keywords{galaxies: active --- galaxies: Seyfert  --- X-rays: galaxies }

\clearpage

\section{Introduction}

Narrow-line Seyfert galaxies are identified by their optical line
properties: H$\beta$ FWHM is $<2000 \rm\,km/s$, the [\ion{O}{III}]
$\lambda 5007$ to H$\beta$ ratio is $<3$ and there are high ionization
lines and frequently strong \ion{Fe}{II} emission present in the
spectrum (Osterbrock \& Pogge\markcite{103} 1985;
Goodrich\markcite{44} 1989).  Although their permitted lines may be
only slightly broader than their forbidden lines, they can be clearly
distinguished from Seyfert 2 galaxies.  For example, the permitted
lines are polarized differently than the forbidden lines, indicating
two emission regions (Goodrich\markcite{44} 1989).  As shown by
Boroson \& Green\markcite{9} (1992) these emission line properties are
strongly correlated; a principal component analysis (PCA) shows that
much of the variance in line properties can be traced to a single
eigenvector, despite the fact that different emission lines are
thought to originate in widely separated parts of the AGN.  Therefore,
these correlations must have an origin in a primary intrinsic physical
parameter.  The narrow-line Seyfert 1 galaxies are located at the
extreme end of the Boroson and Green eigenvector; therefore, as a
class they exemplify an extreme value of this physical parameter.  It
is important to identify this physical parameter and to understand how it
drives these correlations.

This paper is the second of a two part series that presents the first
uniform analysis of the X-ray variability and spectral properties from
{\it ASCA} observations of a sizable sample of Narrow-line Seyfert 1
galaxies.  Part 1 presents the data reduction and the time series
analysis of the {\it ASCA} data as well as a preliminary search for
spectral variability and a general introduction to this subclass of
Seyfert 1 galaxy.  This part presents the spectral analysis of the
{\it ASCA} data and studies of correlations between the spectral and
variability parameters and optical emission line information from the
literature as well as from analysis of spectra in hand.  The results
are discussed in terms of current models for the X-ray emission from
NLS1s and Seyfert galaxies; in particular, effects of orientation and
accretion rate are discussed.  Some of these results have been
previously presented in Leighly\markcite{76} 1998.

{\it ROSAT} observations demonstrated that the soft X-ray spectra from
NLS1s is systematically steeper than those from Seyfert 1 galaxies
with broad optical lines (Boller, Brandt \& Fink\markcite{7} 1996; Forster \&
Halpern\markcite{37} 1996; Laor et al.\markcite{71} 1997a).  {\it ASCA} spectra from NLS1s have
the advantage of a broader band pass and better energy resolution than
the {\it ROSAT} PSPC spectra.  A breakthrough in the understanding of
NLS1s occurred with the observation of RE~1034+39 using {\it ASCA}
(Pounds, Done \& Osborne\markcite{112} 1995).  The {\it ASCA} spectrum
revealed a very strong soft excess component dominating the spectrum
below 1--2 keV, and a very steep hard X-ray power law with photon
index $\sim 2.6$.  This spectrum is very different than the typical
Seyfert 1 spectrum, which tends to have a flatter power law with
photon index typically around 1.7--1.9 and the soft excess is
generally not observed in the {\it ASCA} band pass.  This dichotomy
resembles that observed between Galactic black hole candidates (GBHC)
in the hard (or low) and soft (or high) states (e.g.\
Nowak\markcite{102} 1995), a fact that prompted Pounds, Done \&
Osborne\markcite{112} (1995) to postulate that NLS1s are the
supermassive black hole analogs of Galactic black hole candidates in
the soft state.  Soft state GBHC are thought to be accreting at a
larger fraction of the Eddington limit than hard state GBHC; this
analogy supports the idea that overall behavior of NLS1s originates
from an accretion rate that is a higher fraction of Eddington than in
Seyfert 1 galaxies with broader optical lines.  More recently, Brandt,
Mathur \& Elvis\markcite{11} (1997) presented a collection of photon
indices from {\it ASCA} observations of Seyfert galaxies obtained from
the literature.  They found that the steep hard X-ray spectrum from
RE~1034+39 is not anomalous; NLS1s have generally steeper hard (2--10
keV) X-ray spectra than broad-line Seyfert 1 galaxies.

Narrow-line Seyfert 1 galaxies have been also observed to exhibit
peculiar spectral features in their {\it ASCA} spectra that have not
been observed in Seyfert 1 galaxies with broad optical lines.  Three
NLS1s which have very strong soft excesses were discovered to have
peculiar absorption features near 1~keV (Leighly et al.\markcite{79} 1997a; see
also Otani, Kii \& Miya\markcite{104} 1996; Hayashida\markcite{60} 1996; 
Comastri, Molendi \& Ulrich\markcite{19} 1997).  Absorption by highly
ionized oxygen is common in Seyfert 1 galaxies with broad optical
lines (e.g. Reynolds\markcite{116} 1997).  However, the energies of the features in
these three objects is far too high to be interpreted as absorption by
oxygen in the rest frame of the galaxy.  Citing the similarities
between many aspects of NLS1s and  broad-absorption line quasars, which
are known for their UV absorption lines that indicate high-velocity
outflows, Leighly et al.\markcite{79} (1997a) suggested that these features are due
to relativistically outflowing gas with speeds of 0.2--0.6~c.  More
recently, Nicastro, Fiore \& Matt\markcite{100} 1999 presented an alternative
interpretation: they suggest that these features are due to absorption
by large columns of highly ionized gas, and the features are due to
resonant absorption primarily by Fe~L.  

Other unusual features have also been found.  Fiore et
al.\markcite{34} 1998 discovered a feature near 1~keV in the spectrum
of PG~1244+026 that is best modeled by an emission line.  Comastri et
al.\markcite{18} 1998 discovered that the iron line in the {\it
BeppoSAX} spectrum of Ton~S180 has an energy near 6.7~keV indicating
that the iron is ionized.  Both of these features were interpreted as
evidence for reprocessing in an accretion disk which is highly
ionized, a property predicted if the accretion rate is high (e.g.\
Matt, Fabian \& Ross\markcite{87} 1993a).  However, it has been shown
more recently that 6.7~keV iron lines are not restricted to NLS1s but
occur also in Seyfert 1 galaxies with broad optical lines (e.g.\
Guainazzi et al.\markcite{54} 1998a).

\section{Data Reduction}

\subsection{Spectra}

The data reduction and filtering are described in Part 1, and this
section describes treatment of the data specific to the spectral
fitting.  Dark-frame error and echo corrections were made on data
taken in Faint mode.  As the Rev.\ 2 reprocessing had been performed
on all of the data, no corrections for the gain in GIS3 during later
time periods was made.

To obtain the best signal to noise, a range of three different-sized
extraction regions were used.  For brightest objects, the nominal
extraction regions ($4^\prime$ and $6^\prime$ for SIS and GIS
respectively) were used.  For fainter objects, regions 87.5\% and 75\%
of the nominal ones were used.  Background spectra were accumulated
from source-free parts of the detector.

In order to account for the decline in the energy resolution as a
function of time for the CCD detectors, response matrices appropriate
for the date of the observation were made using the script {\it
sisrmg}.  Ancillary response files (arf files) were made for each
detector using {\it ascaarf}.  Several early observations were made in
4CCD or 2CCD Bright mode and significant flux is spread out over
multiple chips. For the most part, these were reduced and combined as
recommended in {\it The ASCA Data Reduction Guide}.  Spectra and
response files were made for each separate chip. Response files were
averaged together using {\it addrmf} weighted by the number of source
photons in each chip.  A spectrum was extracted from the entire source
region and the arf file was made using that.

Spectral fitting was performed using XSPEC v.\ 10.  The energy bands
over which the spectra were fitted varied.  For observations before
January 1996, the lower limit was constrained only by the presence of
a lower level discriminator applied during the observation.  In fact,
a level discriminator was applied only during the I~Zw~1,
RX~J0439$-$45, 1H~0707$-$495 and Ark~564 observations.  The GIS
spectra were fit above 0.8~keV, since the GIS detectors are not well
calibrated below that energy.  Although Rev.\ 2 reprocessing should
account for the gain changes in GIS3, better behavior of the iron line
was obtained from some spectra when the region around the iron line
was excluded from spectral fitting. After January 1996, differences
between the SIS0 and SIS1 detectors at low energies are marked.  This
is thought to be due to Residual Dark Distribution (RDD) effects
(Dotani\markcite{26} 1998), and this problem most seriously affects
the spectra below about 0.8-1.0~keV. It is thought that this effect
more seriously affects the SIS1 detector because of its average higher
temperature.  Therefore, during observations from this time period,
the SIS1 spectra below 0.8~keV were ignored, and GIS spectra below
1.0~keV were ignored.

The spectra from each observation were fit separately, except for
IRAS~13349+2438.  There were two observations of this object,
separated by about 4 days.  It has previously been shown that no
detectable spectral variability had occurred between these two
observations (Brinkmann et al.\markcite{13} 1996; Brandt et
al.\markcite{12} 1997); this was verified and the data from these two
observations were combined.  For statistical tests, the properties of
NGC~4051 were taken from the second, longer observation because of the
better statistics offered.

\subsection{ROSAT data}

Nearly all of these objects considered here had pointed {\it ROSAT}
observations.  These were reduced using {\it Xselect} and fit between
0.1--2 keV.  The intent of using {\it ROSAT} spectra in this paper is
to investigate the relationship between the slope of the soft X-ray
spectrum and the soft excess observed in the {\it ASCA} spectra.  The
poor energy resolution of {\it ROSAT} prevents robust deconvolution of
complex models.  Therefore, we quote only the results from a power law
fit with Galactic absorption, or with extra absorption if that
improves the fit at greater than 90\% confidence.

Three objects did not have pointed observations: RX~J0439$-$45,
PKS~0558$-$504 and 1H~0707$-$495.  The information for these was taken
from the {\it ROSAT} All Sky Survey data: for the first two,
information from the spectra was available in the literature and for
the last one, an estimation was made for the slope using the hardness
ratios available in the {\it ROSAT} Bright Source Catalog (Grupe 1997
P. comm.). Finally, IRAS~20181$-$2244 was detected in the {\it ROSAT}
All Sky Survey (Boller et al.\markcite{8} 1992; Boller et
al.\markcite{6} 1998) but the flux was too low for it to be included
in the {\it ROSAT} Bright Source Catalog.  Therefore, no photon index
estimate is possible.

\subsection{Emission Line Data}

Table 1 lists some of the optical emission line properties from NLS1s,
obtained both from the literature and from spectra in-hand.  The
properties include the FWHM of H$\beta$, the H$\beta$ equivalent
width, the ratio of \ion{Fe}{II} to H$\beta$, the FWHM of
[\ion{O}{III}] when available, and the ratio of [O~III] to H$\beta$.
Measurement of these parameters is complicated by the fact that NLS1s
often exhibit strong emission from blends of optical \ion{Fe}{II}.
This contaminates the continuum around the H$\beta$/[\ion{O}{III}]
complex and furthermore the multiplet 42 occurs so close to
[\ion{O}{III}]$\lambda 5007$ that it can significantly distort the
profile.  One of the more effective ways to deal with this problem is
to subtract the \ion{Fe}{II} using a scaled and broadened \ion{Fe}{II}
template formed from the spectrum of the quintessential NLS1 I~Zw~1.
This method was used in the seminal paper by Boroson \&
Green\markcite{9} (1992) and has subsequently been used by several
other authors including Grupe\markcite{45} 1996 (also Grupe et
al.\markcite{50} 1999a).  While the I~Zw~1 template method does not
provide a perfect iron subtraction (e.g.\ Boroson \& Green\markcite{9}
1992), it is useful also because it provides a consistent way to
estimate the contribution of \ion{Fe}{II} in the spectra.  Seeking
consistency in Table 1 parameters, the numbers from papers in which
the \ion{Fe}{II} template method was applied are preferentially
quoted.

Spectra were available for analysis from the following objects:
PHL~1092, 1H~0707$-$495, NGC~4051, Mrk~507 (kindly provided by J.
Halpern), RE~1034+39 (kindly provided by D. Grupe), IRAS~13224$-$3809,
IRAS~20181$-$2244 (kindly provided by L. Kay), IRAS~17020+4544
(Leighly et al.\markcite{77} 1997b), Ark~564 (obtained from the {\it
HST} archive), Mrk~142 and Kaz~163.  The I~Zw~1 template method was
used to subtract the \ion{Fe}{II} and the spectra were fitted using
the {\it LINER} spectral fitting package (Pogge \& Owen\markcite{110}
1993).  Deblending of the H$\beta$ line can contribute uncertainty to
the derived parameters.  Some fraction of H$\beta$ originates in low
ionization material from the narrow-line region (NLR); in NLS1s it is
difficult to estimate how much arises from the NLR because the width
of the H$\beta$ is similar to that of the [\ion{O}{III}] lines and
therefore the narrow component cannot be easily recognized (e.g.\
Goncalves, Veron \& Veron-Cetty\markcite{43} 1998).  As it is beyond
the scope of this paper and the signal to noise of many of the spectra
to attempt a detailed deblending, I assume that the NLR contributes
H$\beta$ with $1\over 10$ the flux of [\ion{O}{III}]$\lambda$ 5007,
and I subtract that from the H$\beta$ profile before modeling.  The
assumption is based on the results obtained from narrow-line emission
from Seyfert 2 and intermediate Seyfert galaxies (e.g.\
Koski\markcite{67} 1978; Cohen\markcite{15} 1983).

The [\ion{O}{III}]$\lambda$ 5007 line was modeled using one or two
Gaussians.  Two were required in the following objects; the second
broader component was blueshifted with respect to the narrow
component: RE~1034+39 (see also Mason, Puchnarewicz \&
Jones\markcite{85} 1996 and Goncalves, Veron \&
Veron-Cetty\markcite{43} 1998), NGC~4051, IRAS~17020+4544, Mrk~507 and
IRAS~20181$-$2244 (see also Halpern \& Moran\markcite{59} 1998).
H$\beta$ and [\ion{O}{III}] line fluxes were obtained by integrating
the flux above the continuum, a procedure which has the advantage of
being independent of the line model. The H$\beta$ was finally fit with
a Lorentzian profile, which provided a substantially better
description in most cases than a Gaussian in the NLS1 spectra (also
Goncalves, Veron \& Veron-Cetty\markcite{43} 1998).  In a few cases,
subtraction of strong \ion{Fe}{II} distorted the
[\ion{O}{III}]$\lambda$ 5007 profile, making the FWHM unreliable
(1H~0707$-$495).  These results were merged with those from the
literature to obtain the values in Table 1.

\section{Spectral Properties}

The spectral analysis is limited to the time averaged spectral
properties of the sample.  A preliminary investigation of spectral
variability was discussed in Part 1.  Detailed analysis of spectral
variability has been presented already for Mrk~766 (Leighly et
al.\markcite{78} 1996) and NGC~4051 (Guainazzi et al.\markcite{53}
1996), and will be discussed for Ark~564 in Leighly et al. in prep.
 
The spectra were first fitted with an absorbed power law with the
absorption fixed at the Galactic value, and the ratio of the data to
this model are shown for all the spectra in Figure 1.  Significant
deviations from the power law in nearly all of the spectra can be seen
and the $\chi^2$ for this model are given in Table 2.  An acceptable
$\chi^2$ ($\chi^2_\nu<1.3$) was obtained from only 10 of the 24
spectral fits. Because of the wide range in fluxes represented by this
sample, an acceptable fit may be obtained either when the model is
adequate or when the statistics are poor.  The shape of the residuals
is generally concave upward and thus appears quite different than many
of those from Seyfert galaxies with broad optical lines which tend to
be modified by absorption (e.g.\ Reynolds\markcite{116} 1997).

To investigate the origin of the residuals, initially two phenomenological
models were tried.  These consisted of a power law with a
photon index $\Gamma$ and either a soft excess modeled as a black body
with characteristic temperature $kT$ (the ``soft excess'' model) or an
ionized absorber modeled as two absorption edges at $0.74$ and
$0.87\,\rm keV$ in the galaxy rest frame (the ``warm absorber''
model).  These models have the same number of degrees of freedom. The
$\chi^2$ for these preliminary models are given in Table~2.  The fit
was improved substantially for 20 of the 24 spectra; only spectral
fits to I~Zw~1 and Kaz~163 were not improved at all, and the fit to
IRAS~20181$-$2244 was improved only marginally.  (Note that the
following criteria were used to evaluate the spectral fits: the
improvement was {\it significant} ($P<0.01$) when $\Delta\chi^2>6.34$,
9.21 and 11.3 for 1,2 and 3 additional parameters respectively; the
improvement was marginal ($P<0.05$) when $\Delta\chi^2>3.84$, 5.99,
and 7.82 for 1,2 and 3 additional parameters respectively.)  Nineteen
of the spectra (18 objects) are modeled substantially better by the
soft excess model ($\Delta
\chi^2=10$, corresponding to a likelihood ratio of 140, e.g.
Mushotzky\markcite{92} 1982), and the fit was acceptable
($\chi^2<1.3$) for 18 of these (all but 1H~0707$-$495).  Three objects
(IRAS~13349+2438, IRAS~17020+4544, Mrk~507) were fit substantially
better using the warm absorber model, and IRAS~20181$-$2244 was fit
marginally better ($\Delta\chi^2=6$). 

Some ambiguity remains after this preliminary modeling, as it is
possible that the two-edge model may not be able to
adequately 
describe the warm absorber.  Therefore, a third preliminary model was
applied, consisting of a power law, Galactic absorption and the {\it
absori} model available in {\it XSPEC}; the resulting $\chi^2$ are
listed in Table 2.  The {\it absori} model describes the warm absorber
and is characterized chiefly by two parameters: the column density
$N_w$ and the ionization parameter $\xi$.  The temperature of the
ionized plasma is also an adjustable parameter; however, because the
spectral fitting results are not very sensitive to this parameter, the
temperature was fixed at either $3
\times 10^4\rm\, K$ or $1 \times 10^6\rm\, K$, and the lower value of
$\chi^2$ is quoted. Note that the {\it absori} model does not include
resonance line absorption.  In 18 spectra from 17 objects, the $\chi^2$
for the {\it absori} model is more than 10 less than the $\chi^2$ for
the two-edge model, indicating a likelihood ratio $>140$.  In many of
these objects, detailed analysis indicates the presence of a warm
absorber, as discussed below.  Alternatively, the ionization parameter
becomes very low and neutral absorption is modeled (Mrk 507;
IRAS~20181$-$2244).  However, for the others, the ionization parameter
becomes pegged at the maximum value of $\xi=5000$, where the model is
not considered to be very accurate (P.\ Magdziarz 1997, P.\ comm.).
When this is the case, the {\it absori} model appears to be trying to
model the continuum.  However, a soft excess models the continuum
better. In nearly all cases, the difference in $\chi^2$ between the
soft excess and {\it absori} models is larger than 10, indicating a
likelihood ratio of $>140$.   It is less than 10 only for
Mrk~142; this is likely to be the fault of poor statistics, as the
source is relatively faint and the exposure for the SIS0 is only
19~ks.  Further evidence that the complexity of the spectrum is better
described by a soft excess model is given by the lower panels in
Figure 1.  These show the ratio of the data to a power law (plus iron
line as necessary) model fit above 2--3 keV, after the iron line
component is removed from the model.  The signature of absorption
should be characterized by a deficit or localized features upon the continuum;
in contrast, in many of the objects, a broad continuum excess is seen.

These preliminary results show that a soft excess is generally present
in narrow-line Seyfert 1 galaxy {\it ASCA} spectra.  This situation
contrasts with that from Seyfert galaxies with broad optical emission
lines; Reynolds\markcite{116} (1997) finds that the warm absorber
fitted most of the spectra adequately, and a soft excess was required
in only 3 cases.

These simple models do not adequately model all of the spectral
residuals; evidence is also present for iron $\rm K\alpha$ lines and
additional absorption.  I follow Reynolds\markcite{116} (1997) and
proceed to fit with a phenomenological model with the following
components, where all parameter values are obtained in the rest frame:

\begin{itemize}
\item a power-law with photon index $\Gamma$ which models the high
energy continuum;
\item a soft excess modeled by a black body with temperature $kT$,
which provides a generic model of the soft excess;
\item two absorption edges with energies 0.74 and 0.87~keV
in the rest frame of the galaxy, which represent absorption by O~VII
and O~VIII in ionized gas;
\item Galactic absorption as listed in Table 3;
\item intrinsic neutral absorption in the rest frame of the galaxy;
in some cases, information from nonsimultaneous {\it ROSAT} spectra
is used to constrain this parameter.
\item an iron $K\alpha$ line, represented by a Gaussian with central
energy $E$ and width $\sigma$.  
\end{itemize}

The results are listed in Tables 3a, b, and c.  Note that the values of
the parameters and their errors ($\chi^2=2.71$, that is, 90\% for one
parameter of interest) are listed only when there is a significant or
marginal detection of that component.  For the absorption edges and
iron lines, upper limits are also listed. 

Because the two-edge model may not adequately model the warm absorber
in all cases, a complementary set of fits using the {\it absori} model
was performed, and the results are listed in Table 4.  These results
are discussed in Section 3.3.

After applying this simple phenomenological model uniformly to the
spectra, evidence for further complexity was discovered in 5 objects. 
There is additional evidence for an edge-like structure near 1~keV in
1H~0707$-$495, IRAS~13224$-$3809 and PG~1404+226.  These features were
discussed extensively in Leighly et al.\markcite{79} 1997a; these
features were modeled as edges here and will not be discussed.  An
additional soft X-ray emission line feature is seen around 1~keV in
PG~1244+026 and Ark~564.  These features will be discussed in Section 3.4.

\subsection{The Photon Index in NLS1s}

The photon index from narrow-line Seyfert 1 galaxies appears to be on
average steeper than that from Seyfert galaxies with broad optical
lines.  This result was first reported by Brandt, Mathur \&
Elvis\markcite{11} (1996), and it is confirmed by the spectra in this
sample.  Histograms of the photon indices for the NLS1s and for
broad-line Seyfert 1 galaxies taken from Reynolds\markcite{116} (1997)
are plotted in Figure 2.  The broad-line radio galaxies and 3C~273 are
excluded; radio-loud objects often have flatter spectra than
radio-quiet objects, possibly as the result of a flat component from the
jet.  Note that all of our narrow-line Seyfert 1 galaxies are radio
quiet except PKS~0558$-$504 (Remillard et al.\markcite{115} 1986).
Also note that with or without the radio-loud objects, the
distributions of the 2--10 keV luminosities of the broad and
narrow-line AGN are consistent as determined by a two-sample KS test.
Figure 2 shows clearly that the photon indices are consistently
steeper for the NLS1s.  To test whether the photon indicies could have
been drawn from the same parent population, the statistical package
{\it ASURV} Rev. 1.2 (Isobe \& Feigelson\markcite{61} 1990) which
implements methods presented in Feigelson \& Nelson\markcite{33} (1985)
was used.  A variety of tests are available in this package which make
different assumptions about the underlying distributions.  The
differences were found to be highly significant ($P<0.0001$)
regardless of the assumed distribution.

The histogram shows an apparent overlap and spread in the values for
both types of objects; however, it is difficult to determine whether
this is significant from the histogram alone since the uncertainties
in the photon indices are not taken into account.  Therefore, the
maximum likelihood method was used to determine jointly the best
estimate of the mean and dispersion of both sets of indices (Figure 2;
e.g.\ Maccacaro et al.\markcite{82} 1988).  These were $2.19 \pm 0.10$
and $0.30^{+0.07}_{-0.06}$ for the NLS1s and $1.78 \pm 0.11$ and
$0.29^{+0.09}_{-0.07}$ for the broad line objects (errors are 68\% for
two parameters of interest).  In both cases the dispersions are
significantly greater than zero (99\% confidence $>0.19$ and $0.17$
for the narrow and broad-line objects respectively; Figure 2).  A
significant spread in Seyfert galaxy X-ray 2--10 keV photon indices
has been recognized previously by several investigators (e.g.\ Turner
\& Pounds\markcite{133} 1989; Nandra \& Pounds\markcite{96} 1994).
The contours representing the mean and dispersion exclude each other
at greater than 90\% significance. The validity of the maximum
likelihood test depends on the assumption that the photon indices are
normally distributed. This assumption is justified if the
distributions are symmetric; however, the distribution of the
broad-line objects appears to be significantly skewed ($S=-0.82\pm
0.50$). There is an interesting suggestion that high values are
preferred for both types of objects, and the distributions tail toward
the lower values.

\subsection{The Soft Excess in NLS1s}

A soft excess component is required to model 18 of the 24 spectra (17
of 23 objects).  The most striking difference is in the degree of
prominence of the soft excess in the spectra.  This can be seen in the
lower panels of Figure 1, which show the ratio of the data to a power
law plus iron line (as necessary) model above 2 keV, and extrapolated
to lower energies.  This figure shows that the soft excess component
in NLS1s stretches well into the {\it ASCA} band pass and that it can
dominate the spectrum up to at least 1~keV.  This situation contrasts
with that from broad-line Seyfert galaxies; in that case, the soft
excess is seen in only in the lowest channels in the {\it ASCA} band,
if at all (e.g.\  Guainazzi et al.\markcite{52} 1994).

The measured temperatures of the soft excess component in the galaxy
rest frame when modeled by a black body range from 0.1 to 0.25 keV.
The maximum likelihood method gives an average of $0.15\pm 0.05 \rm\,
keV$ and dispersion of $0.12^{+0.09}_{-0.05}\rm\, keV$ (uncertainties
are 68\% confidence).  The dispersion is significantly greater than
zero at $>99$\% confidence; however, it must be kept in mind that
there may be some model dependence in the temperatures.  The measured
temperature is likely to be correlated with the energy of the lowest
spectral channel, which is not the same for all of the spectra.
Observations from later in the mission generally had a higher event
threshold than observations from early in the mission, and some
observations were performed with a level discriminator.  Also, the
objects represent a range of redshifts.  Furthermore, application of
warm absorber edges and additional absorption can alter the measured
temperature.  Finally, physical models for the soft excess (e.g.\ an
accretion disk spectrum) are likely to be characterized by a range of
temperatures, although the inner edge temperature may be well defined.

To study the variation in strength of the soft excess, I borrowed the
hardness ratio plot technique used in the study of Galactic X-ray
sources.  Figure 3 shows the ratio from the summed SIS0 and SIS1
spectra of the medium (1.0--2.0~keV) to soft (0.6--0.9~keV) energy
bands on the x axis, and the ratio of the medium to the hard
(2.0--6.0~keV) on the y axis.  There are two problems with applying
this method directly.  The first is that the Galactic column densities
differ among the sample enough to seriously affect the flux in the
soft band.  Therefore, before accumulating the ratios, the spectra were
corrected channel by channel for Galactic absorption using the
analytic formulation from Morrison \& McCammon\markcite{90} (1983).  The
second problem is that the redshifts of the sources differ
significantly.  This can be corrected using the {\it ASCA} SIS0
effective area curve; however, the correction becomes increasingly
less accurate as the redshift increases as large changes in
the effective area become shifted from one band to the other.  The
accuracy of this method is checked by simulating the spectra using the
best fit models shifted to zero redshift.  The result is that the
correction gives acceptable results for all but the two highest
redshift objects (PHL~1092 and RX~J0493$-$45, with redshifts 0.396 and
0.224 respectively); therefore they are not included in the top panel
of Figure 3.  Finally, the hardness ratios are made from the
luminosities predicted by the best fit models with the Galactic $N_H$
removed.  These are shown in the bottom panel of Figure 3.  The
advantage here is that no $N_H$ or redshift correction is necessary;
the disadvantage of this is that the luminosities are model dependent.
Thus the top and bottom panels are complementary.

In Figure 3, the x axis, being the ratio of the medium to soft
X-rays, displays the strength of the soft excess, where objects with
{\it strong} soft excesses are on the {\it left} side of the plot and
correspond to {\it lower} values of the ratio. Henceforth, this ratio
will be referred to as the ``color ratio''. Six points marked by
stars have very strong soft excesses: PHL~1092, RX~J0439$-$45,
1H~0707$-$495, RE~1034+39, IRAS~13224$-$3809 and PG~1404+226.  This
group will be collectively referred to a ``strong soft-excess
objects''. In the middle of the plot marked by circles are all the
other objects with detectable soft excesses; note that NGC~4051 (2
observations) and Mrk~766 are separately marked with open circles.
These will be referred to as ``weak soft-excess objects''.  Note that
although the strength or contrast of the soft excess is weak in these
objects, the soft excess is required with high confidence in the
spectral fit. On the left side of the plot marked by triangles are
objects with no detected soft excess: I~Zw~1, IRAS~13349+2438,
IRAS~17020+4544, and Kaz~163.  Note that absorption may be hiding a
weak soft excess in some of these objects.  Finally, the more heavily
absorbed objects Mrk~507 and IRAS~20181$-$2244 are off the plot toward
the right.

The y axis on this graph, displaying the ratio of the 1.0-2.0~keV
medium band to the 2.0--6.0~keV hard band, illustrates the steepness
of the hard X-ray spectrum or power law, where large values correspond
to steeper spectra.  This assertion was verified by comparing the
ratios with the measured photon indices.  This graph shows that the
strong and weak soft excess objects have almost the same distribution
of photon indices.  This means that the large range of soft excess
strengths cannot be attributed to differences in the slope of the
power law, but rather must be due to differences in the relative
normalizations of the soft excess component and the power law.

The color ratio, as defined above, provides a measure of the strength
of the soft excess.  However, since only the effect of the Galactic
absorption has been removed, the color ratio is also sensitive to
intrinsic absorption, either ionized or neutral.  Two other
complementary parameters to measure the strength of the soft excess
can be defined.  These have the advantage that they should not be
affected by the presence of intrinsic absorption; the disadvantage is
that they are model dependent.   The first, called the ``black body
parameter'' is defined as the model flux color ratio between the
black body component and the power law at 0.45~keV plus the
measured temperature of the black body in the rest frame
[$(F(pl)-F(bb))/(F(pl)+F(bb))$].   This parameter describes the
prominence of the black body over the power law and it should be
approximately directly proportional to the color ratio in the case
where there is no intrinsic absorption.  The values of the black
body parameter are listed in Table 3.  Because this parameter
potentially depends most sensitively on the power law index, the
uncertainties were evaluated by fitting the spectra with the power law
index fixed at its 90\% error limits.  Values near $-1$ and $1$ imply
very strong and very weak soft excesses, respectively.  The black body
parameters range from $-0.72$ for IRAS~13224$-$3809 to $0.77$ for
PKS~0558$-$508 (not counting objects with no detectable soft excess,
for which the black body parameter equals 1).  These values imply that
the black body component is a factor of $\sim 50$ stronger in
IRAS~13224$-$3809 than in PKS~0558$-$508.  

The final parameter is $\alpha_{xx}$.  Defined as the slope between
0.7 and 4~keV in the rest frame, it measures the overall steepness of
the continuum spectrum.  This slope is computed from the modeled
continuum and therefore should also be independent of intrinsic
absorption; however, it is model dependent.  This parameter
indirectly measures the strength of the soft excess: when there is no
soft excess or it is weak, $\alpha_{xx}$ is close to the photon index;
when the soft excess is strong, $\alpha_{xx}$ is much larger than the
photon index.

\subsection{Neutral and Ionized Absorption in NLS1s}

Evidence for an ionized ``warm'' absorber was found in more than half
of a sample of {\it ASCA} spectra from bright, predominantly
broad-line Seyfert galaxies (Reynolds\markcite{116} 1997; also George
et al.\markcite{40} 1998), a result which supports the idea that
absorption by highly ionized material may be ubiquitous in Seyfert
galaxies.  In contrast, significant detections of a warm absorber as
modeled by two absorption edges with fixed energies were found in only
6 spectra from 5 of the 23 objects; marginal detections were obtained
from 5 others.  It is conceivable that this is a consequence of the
low flux level of our objects; however, there is also the impression
that when a warm absorber is present, it has a lower optical depth,
especially in the higher ionization, O~VIII edge.  The detections and
upper limits are plotted in Figure 4, along with the results from
broad-line AGN given in Reynolds\markcite{116} (1997).  Note that
several of Reynold's detections were changed to upper limits for
consistency with the detection criteria adopted here.  This figure
shows that many of the O~VIII upper limits from the NLS1s are below
the detections from the broad-line objects.

The possibility of a difference in warm absorber parameters for the
broad and narrow-line objects was investigated by running statistical
tests on the measured edge depths and upper limits from the sample of
NLS1s presented here and from the comparison sample of broad-line AGNs
from Reynolds\markcite{116} (1997).  Note that because the warm
absorber is not detected in all objects, the upper limits should be
taken into account properly using survival analysis statistical
methods (e.g.\ Feigelson \& Nelson 1985); and here they were accounted
for using the Kaplan-Meier estimator as implemented in the {\it ASURV}
package.  The resulting average optical depths for O~VII and O~VIII
are $0.19\pm 0.04$ and $0.053 \pm 0.020$ respectively for the NLS1s,
and $0.29\pm 0.07$ and $0.18 \pm 0.06$ for 21 broad-line objects.
Note that in the algorithm when the minimum optical depth is an upper
limit, it must be changed to a detection; this occurred in both data
sets so the mean could be biased.  Nevertheless, these results suggest
that the O~VII optical depths are consistent between broad and
narrow-line objects, while the O~VIII optical depths are smaller for
the narrow-line objects compared with the broad.  To test this
further, {\it ASURV} two-sample tests which assume a variety of
different intrinsic distributions were run.  No difference between
O~VII optical depths was found (probability of a significant
difference of 29--43\%; a range is found for the different assumed
intrinsic distributions), but a difference between the O~VIII optical
depths at 97--98\% confidence was found.  Reynolds\markcite{116}
(1997) found some evidence for a difference between high and low
luminosity objects that could possibly be attributed to different
behavior of radio-loud objects.  If the radio-loud objects are
excluded from the broad-line sample, there is still no difference in
$\tau_{O\,VII}$, but the significance of the difference in
$\tau_{O\,VIII}$ becomes $>99$\%.  Finally, warm absorber in the
broad-line Seyfert 1 galaxy NGC~3783 has a very high optical depth; it
is possible that this galaxy skews the results.  If it is excluded, no
difference in $\tau_{O\,VII}$ is obtained, but there is still a
significant difference in $\tau_{O~VIII}$ of $>98$\%.

The fact that the average O~VIII optical depth is lower, that many of
the upper limits on the NLS1 $\tau_{O~VIII}$ are smaller than the
broad-line object detections, and that there is an apparently robust
significant difference in the distribution of $\tau_{O~VIII}$ suggests
that the ionization state in NLS1s is lower on average than in
broad-line Seyfert galaxies.  A better way to test this supposition
would be to compare ratios of edge optical depths; however, it is
difficult to determine a sensible way to derive the ratio of the edges
when both values are upper limits.  However, the joint distributions
of $\tau_{O~VII}$ and $\tau_{O~VIII}$ can be compared using a two
sample test for multivariate data containing upper limits (Makuch,
Escobar \& Merrill\markcite{84} 1991).  Testing against the
hypothesis that the joint $\tau_{O~VII}$ and $\tau_{O~VIII}$
distributions are equal, I find mixed results.  If the generalized
Gehan distribution is appropriate, the warm absorber properties from
the NLS1s are not significantly different from those of the broad-line
Seyferts at 84--91\% confidence.  However, if the logrank distribution is
appropriate, indications are that the difference may be significant
(95--98\%). Since it is not clear which distribution is
appropriate, I conservatively conclude that there is a suggestion rather
than a discovery that the distribution of warm absorber properties
{\it as a whole} is different in NLS1s than broad-line AGN. However,
the difference in distribution of the \ion{O}{VIII} optical depths
remains significant.

All of the spectra were also fit with warm absorber modeled using the
{\it absori} model in {\it XSPEC}, and the results are listed in Table
4.  A warm absorber was detected using this model in all of the
objects in which one was detected using the two-edge model.  A warm
absorber was detected with low significance in several other objects
as well (marginal detections in PG~1244+026, IRAS~13224$-$3809,
Kaz~163; $\Delta\chi^2=10$ in Ark~564).

Additional neutral absorption in the rest frame of the galaxy was
detected in 4 objects: I~Zw~1, IRAS~17020+4544, Mrk~507 and
IRAS~20181$-$2244.  I~Zw~1 is a moderately bright object, and thus
since the detection of absorption is not very significant, the
necessity of this component is somewhat doubtful. Furthermore there is
marginal detection of a warm absorber and the two components are
probably coupled in the spectral fitting.  Significant absorption is
indicated in the {\it ROSAT} spectrum, but at a much lower level
($N_H=2.3\times 10^{20}\rm\,cm^{-2}$).  Further broad band
observations are are necessary to determine whether this is
significant or not.  The absorption in Mrk~507 has been discussed by
Iwasawa, Brandt \& Fabian\markcite{62} (1998); it is likely to be the
reason that the {\it ROSAT} photon index is so flat in this object.
Significant neutral absorption as well as ionized absorption was
reported in IRAS~17020+4544, and has been linked to the high optical
polarization in this object (Leighly et al.\markcite{77} 1997b).  The
absorption in IRAS~20181$-$2244 has been discussed by Halpern \&
Moran\markcite{59} (1998), and is probably the reason that this object
was so faint in the {\it ROSAT} All Sky survey. 

It is important to note that soft excesses were not detected in
several of the objects with warm or neutral absorbers. This may be
because weak soft excesses cannot be distinguished when there is also
absorption.  Thus, soft excesses may be even more prevalent in NLS1s
than is indicated here.  It is possible that the presence of a soft
excess can also make detection of a warm absorber more difficult,
since the absorption structure may be easier to see when the continuum
is a power law.  However, the moderate energy resolution provided by
{\it ASCA} should be adequate to allow us to detect features from the
warm absorber even if they fall upon a multicomponent continuum.

Leighly et al.\markcite{77} 1997b discovered that highly polarized
Seyfert 1 galaxies almost always had a warm absorber.  This result
supports the hypothesis of dusty warm absorbers (Brandt, Fabian \&
Pounds\markcite{10} 1996; Reynolds et al.\markcite{118} 1997); that is,
dust which reddens and polarizes the optical emission is coincident
with ionized gas.  Several of the NLS1s discussed here have high
optical polarization, including I~Zw~1 (Smith et al.\markcite{127}
1997), NGC~4051, Mrk~766 (Goodrich\markcite{44} 1989), IRAS~13349+2438
(Wills et al.\markcite{138} 1992), IRAS~17020+4544 (Leighly et
al.\markcite{77} 1997b), Mrk~507 (Goodrich\markcite{44} 1989) and
IRAS~20181$-$2244 (Kay et al.\markcite{64} 1999).  All of these
objects show evidence for either neutral or ionized absorption.  The
following objects have weak or no polarization: Mrk~335, Mrk~142,
PG~1211+143, PG~1244+026, PG~1404+226, Mrk~478 (Berriman et
al.\markcite{5} 1990), IRAS~13224$-$3809 (Kay et al.\markcite{64}
1999), and Ark~564 (Goodrich\markcite{44} 1989).  There is either no
or only marginal evidence for a warm absorber in these, except for
Mrk~478.  Polarization properties of the following objects have not
yet been reported: Ton~S180, PHL~1092, RX~J0439$-$45, NAB~0205+024,
PKS~0558$-$504, 1H~0707$-$495 and Kaz~163.  Thus the NLS1s appear to
show the same trend of absorption and polarization as we found for
Seyfert galaxies in general (Leighly et al.\markcite{77} 1997b).

\subsection{Soft X-ray Features in NLS1s}

Features resembling absorption edges near 1~keV were found in the {\it
ASCA} spectra from the three NLS1s: 1H~0707$-$495, IRAS~13224$-$3809
and PG~1404+226. This work is discussed in detail in Leighly et
al.\markcite{79} 1997a.  These features could be fit either by a one
or two absorption edges or two or three unresolved absorption lines.
Oxygen offers the primary opacity in photoionized gas; however, the
energies of the absorption features are much too high to be attributed
to oxygen absorption.  If interpreted as absorption due to oxygen, the
energies of these features imply a high blueshift of the absorbing
material: 0.2--0.3~c for the edge model and near 0.57~c for the line
model.  It is difficult to accelerate ionized gas to such high
velocities; however, the notable similarities between NLS1s and
low-ionization broad absorption line quasars (BALQSOs) prompt us to
take this hypothesis seriously.  Both NLS1s and BALQSOs show strong or
extreme \ion{Fe}{II} and weak [\ion{O}{III}] emission; many NLS1s and
low-ionization BALQSOs have red optical spectra and strong infrared
emission; finally, both classes are predominantly radio-quiet.  See
Leighly et al.\markcite{79} 1997a for more details.  An alternative
interpretation in terms of a highly ionized warm absorber dominated by
resonance absorption lines dominated by Fe L has been proposed by
Nicastro, Fiore \& Matt\markcite{100} 1999.  For this model, it is
probably relevant that these absorption features near 1~keV are found 
exclusively in the objects with the strongest soft excesses in the
sample.  

The spectra of two other NLS1s, PG~1244+226 and Ark~564 revealed
excess emission near 1~keV (Figure 1).  The presence of this feature
in PG~1244+226 has been discussed by Fiore et al.\markcite{34} 1998.
The residual can be most simply modeled as an unresolved Gaussian
emission line.  The features are clearly significant {\it in addition}
to a weak soft excess component.  This is demonstrated by energy
versus flux $\chi^2$ contours shown in Figure 5, where the narrow
width has been fixed at the best fit value to better constrain the
fit; when the line width is left free, it becomes confused with the
soft excess when at large fluxes and widths.

Fiore et al.\markcite{34} 1998 discuss possible origins of this line.
Emission in an optically thin thermal plasma is  highly
unlikely because the large luminosity requires a large amount of
gas, and the rapid variability implies a small emission region. The
combination of these two constraints implies the gas must be optically
thick, in contrast with the assumption that the gas is optically thin.
Rather, they support the hypothesis that this feature may arise from
reflection in a photoionized accretion disk.  This hypothesis is
supported by the idea that NLS1s are accreting at a higher fraction of
the Eddington rate, since under that condition, the accretion disk is
expected to become ionized (e.g.\ Matt, Fabian \& Ross\markcite{87}
1993a).  The energy near 1~keV supports interpretation as a blend of
\ion{Ne}{IX} and \ion{Ne}{X} at 0.92 and 1.02 keV respectively, as
well as \ion{Fe}{L} lines at around 0.8 keV.  The lack of a \ion{O}{VIII}
K$\alpha$ line near 0.65 keV is puzzling however as it is expected to
be strong in photoionized plasmas (e.g.\ Netzer\markcite{98} 1996).

Fiore et al.\markcite{34} 1998 also note that in the Matt et
al.\markcite{87} (1993a) models, the disk ionization parameter $\xi$
depends on the accretion rate as $\dot m^3$.  The sensitivity of this
dependence means that while the accretion rate may be large in many
NLS1s and therefore while the disk may be ionized in many NLS1s,
observation of these features would not necessarily be expected in all
of them.  Note that the $\xi \propto \dot m^3$ dependence is based on
several assumptions; two primary ones are that the fraction of
accretion energy going into X-rays which illuminate the disk remains
the same as the accretion rate increases and that the geometry also
remains the same.

\subsection{The Iron Line in NLS1s}

An iron K$\alpha$ line with energy consistent with emission from
neutral or low-ionization-state iron is ubiquitous in {\it ASCA}
spectra from low-luminosity broad-line Seyfert 1 galaxies (e.g.\ Nandra
et al.\markcite{94} 1997a; Reynolds\markcite{116} 1997).  Frequently the
line is found to be significantly broader than the instrument
resolution (e.g.\  Mushotzky et al.\markcite{93} 1995).  The signal to
noise is sufficient in at least one observation that the line profile can
be examined; a clearly identified red wing provides the best evidence
so far for emission from a relativistic accretion disk near a black
hole (MCG--6$-$30$-$15; Tanaka et al.\markcite{131} 1995).  At higher
luminosities, however, the iron line appears to be less frequent, and
there is some indication that the characteristic ionization state of
the iron is observably higher (Nandra et al.\markcite{95} 1997b).

It is important to compare the iron line properties from narrow-line
Seyfert 1 galaxies with those from broad-line Seyfert galaxies in
order to place constraints on the geometry and importance and
conditions of reprocessing.  This is a difficult task using these
data, however.  The low flux combined with the steepness of the
spectrum means that the signal to noise at the iron line is generally
very poor.  In fact, some of the sample were not significantly
detected above 6.0 keV.  For this reason, no useful constraints could
be obtained from the following: 1H~0707$-$495, PG~1404+226 and
RE~1034$-$3809.

To examine the iron line in the remaining 20 objects, the spectra
above 2~keV were fit, rather than the whole band pass, so that
potential broad band spectral curvature would minimally
affect the results.  Some evidence for an iron line was found in 13 of
the 20 objects detected in the iron line energy range.  An iron line
was not statistically necessary in the models for PHL~1092,
RX~J0439$-$45, NAB~0205+024, Mrk~142, Mrk~507, Kaz~163 and
IRAS~20181$-$2244.  It is interesting to note that in PHL~1092 and
IRAS~20181$-$2244 an iron line is found just below the detectability
threshold (PHL~1092: $\Delta\chi^2=5.3$; IRAS~20181$-$2244:
$\Delta\chi^2=5.6$) and in both cases the best fitting line energy is
high, near 6.8--6.9~keV. An upper limit was obtained from the
nondetections by fitting with an narrow Gaussian ($\sigma=0.05\rm\,
keV$).  The rest frame energy was fixed at either the best fitting
energy or at $6.4$ and $6.7$ keV and the result from the better fit
was reported.  These results show that the upper limits on the
equivalent widths for these 7 objects are commensurate with those from
the detections, and the line fluxes are on the low end of the
distribution.  This implies that lines were not detected in these
objects simply because the statistics are poor.

In 6 spectra from 6 objects the uncertainty on the width of the line
excludes zero at 90\% confidence for 1 parameter of interest.
$\chi^2$ contours were constructed from fits in which the line energy,
width and flux were allowed to vary and a large region of parameter
space was explored.  These show that for two parameters of interest,
the iron line is broad at the 90\% confidence level in all 6 objects:
I~Zw~1, Ton~S180, NGC~4051, Mrk~478, IRAS~17020+4544 and Ark~564.  Not
surprisingly, those are among the brightest objects at high energies.
Interestingly, the line was unresolved in Mrk~335 and Mrk~766, both
bright objects.  Evidence for both broad and narrow iron lines was
found in the first observation of NGC~4051.

In 6 of the spectra, the best fit line energy and 90\% confidence
interval for one parameter of interest exclude neutral iron emission
at 6.4~keV.  However, note that the line detection is marginal or
nearly marginal in 4 of these.  $\chi^2$ contours show that, for two
parameters of interest, the line energy excludes 6.4~keV in only two
of the objects (I~Zw~1 and Ton~S180) at greater than 90\% confidence
level when the line was allowed to be broad.  The line energy in
Ton~S180 excludes 6.4~keV at 99\% confidence (Figure 6; see below).
The line energy excludes 6.4~keV when the line width was constrained
to be narrow in three other objects: PKS~0558-504, PG~1244+026, and
IRAS~13349+2438.

The average equivalent width, evaluated using the Kaplan-Meier
estimator in {\it ASURV} which takes into account upper limits, was
computed using the results from the 20 objects in which the iron line
properties could be investigated.  The resulting average was $352 \pm
77\rm\, eV$.  This is consistent with the average of $362
\pm 56 \,\rm eV$  for  radio-quiet broad-line objects
from Reynolds\markcite{116} (1997).  The distributions of equivalent
widths are not found to be significantly different using two sample
tests implemented in {\it ASURV}.  There are some difficulties
with a direct comparison, however.  A difference in equivalent width
should be expected based solely on the difference in photon index: the
equivalent width of the line is expected to be $\sim 20$\% larger when
the photon index is 1.9 compared with 2.3, assuming a face-on viewing
angle, because there are relatively fewer photons above the iron
K-edge in the steeper spectrum (George \& Fabian\markcite{39} 1991).
The equivalent widths of detected lines may be artificially large in
the NLS1s because the statistics are poorer in those spectra (a
Malmquist bias).  The energies of some of the lines are consistent
with emission from ionized iron and under these conditions the
fluorescence yield can be either enhanced or depressed depending on
the ionization state (e.g.\ Matt, Fabian \& Ross\markcite{86} 1993b).

Since the equivalent width relative uncertainty is approximately the
same as the line flux relative uncertainty (Yaqoob\markcite{138}
1998), a maximum likelihood test on the 12 significant and marginal
detections could be done.  This revealed a significant dispersion on
the equivalent widths at 90\% confidence for two parameters of
interest: the mean and dispersion are $270^{+140}_{-110}\,\rm keV$ and
$190^{+130}_{-110}\rm\, keV$, respectively.

The inclination of the disk is also important.  The five of the six
objects in which a broad line was detected with 90\% confidence were
fit with a model representing a emission line emerging from a planar
geometry near a Schwarzschild black hole (Fabian et al.\markcite{32}
1989); note that the detection in Mrk~478 was too nearly marginal to
provide any useful constraints).  Poor statistics prevented many of
the parameters from being constrained; therefore, the inner and outer
radii were fixed at 6 and 1000 gravitational radii, respectively, and
the emissivity parameter was fixed at its best fit value between $-2$
and $-3$.  The results shown in Figure 1 illustrate that useful
constraints could only be obtained for NGC~4051.  Since the
inclination is determined by the presence of a blue wing, the steep
spectrum, conspiring with the low flux, make the inclination
especially difficult to constrain in {\it ASCA} data from NLS1s.

Two objects present particularly interesting iron line behavior.
I~Zw~1 has a strong hard excess in its spectrum (Figure 7) which
appears to be a very large equivalent width iron line.  This result is
interesting in light of the fact that I~Zw~1 is the prototype
narrow-line quasar with very strong optical \ion{Fe}{II} emission
(e.g.\ Phillips\markcite{106} 1976).  The iron emission in the UV is
also very strong (Laor et al.\markcite{72} 1997b).  Thus it may not be
surprising that the iron emission in X-rays is also strong, although
if it is connected with the optical and UV results it may imply a
nucleus-wide overabundance of iron, since the emission mechanisms for
\ion{Fe}{II} and iron K$\alpha$ are much different.  Ton~S180 has the
clearest example of a broad, ionized iron line among the objects in
our sample.  Such a line was previously detected in the {\it SAX} data
on this object (Comastri et al.\markcite{18} 1998).  The $\chi^2$
contours of the energy versus the width are shown in Figure 6.

In summary, it appears that the equivalent width of iron line is the
same in NLS1s as in broad-line objects; this would imply that the
covering fraction of optically thick material is the same in both
types of objects.  This result is severely limited by the statistics
of the spectra, however.

\subsection{Correlation Analysis}

This paper and Part 1 present the X-ray spectral and variability
results as well as some optical emission line properties from a
sample of narrow-line Seyfert 1 galaxies observed by {\it ASCA}.  It
is appropriate to search for correlations among these properties.
Nonparametric correlation coefficients are more robust against
outliers and more appropriate for these data which are not expected to
be distributed normally.  The Kendall's Tau statistic was used (e.g.
Press et al.\markcite{113} 1992), primarily because there is an
implementation with reliable confidence estimates available in {\it
ASURV}.  Tests for correlations were also done using the more familiar
Spearman rank statistic, and the results were not substantially different.

Two sets of correlations were performed.  One set included data from
all 23 objects, and the other used data from a subsample consisting of
only the 17 objects that possessed an observable soft excess.  There
were two observations of NGC~4051, and the data from the second
observation was used for the correlations owing to the longer exposure
and better statistics. The correlation matrix, listing the
probabilities that the data are uncorrelated, is given in Table 5.
Plots of some of the correlations are given in Figure 8.  I refer to
those correlations having probabilities of no correlation of $<0.06$
to be {\it significant} or suggestive, and those with probabilities of
$<0.01$ to be {\it strong}.  Significant and strong correlations are
marked in Table~5 by open and solid circles respectively.

Correlations between the optical emission line properties, including
H$\beta$ FWHM, H$\beta$ equivalent width, the ratio of \ion{Fe}{II} to
the H$\beta$ emission and the ratio of [\ion{O}{III}] to H$\beta$
emission, were examined.  Note that detection of a significant
correlation between H$\beta$ FWHM and any parameter is somewhat
surprising, considering that the sample of objects is selected on the
basis of small H$\beta$ FWHM and therefore the dynamic range for this
parameter is small.  A strong correlation was found between the
H$\beta$ FWHM and the H$\beta$ equivalent width.  This correlation has
been discussed by Goodrich\markcite{44} (1989) and references therein,
and indicates a connection between the kinematics of the broad-line
region and the physical conditions within the emitting gas.  This
correlation was also found to be strong in a sample of optically
selected NLS1s observed by {\it ROSAT} that was considered by
Forster\markcite{36} (1998).  Correlations were also performed between
the \ion{Fe}{II} equivalent width, which can be approximated as the
product of the
\ion{Fe}{II}/H$\beta$ ratio and the H$\beta$ equivalent width,
assuming the continuum is approximately flat. An anticorrelation
between \ion{Fe}{II} equivalent width and H$\beta$ FWHM has been
previously reported by many investigators including Zheng \&
O'Brien\markcite{141} 1990, Puchnarewicz et al.\markcite{114} 1992,
Grupe et al.\markcite{50} 1999a, and Boroson
\& Green\markcite{9} 1992.  Interestingly, I find a positive {\it correlation}
between these properties.  Forster\markcite{36} (1998) also finds this
result; specifically, he finds that for low H$\beta$ FWHM, there is a
correlation with
\ion{Fe}{II} equivalent width which shifts to an anticorrelation for
higher H$\beta$ FWHM.

Correlations were sought between the {\it ASCA} and {\it ROSAT} photon
indices and the optical line properties.  A significant correlation
was found between the H$\beta$ FWHM and the {\it ASCA} hard X-ray
photon index (Figure 8).  There is also a strong correlation between
H$\beta$ equivalent width and the {\it ASCA} photon index which is
probably a consequence of the strong correlation between the H$\beta$
parameters discussed above.  These correlations are important because
they imply a direct connection between the conditions in the central
engine and the kinematics of the broad-line region.  This result is
also important because it challenges some models proposed for the
production of the strong \ion{Fe}{II} characteristic of NLS1s, as has
been noted previously.  The large
\ion{Fe}{II} to H$\beta$ ratio is generally difficult to understand
because both emission lines are predicted to be produced under the
same conditions in photoionization models and those models generically
underestimate \ion{Fe}{II} (for a review, Joly\markcite{63} 1993).  It
has been proposed that strong
\ion{Fe}{II} emission would be produced if the X-ray spectrum is flat
because hard X-ray photons can penetrate deeper into emission line
clouds producing heating and leading to strong \ion{Fe}{II} emission.
However, an anticorrelation between H$\beta$ FWHM and the X-ray photon
index has always been found (Puchnarewicz et al.\markcite{114} 1992,
Boller, Brandt \& Fink\markcite{7} 1996, Grupe\markcite{45} 1996 and
Grupe et al.\markcite{50} 1999a, and Brandt, Mathur \&
Elvis\markcite{11} 1997, Wilkes, Elvis \& McHardy\markcite{135} 1987).
Most of these studies used soft X-ray spectral indices obtained from
{\it ROSAT} or {\it Einstein} IPC.  Therefore it is interesting that
the correlation between H$\beta$ FWHM and the soft X-ray photon index
measured using {\it ROSAT} is not significant.  Conceivably this could
be an effect of absorption which would tend to flatten the soft X-ray
spectrum.  However, the subsample of objects that have detectable soft
excesses are minimally affected by absorption, and a significant
correlation is not found within that sample either.  Figure 8 shows
that the {\it ROSAT} photon index is clustered around 3 for most of
the sample except for the objects with the strongest soft excesses,
and those with significant absorption.

No correlation is observed between the {\it ASCA} and {\it ROSAT}
photon indices, even for the relatively unabsorbed, soft excess
subsample. Again, this is apparently because $\Gamma_{ROSAT}$ is
clustered between 3 and 4 for a broad range of {\it ASCA} photon
indices.  The lack of correlation between soft and hard photon indices
is interesting because it appears to not support the most direct
interpretation of the model proposed by Pounds, Done \&
Osborne\markcite{112} (1995) for NLS1s.  They proposed that the steep
hard X-ray photon index is produced when the Comptonizing electron
cloud is cooled by strong soft excess emission, in analogy with
Galactic black hole candidates in the high state (e.g.\
Nowak\markcite{102} 1995). The steep {\it ROSAT} spectrum implies a
strong soft excess (see below); therefore, one might expect that the
abundance of soft photons available when the soft excess is strong
would predict that the strongest soft excesses would be associated
with the steepest hard X-ray spectra.  However, the soft photons
necessary to cool the electrons may be emitted in the unobservable EUV
and therefore a correlation between these indices is not required for
the model to be viable.

Correlations between the overall shape of the {\it ASCA} spectra and
the strength of the soft excess with other parameters may be important
to examine.  Three complementary parameters as defined in Section 3.2
(the color ratio parameter, the black body parameter and
$\alpha_{xx}$) were used to to describe the overall shape of the {\it
ASCA} spectra.  These parameters are very strongly correlated, as
expected.  A strong correlation was found between $\alpha_{xx}$ and
the {\it ASCA} photon index.  This correlation is expected since these
parameters are not independent: $\alpha_{xx}=\Gamma_{ASCA}-1$ when
there is no soft excess.  There is no correlation between the color
ratio and the {\it ASCA} photon index; this appears to be because the
color ratio includes intrinsic absorption, which would not be expected
to be correlated with intrinsic continuum parameters.  The correlation
is weak between the black body parameter and the {\it ASCA} photon
index; again, this supports the idea that the hard X-ray photon index
is independent of the strength of the soft excess in the {\it ASCA}
band.  The color ratio and $\alpha_{xx}$ are both correlated with the
{\it ROSAT} photon index, a result that may originate in two effects.
The color ratio correlation appears to be dominated by absorption;
that is, objects with high color ratio and low {\it ROSAT} photon
index are also the ones that were found to have warm or neutral
absorption in their {\it ASCA} spectra.  On the other hand,
$\alpha_{xx}$ should be largely independent of absorption and thus
this correlation is due to the fact that objects with steeper {\it
ROSAT} spectra have overall steeper {\it ASCA} spectra.  Figure 8
shows that most of the strong soft excess objects are located in the
upper right corner of the $\alpha_{xx}$--$\Gamma_{ROSAT}$ plot.

Correlations between the variability parameters and other parameters
are also important; however, they must be examined carefully. Figure 3
in Part 1 shows that the excess variance appears correlated with the
luminosity.  However, the measured correlation is weak, a result which
may be caused by the large scatter.  Recall from Part 1 that the skew
parameter is defined as the number of sigma from the predictions of a
very specific model; it is therefore more reliable for high values
rather than low values which would be dominated by fluctuations.

A most interesting result is the evidence for correlations between the
variability parameters and the overall shape of the {\it ASCA}
spectra.  Consistently, more variable objects have steeper spectra.
The parameter $\alpha_{xx}$ was chosen to be most representative of
the intrinsic steepness and is plotted versus excess variance and skew
in Figure 8.  This figure illustrates that a primary reason for this
correlation is that the objects with strong soft excesses are more
variable than the ones with weak soft excesses.  To investigate this
further, the distribution of properties shown in Figure 8 for the
strong (6 objects) and weak (9 objects) soft excess objects were
investigated using several two-sample tests implemented in {\it
ASURV}.  The emission line parameters, including H$\beta$ FWHM,
H$\beta$ equivalent width, \ion{Fe}{II} equivalent width and ratio of
\ion{Fe}{II} to H$\beta$, are distributed equally among the strong and weak
soft excess objects.  This is important because it suggests that the
correlation between the overall shape of the {\it ASCA} spectra and
the variability properties may not be associated with the Boroson \&
Green (1992) Eigenvector 1.  There is no difference in distributions
of the {\it ASCA} photon index or 2--10 keV luminosity either.  There
are only strong differences in the distributions of the variability
parameters, as indicated by the plots.

Correlations between the iron line equivalent width and all of the
other parameters listed in Table 5 were also considered.  Only one
significant correlation was found: an anticorrelation between color
ratio and the equivalent width for the soft excess objects which was
significant at 0.058. 

Several other significant correlations were found.  The color ratio is
strongly correlated with the optical depth of O~VII.  This is
expected, based on the sensitivity of the color ratio to absorption.
Similarly, the {\it ROSAT} photon index is correlated with
$\tau_{O~VII}$, a further indication that the measured {\it ROSAT}
photon index is influenced by unmodeled absorption.  The temperature
of the black body model for the soft excess is not significantly
correlated with any parameter.  It is weakly anticorrelated with the
black body parameter, a result that may reflect model dependence.
There is an intriguing correlation between the
\ion{Fe}{II} to H$\beta$ ratio and the variability parameters for the
soft excess objects.  This appears to be due to the fact that some of
the strong soft excess objects have high \ion{Fe}{II} to H$\beta$
ratios.

In summary, the most interesting correlations are as follows: a
significant correlation between {\it ASCA} photon index and H$\beta$
FWHM was found; however, no correlation between {\it ROSAT} slope and
H$\beta$ FWHM was found.  The strength of the soft excess was not
correlated with the hard X-ray photon index, as would be naively
expected from Compton cooling models, but is correlated with the
variability parameters.  This is especially interesting because the
latter correlation does not appear to be associated with the Boroson
\& Green\markcite{9} (1992) Eigenvector 1.

\section{Discussion}

The discussion is organized as follows.  First, the constraints
imposed by the results on current models for X-ray emission and
absorption are discussed in a general way.  Then, the consistency of
the results on general models based on the orientation and accretion
rate are discussed.

\subsection{The Photon Index}

The photon indices in the sample of NLS1s presented here was found to
be systematically steeper than those in the broad-line Seyfert 1
galaxies analyzed by Reynolds\markcite{116} (1997).  A maximum
likelihood analysis showed that the difference is significant at 90\%
confidence, and there is significant dispersion among the photon
indices of both types of objects.  Brandt, Mathur \&
Elvis\markcite{11} (1997) previously reported this result.  The
advantage of the analysis presented here is that it is uniform,
whereas Brandt et al. took their results from the literature and
therefore the continuum model or fitting range could be different from
object to object.  This is conceptually important, because the soft
excesses in some objects are strong enough and extend to high enough
energy that they could modify the spectrum above 2~keV and skew the
photon index obtained from a 2--10 keV fit.  In practice, however, the
results are probably within the errors obtained from both approaches.

\subsubsection{Thermal Comptonization}

Pounds, Done \& Osborne\markcite{112} (1995) proposed that, in analogy
with the soft state of Galactic black hole candidates, the steep hard
X-ray photon index in the NLS1 RE~1034+39 can be understood in terms
of a thermal Comptonization model.  The energetic electrons are cooled
by the copious soft photons from the soft excess, and the resulting
cooler temperature leads to a steeper spectrum.  This scenario was
also adopted by Brandt, Mathur \& Elvis\markcite{11} (1997) to explain
the anticorrelation between H$\beta$ FWHM and {\it ASCA} photon index.

However, the simplest thermal Comptonization models (e.g. \ Rybicki \&
Lightman\markcite{120} 1979), will not work for AGN, because they
predict a far more sensitive photon index dependence on the electron
temperature and optical depth than is observed (e.g.\ Haardt \&
Maraschi\markcite{56} 1991).  Feedback between the soft photon source
(cold phase) and the energetic electrons (hot phase) is required to
stabilize the models and produce photon indices fairly narrowly
distributed around a single value (e.g.\ Figure 2).  Feedback models
have been developed by Haardt
\& Maraschi\markcite{56} 1991,\markcite{57} 1993, Ghisellini \&
Haardt\markcite{41} 1994, and Pietrini \& Krolik\markcite{109} 1994.
Generally, the feedback consists of reprocessing of radiation from the
hot phase by the cool phase, which increases the luminosity of the
cool phase, as well as pair creation in the hot phase which alters the
optical depth.  Pietrini \& Krolik\markcite{109} (1994) find that the
plasma parameters can be described by two scaling laws which are weak
functions of the ratio of hard and soft luminosity: $\theta \tau_T
\simeq 0.1 (l_h/l_s)^{1/4}$ and $\alpha \approx 1.6(l_s/l_h)^{1/4}$,
where $\theta=kT/m_e c^2$ is the plasma thermal energy in units of
electron rest energy, $\tau_T$ is the scattering optical depth,
$\alpha$ is the power law energy index and the compactness is
$l=(\sigma_{T}/m_e c^3)(L/R)$.  A difficulty with these feedback
models is that if the hot phase completely covers the cool phase, and
if all of the dissipation occurs in the hot phase, the predicted
photon index will be far steeper than observed in broad-line Seyfert 1
galaxies.  This problem has been addressed by proposing that covering
is not complete; instead, the hot phase is confined to isolated hot
spots (Haardt, Maraschi \& Ghisellini\markcite{58} 1994; Stern et
al.\markcite{128} 1995).  If the spots hug the surface of the
accretion disk, then the hot phase will intercept few photons from the
cool phase, and the resulting photon index will be flat.  If instead
the hot spot is tall and thin, many soft photons will be intercepted,
and the resulting photon index will be steeper.

The scaling law can be used to estimate the difference in $l_s/l_h$
ratio in NLS1s and broad-line Seyfert 1 galaxies.  Assuming that the
intrinsic indices are $\sim 2.4$ and $\sim 2.0$ for NLS1s and
broad-line Seyfert 1 galaxies, respectively, means that $l_s/l_h$
should be a factor of 3.8 larger in NLS1s.  Pietrini \&
Krolik\markcite{109} (1994) point out that $l_s$ can be interpreted as
a combination of intrinsic soft X-ray luminosity and of geometry
(efficiency of reprocessing); therefore, if the geometry and therefore
reprocessing is the same in NLS1s and broad-line Seyfert 1s, then the
intrinsic soft X-ray luminosity must be higher to explain the
difference in index.  However, the amount of reprocessing and
therefore geometry is not well constrained in NLS1s. The iron line
equivalent width potentially gives some information about the amount
of reprocessing.  In the sample of NLS1s presented here, the iron
emission line has approximately the same equivalent width as the lines
in broad-line Seyfert 1s, overall, arguing that the covering fraction
of reprocessing material in NLS1s is about the same as in broad-line
objects.  However, the iron line measurements have very large
uncertainties.  Also, there is evidence for an ionized iron line in a
few objects; ionization of iron can result in either enhanced or
reduced equivalent widths compared with neutral iron, depending on its
ionization state (e.g.\ Matt, Fabian \& Ross\markcite{86} 1993b).
Because there is very little information above 10~keV from NLS1s, the
amount of Compton reflection cannot be determined.  NGC~4051 and
Mrk~335, the only NLS1s considered by Nandra
\& Pounds\markcite{96} (1994), both have evidence for a reflection component in
their {\it Ginga} spectra. 

Spectral variability provides a potential diagnostic for this model.
Assuming the geometry and plasma parameters remain the same, because
of the shallow dependence of $\alpha$ on $l_s/l_h$, little spectral
variability is expected for moderate changes in flux.  It is also
worth noting that these thermal comptonization models cannot explain
the spectral variability exhibited by Mrk~766 as a change in $l_s/l_h$
(Leighly et al.\markcite{78} 1996).  In this case, a change in
geometry would be required also.

A simple interpretation of the Pietrini \& Krolik\markcite{109} (1994)
scaling law indicates that there should be a correlation between the
strength of the soft excess, if it represents $l_s$, and the hard
X-ray photon index, if it represents $l_s/l_h$.  The results presented
here do not support this, as there is no significant correlation
between the hard X-ray photon index and any of the soft excess
indicators (except $\alpha_{xx}$ which is not independent of the
photon index).  Figure 8 shows that strong soft excess objects are
distributed more or less equally above and below the maximum
likelihood average photon index of 2.19, and there are several weak
soft excess objects which have very steep hard X-ray spectra.  On the
other hand, the soft photons which comprise $l_s$ may be emitted
predominately in the unobservable EUV, and may not be directly related
to the {\it ASCA} band soft excess.  It is interesting to recall that
the EUV photons are those that should be responsible for
photoionization of the broad-line region clouds, rather than the soft
excess photons seen in the {\it ASCA} or {\it ROSAT} band.  If we
accept the model proposed by Wandel \& Boller\markcite{134} (1998),
that the narrowness of the lines is a consequence of a very strong but
unobservable EUV emission, the observed correlation between hard X-ray
photon index and H$\beta$ FWHM would support a difference in $l_s$ as
the origin of the steep hard photon index, rather than in geometry of
the Comptonizing electrons.

A yet unresolved point is the fact that soft X-ray selected samples
are composed of half NLS1s but the other half are broad-line Seyfert 1
galaxies (Grupe et al.\markcite{50} 1999a).  The presence of
broad-line Seyfert 1 galaxies in soft X-ray selected samples means
that they too have strong soft excess components.  It is not yet known
whether or not these the soft X-ray selected broad-line objects have
systematically steeper or flatter hard X-ray spectra than hard X-ray
selected Seyfert galaxies; however, there is evidence that at least
some of them have very flat hard X-ray spectra (e.g.\ 1H~0419$-$577
has a hard X-ray photon index $\sim 1.5$; Guainazzi et
al.\markcite{55} 1998b).  Again, this may mean that the bulk of the
soft photons are emitted in the unobservable EUV and we don't know how
strongly peaked the spectrum is in that region.  Some information may
come from examination of UV line ratios from soft X-ray selected AGNs
as they may be sensitive to the shape of the EUV ionizing continuum
(Krolik \& Kallman\markcite{68} 1988; Zheng, Kriss
\& Davidsen\markcite{140} 1995; Korista, Ferland, \&
Balwin\markcite{66} 1997).  This work must be done carefully, however,
as difference in radial or density distribution of emitting clouds can
also modify line ratios.

\subsubsection{Nonthermal Comptonization}

It has recently been discovered that in the soft state, the gamma-ray
spectrum of the Galactic black hole candidate Cyg X-1 extends as a
power law to at least 1~MeV without a break (Phlips et
al.\markcite{107} 1996).  Such high energy emission is unlikely to be
produced in a thermal plasma, 
it has therefore been asserted that this is evidence for nonthermal
Comptonization in the soft state of Cyg X-1 (Gierli\'nski et
al.\markcite{42} 1998).  Following the analogy that NLS1s may be
Seyfert galaxies in the soft state implies that nonthermal
Comptonization may be occurring in NLS1s.

Gierli\'nski et al.\markcite{42} (1998) apply a hybrid
thermal/nonthermal model to Cyg X-1 (see also Coppi\markcite{20} 1999
for a review of hybrid models). The nonthermal part is comprised of
electrons with a steep power law distribution, and the optical depth
is sufficiently low that the soft photons experience at most 1
scattering.  Therefore, the photon index of the radiation will be the
same as that of the electrons, and in principle a photon index of
$2.4$ can be produced easily.  The required electron distribution may
be natural as it is the same one that cosmic rays have. This model is
to be distinguished from the saturated nonthermal Comptonization,
which produces photon indices $\leq 2$ (e.g.\ Swensson\markcite{144}
1994).  Thermal electrons can be present also, if the time scale for
energy loss due to Coulomb interactions is shorter than the
Comptonization time scale.  They can also Compton-scatter soft photons
and produce the usual thermal spectrum.

\subsubsection{Bulk Motion Comptonization}

A final model that has been suggested for NLS1s is that of bulk motion
Comptonization (Shrader \& Titarchuk\markcite{126} 1998).  This model
assumes a three part accretion flow consisting of a cold Keplerian
optically thick accretion disk and a sub-Keplerian optically thin
corona, both of which suffer a shock at 10--30 $R_S$.  Interior to the
shock, the flow is hot and the optical depth is $\sim 1$.  The hot
plasma interior to the shock will have a high inflow velocity.  It
will intercept soft photons from the optically thick material exterior
to the shock and upscatter them by transferring kinetic energy rather
than thermal energy.  The predicted hard X-ray spectrum will have a
steep slope with $\alpha \sim 1.5$ (Chakrabarti \&
Titarchuk\markcite{14} 1995; also Ebisawa, Titarchuk \&
Chakrabarti\markcite{28} 1996).  Bulk motion Comptonization is also
discussed by Colpi\markcite{17} (1988).  A diagnostic of this model is
that the spectrum should not extend beyond 511~keV; thus it cannot be
applied to Cyg X-1 in the soft state because the observed spectrum extends to
1~MeV.

In this model, it is asserted that the hard X-ray spectrum will be
weak.  This is at least partially because the intensity of the soft
X-ray photons that the infalling electrons see will be decreased by
special relativistic effects (e.g.\ Pietrini \& Krolik\markcite{109}
1994).  Also, some of the upscattered photons may possibly be trapped
by the flow and not be able to diffuse out (e.g.\ Colpi\markcite{17}
1988). This scenario may not be appropriate for NLS1s; Grupe et
al.\markcite{48} (1998a) finds that soft X-ray selected AGN, which are
comprised of 50\% NLS1s, are stronger in soft X-rays rather than being
weaker in hard X-rays.  However, there is some evidence that the hard
power law is weaker in optically selected samples (Laor et
al.\markcite{71} 1997a).

\subsection{The Soft Excess}

This paper is the first to systematically study the soft excess in a
moderate number of {\it ASCA} observations from NLS1s.  Evidence for a
soft excess was found in 17 of the 19 objects without significant
absorption, implying that it is nearly ubiquitous in NLS1 {\it ASCA}
spectra.  The temperature of the black body used to parameterize the
soft excess was nearly the same for all objects.  However, an
important result is that there is a wide range of soft excess
strengths.  At 0.6~keV, a factor of 50 is spanned, as measured by the
black body parameter (Section 3.2).  Note that in principle this
result could be the effect of a hard X-ray power law with a range of
strengths, or an inverse correlation between strength of the soft
excess and weakness of the hard power law.  These possibilities cannot
be distinguished on the basis of the {\it ASCA} data alone.  Another
important result is that the strength of the soft excess is correlated
with measures of X-ray variability: objects with strong soft excesses
tend to have higher excess variance and skew parameter than objects
with weak soft excesses.  There appears to be no correlation between
the soft excess strength and optical emission line parameters,
suggesting that this behavior is not connected with Boroson \&
Green\markcite{9} (1992) Eigenvector 1.

\subsubsection{The Soft Excess and the Optical Emission Lines}

A strong soft excess is predicted to affect the thermal balance in the
two phase model of the broad-line region, resulting in a lower
temperature of the hot intercloud medium (Fabian et al.\markcite{31}
1986).  They predict that this will result in a very small region of
the thermal stability curve being amenable to two phases.
While a stationary two-phase medium in pressure balance may not be a
completely appropriate description for the broad-line region clouds,
this result may be partially responsible for the lower equivalent
widths of permitted emission lines (e.g.\ Goodrich\markcite{44} 1989)
and the correlation between photon index and equivalent width observed
in NLS1s.

The strong soft excess may also affect the location of the broad line
region.  Wandel \& Boller\markcite{134} (1998) note that if the steep
soft {\it ROSAT} spectra are extrapolated into the unobservable EUV, a
stronger ionizing continuum is inferred.  If the permitted lines form
at a particular ionization parameter, a larger radius for the
broad-line clouds would result.  The Keplerian velocities at larger
radii will be smaller, and therefore the lines will be narrower.
Alternatively, the densities in the broad line region clouds could be
larger.  Such an idea may explain correlation of the
density-indicating ratio
\ion{Si}{III}]/\ion{C}{III}] with \ion{Fe}{II}/H$\beta$ recently
presented by Wills et al.\markcite{137} (1999).

\subsubsection{The Soft Excess Temperature}

The soft excess is conventionally thought to be the high energy tail
of the accretion disk spectrum. If so, a soft excess at higher
temperatures is expected for larger accretion rates and
smaller black hole masses (Ross, Fabian \& Mineshige\markcite{119}
1992).  Drawing upon an analogy with Galactic Black Hole candidates in
their high state, and using RE~1034+39 for their example, Pounds, Done
\& Osborne\markcite{112} (1995) claim that the soft excess could be
primary emission from an accretion disk at near Eddington luminosity.
They note that the relation of inner disk temperature and black hole
mass for two objects radiating at near the Eddington limit is
$T_1/T_2=(M_2/M_1)^{1 \over 4}$, and that the observed temperature ratio is
consistent with plausible differences in the black hole mass of
RE~1034+39 and the BHC Cyg~X-1.

As noted above and in Section 3.2, the temperature of the soft excess
measured among NLS1 is nearly constant, despite large differences in
luminosity.  RE~1034+39 and PHL~1092 have, respectively, the lowest
and highest luminosities of the six strong soft excess objects in this
sample. The difference in luminosity is a factor of 30, suggesting
that the black hole mass is a factor of 30 higher in PHL~1092.  If so,
according to thin disk models, the temperature at the inner edge of
the accretion disk should be a factor of 2.3 lower in PHL~1092
compared with RE~1034+39; in fact, it is slightly higher.  A soft
excess with $kT$ a factor of 2.3 lower would not appear in the {\it
ASCA} band pass.  However, it must be noted that there could be
systematic uncertainties with the measurement of the temperature of
the soft excess because it is a continuum component appearing at the
edge of the {\it ASCA} bandpass, and there could be some model
dependence.  Furthermore, simulations show that there can be a
tendency for very weak soft excesses to have a higher measured
temperature.

This temperature constraint becomes even stiffer when {\it ROSAT}
results are considered.  In this study, a significant correlation was
found between the {\it ROSAT} photon index and the soft excess
strength, implying that objects with strong soft excesses generally
have very steep {\it ROSAT} spectra.  Grupe \markcite{45} (1996) found
a significant correlation between the photon index in a soft X-ray
selected sample and the redshift.  High redshift objects are more
luminous in count rate limited samples.  All of may imply that there
could be objects which are even more luminous than PHL~1092 which have
soft excesses observable in the {\it ASCA} band; however,
spectroscopic observations would be necessary to confirm the
temperature.

\subsubsection{Normalization of the Soft Excess}

The soft excesses were modeled using a blackbody.  Assuming that the
emission originates in an accretion disk, one may ask whether the size
of the emission region implied by the strength of the black body is
commensurate or at least smaller than the source size inferred using
variability arguments.

Using $H=50\rm km/s/Mpc$, the area of the emission region was computed
from the black body model fits for all of the objects which had
detectable soft excess component.  The areas ranged from
$10^{+20.7}\rm \,cm^{-2}$ for NGC~4051 to $10^{24.8}\rm \,cm^2$ for
RX~J0439$-$45. The areas were also quite large for PHL~1092,
PG~1211+143 and PG~1404+226, which were among the most luminous
objects in the sample.  Rapid variability was observed from
RX~J0439$-$45: a change in flux by a factor of $>2$ was detected in
about 2700 seconds.  If the emission occurs within $7\,R_S$, then the
upper limit on the black hole mass would be $4\times 10^7 M\odot$.
However, this exercise implicitly assumes that the emission is not
coming from the entire central region but rather a localized emission
region, and therefore, the light crossing time of the source may
overestimate the size of the emission region.  Nevertheless, for
illustration, consider the emitting region for the {\it ASCA} soft
excess to be between 7 and 8 $R_S$.  Then the area of that emitting
region for a $4\times 10^7 M\odot$ black hole would be $6.8\times
10^{27}\rm \, cm^2$.  This means that the time scale of variability
for RX~J0439$-$45, combined with its soft excess normalization, imply
that only 0.1\% of the region need be emitting if the whole thing is
optically thick.

However, in the classical the thin disk model, the temperature of the
disk at $7\,R_S$ should be much cooler than the temperatures of the
soft excesses observed here.  This may be a consequence of the fact
that the temperatures in the thin disk model are obtained by averaging
in the vertical direction.  When realistic radiative transfer is taken
into account, it turns out that the underlying disk emission will be
Compton upscattered by the hotter upper layers of the disk and the
observed emission should be hotter (Czerny \& Elvis\markcite{23} 1987;
Ross, Fabian \& Mineshige\markcite{119} 1992; Shimura \&
Takahara\markcite{125} 1993).  Compton scattering will smear the local
disk black body over a larger band pass, and therefore a larger area
will be required to emit the observed soft excesses.  An estimation of
the increased area needed can be obtained from Figure 6 in Ross,
Fabian \& Mineshige\markcite{119} (1992), which shows the fraction of
total disk emission in soft X-rays above 150~eV as a function of black
hole mass and accretion rate.  For a $4\times 10^7\, M_\odot$ black
hole, the fraction of flux in soft X-rays is order 10\% if the
Eddington fraction is as high as 1/3, implying that an area at least
ten times larger than the previous estimate is required; this would
still be only 1\% of the 7--8$R_S$ annulus.  For larger black hole
masses and smaller Eddington fractions, the fraction emitted in soft
X-rays is smaller.

\subsubsection{Models of the Soft Excess}

Accretion disk models predict more intense X-ray emission as well as
higher inner edge temperature when the accretion rate is high (e.g.
Shakura \& Sunyaev\markcite{122} 1973).  However, while standard
models of accretion disk emission find a good fit to the optical-UV
continuum, they predict far too cool temperatures at the inner edge to
explain the soft X-ray emission.  Many of these models ignore the
vertical structure in the accretion disk and the radiation transfer.
More sophisticated treatments of these corrections indeed do find that
the accretion disks can emit strongly in soft X-rays, especially when
the accretion rate is high and/or the black hole mass is low (e.g.\
Czerny
\& Elvis\markcite{23} 1987; Ross, Fabian \& Mineshige\markcite{119}
1992; Shimura \& Takahara\markcite{125} 1993).  A shifted and
strengthened accretion disk spectrum can explain the steeper {\it
ROSAT} spectra observed in NLS1s, and therefore this was interpreted
as evidence for a high accretion rate (e.g.\ Boller, Brandt \&
Fink\markcite{7} 1996; Laor et al.\markcite{71} 1997a; Grupe et
al.\markcite{50} 1999a).  This can also generally explain the high
temperatures and higher frequency of X-ray soft excesses observed in
the {\it ASCA} spectra from NLS1s (e.g.\ Pounds, Done \&
Osborne\markcite{112} 1996).

More realistic accretion disk models can also potentially explain the
fact that the temperature of the X-ray soft excess remains constant
over a large range of source luminosities and inferred black hole
masses (Section 4.2.2).  Examining a wide range of parameters, Shimura
\& Takahara\markcite{125} (1993) solve for hydrodynamic equilibrium
and radiative transfer self-consistently .  They find that the
emission spectrum gradually shifts toward the soft X-rays as the
accretion rate increases, confirming the results of Ross, Fabian \&
Mineshige\markcite{119} (1992).  However, when the accretion rate is
very large, nearly the Eddington rate, the derived temperature, near
100--200 eV, has only a weak dependence on the black hole mass.  This
constancy in spectral shape occurs because at such high accretion
rates, the scattering optical depth becomes independent of black hole
mass, Comptonization is the dominant process at the inner edge of the
disk, and the Compton $y$ parameter in the upper layers of the disk
becomes larger than 1 and nearly independent of black hole mass.  This
behavior is not expressed for lower accretion rates, where the soft
X-ray cutoff is a stronger function of temperature.

The large range in soft excess strengths may suggest that the soft
excess does not have a uniform origin in NLS1s.  Another possible
origin besides primary disk emission is suggested by the discovery of
soft X-ray line emission near 1~keV in two of the weak soft excess
objects, PG~1244+026 (Fiore et al.\markcite{34} 1998) and Ark~564, as
well as the ionized Fe K$\alpha$ line in Ton~S180.  The soft excess in
these objects could be due to reflection from an ionized accretion
disk.  When the surface of the accretion disk becomes ionized, the
reflectivity to soft X-rays is increased as the light elements are
ionized and the absorption in the upper layers is decreased.  This
situation has been explicitly modeled by Matt, Fabian \&
Ross\markcite{87} (1993a) and \.Zycki et al.\markcite{142} (1994) and
discussed by Czerny \& \.Zycki\markcite{24} (1994) and Fiore, Matt \&
Nicastro\markcite{35} (1997).  Fiore, Matt \& Nicastro\markcite{35}
(1997) mention that the effect of the reflection by ionized material
would be an increase in the steepness of the 0.2--2~keV spectrum by
about $\Delta \Gamma \approx 0.3$.

Another possibility is that the weak soft excesses seen in some NLS1s
is the signature of Compton scattering of blackbody emission (e.g.
Nishimura, Mitsuda \& Itoh\markcite{101} 1986).  Evidence for this
feature was found in the high state spectrum from Cyg X-1 (Cui et
al.\markcite{22} 1998; Gierli\'nski et al.\markcite{42} 1999).  The
spectrum was not adequately modeled with a black body plus power law;
an additional soft excess, a signature of the fact that a black body
is the source of photons, was required for intermediate energies.  If
the weak soft excess observed in some objects is in fact Comptonized
blackbody emission, then the primary black body must not be very far
below the {\it ASCA} bandpass.

\subsection{The Warm Absorber}

This paper presents the first systematic survey of the warm absorber
properties of NLS1s.  The first result is that there is evidence that
the incidence of warm absorbers, as indicated by the presence of
\ion{O}{VII} and \ion{O}{VIII} edges, is lower in NLS1s.   The second result is
that when warm absorbers are detected, they appear to have a lower
ionization state than those in Seyfert 1 galaxies with broad optical
lines.  Both of these conclusions, however, are subject to
restrictions based on the methods used to search for the warm
absorber.

\subsubsection{Biases}

Neither the narrow-line Seyfert 1 sample considered here, nor the
broad-line Seyfert 1 comparison sample presented in
Reynolds\markcite{116} 1997, can be considered complete in any way.
Therefore, it is important to determine whether biases in the sample
selection could influence the result.

The NLS1s discussed here are selected in several ways, all of which
may be biased against discovering highly absorbed objects.  Soft X-ray
selected objects present the strongest bias.  They are drawn from
Grupe's {\it ROSAT} soft X-ray selected AGN (e.g.\ Ton~S180,
RX~J0439$-$45, RE~1034+39, IRAS~13349+2438); the {\it ROSAT} spectra
from these objects indicates little evidence for heavy absorption
(Grupe et al.\markcite{49} 1998b).  They are also drawn from the Moran, Halpern \&
Helfand\markcite{89} (1996) {\it IRAS}/{\it ROSAT} cross correlation
sample (e.g.\  IRAS~13224$-$3809, IRAS~17020+4544, IRAS~20181$-$2244).
Two objects were selected by the {\it HEAO} A-1 experiment
(PKS~0558$-$504 and 1H~0707$-$495).  Other objects are drawn from PG
quasars and Markarian galaxies, including Mrk~335, I~Zw~1, Mrk~142,
PG~1211+143, Mrk~766, PG~1404+226 and Mrk~478.  These objects are
chosen by their blue optical spectra, which means that objects with
moderate reddening may be present, but objects with very heavy
absorption should not be.

In contrast, many of the broad-line Seyfert 1 galaxies considered by
Reynolds\markcite{116} 1997 are members of the HEAO-A2 Piccinotti
sample (Piccinotti et al.\markcite{108} 1982).  HEAO-A2 was sensitive
between 2 and 10 keV, so many more absorbed sources are present in the
Piccinotti sample.  Thus the difference in selection could lead to a
bias in the warm absorber result.  On the other hand, no NLS1s are
present in the Piccinotti sample, suggesting that hard X-ray selection
is biased against these objects, whether or not they are absorbed.
That may be because the steeper hard X-ray spectrum of NLS1s make them
somewhat weaker in the 2--10 keV band compared with broad-line Seyfert
galaxies.

There is also a possibility of a spectroscopic bias.  It is possible,
although it has not yet been demonstrated, that warm absorbers and
associated polarization is found more often among Seyfert 1.5 galaxies
compared with Seyfert 1s.  Intermediate-type narrow-line Seyfert 1
galaxies might be difficult to identify without good signal-to-noise
spectra.  A case in point is IRAS~20181-2244; it was thought to be a
Seyfert 2 galaxy until good signal to noise spectroscopy allowed clear
identification of the \ion{Fe}{II} emission (Halpern \&
Moran\markcite{59} 1998).

It is difficult to determine what kind of sample would be the best one
in which to study the absorption properties in NLS1s and broad-line
objects in an unbiased way.  Hard X-ray selected samples should be
ideal, since they choose objects whether or not they are absorbed.
However, the lack of NLS1 in the Piccinotti sample make this
impossible.  Spectroscopically identified Seyfert galaxies may present
the best sample, since both absorbed and unabsorbed Seyfert 1s and
NLS1s are present.

\subsubsection{The Frequency of the Warm Absorber in NLS1s}

Despite the potential selection bias, there are also some physical
reasons to explain why a warm absorber may be less frequent in NLS1s.
The simplest explanation is that it is a consequence of the steep
X-ray spectrum.  There will be fewer photons above the threshold
energy for oxygen ionization if the spectrum is steep, and therefore,
fewer appropriate ions may be created.  Note that this may not be the
whole story; if the warm absorber occurs in equilibrium gas, then the
thermal and ionization balance would be more important for determining
the properties of the warm absorber.

\subsubsection{The Ionization State of Warm Absorbers in NLS1s}

The potential biases explored above should not affect the second
result of this study, that the ionization appears to be lower in warm
absorbers in NLS1s compared with Seyfert 1 galaxies with broader
optical lines.  Recall that this inference is based on the comparison
of distributions of \ion{O}{VII} and \ion{O}{VIII} edges between the
NLS1s presented here and a sample of broad-line objects from
Reynolds\markcite{116} (1997): while the distributions of \ion{O}{VII}
edge depths were indistinguishable, detection of significant optical
depth in the
\ion{O}{VIII} edge was less frequent in the NLS1s.  This can be
inferred to be evidence that the ionization parameter is lower in
NLS1s.  However, there are several weaknesses in this argument.  In a
single-zone model, the oxygen edges are not independent. Therefore, it
is conceivable that a detailed study of the fractional ionization of
the gas as a function of ionization parameter will reveal that for
NLS1s the $\rm O^{+7}$ inhabits only a very narrow range of ionization
parameter, and the gas is more typically characterized by $\rm O^{+6}$
and fully ionized oxygen.  Furthermore, assuming that the gas is in
multiphase equilibrium, it may happen that $\rm O^{+7}$ is the
dominate ionization state under conditions when the gas is unstable.
Both of these questions require computations beyond the scope of this
paper to address. Finally, the two-edge model will fail to detect very
high ionization warm absorbers which may be dominated by Fe~L
resonance absorption and have been postulated to be present in several
strong soft excess NLS1s by Nicastro, Fiore \& Matt\markcite{100}
1999.

Among Seyfert 1 galaxies with broad optical lines, there is some
evidence that the ionization of the warm absorber is about the same
for all objects.  George et al.\markcite{40} 1998 found, in their
study of a sample of 23 {\it ASCA} observations of 18 objects
dominated by broad-line Seyfert 1 galaxies, that the ionization
parameter appears to be strongly peaked around one value, although a
maximum likelihood analysis indicates a significant dispersion.  This
result may be due either to a sample selection effect or a physical
selection effect.  Possible reasons why it may be a selection effect
were argued by George et al.  They postulate that if the absorbing gas
with lower ionization parameter happens to always be coincident with
dust, the dust may block the view to the broad line region clouds
causing the objects with low ionization parameters to be classified as
Seyfert 2s which would have been excluded from the George et al.\
sample.  The results presented here suggest that Seyfert 1 galaxies
with less ionized warm absorbers exist.

The typically steeper X-ray spectrum may affect the conditions of the
warm absorber in NLS1s.  Reynolds \& Fabian\markcite{117} (1995)
consider models in which the warm absorber is in thermal equilibrium
with the broad-line region clouds.  They show that the presence of a
soft excess or a steeper X-ray spectrum appears to modify the thermal
stability curve such that the warm absorber is stable over a broader
range of parameters.  Everything else being equal, this would predict
that warm absorbers should be more common in NLS1s.  At a given
temperature, the values of $\xi/T$ are generally larger for steeper
spectra and stronger soft excesses (see Reynolds \&
Fabian\markcite{117} 1995), suggesting that the ionization should be
larger in NLS1s rather than smaller.  This is in contrast to what is
inferred from the oxygen edge measurements but may be consistent with
the presence of a very highly ionized warm absorber.

A similar result is presented by Shields, Ferland \&
Peterson\markcite{124} (1995).  They investigated the contribution of
optically-thin broad-line region gas in photoionization but not
thermal equilibrium in Seyfert 1 galaxies, and find that optically
thin gas may contribute substantially to the high ionization line
emission in AGN.  The predicted opacity and some ionization states of
this gas is also appropriate to produce the warm absorber phenomenon.
Shields et al.\ find that when there are more soft X-rays relative to
UV, the optically thin material is in general more highly ionized and
will more likely appear as a warm absorber.  Again, this predicts a
more prevalent warm absorber in NLS1s, in contrast to what is inferred.

These models assume that the warm absorber gas is in ionization and
sometimes also thermal equilibrium.  That may not be the case.  Krolik
\& Kriss\markcite{69} (1995) and Nicastro et al.\markcite{99} (1999)
consider in some detail the situation in which the warm absorber gas
is subject to a time variable photoionizing source.  Since each ion
has a different recombination rate, the details can be complicated.
However, when the time scale of variability is shorter than the
recombination time scale, the warm absorber will be overionized with
respect to ionization equilibrium, generally speaking.  This would
predict an even higher ionization state than predicted in equilibrium
models.

This discussion shows that many simple one-zone models cannot easily
explain the lower ionization inferred from the oxygen edge
measurements in the sample of NLS1s.  However, two-zone models may be
able to explain this result naturally.  Evidence for more than one
zone of ionized material has been found in a few well-studied
broad-line Seyfert galaxies.  Difference in variability behavior of
the \ion{O}{VII} and \ion{O}{VIII} edges in response to flux changes
in MCG--6$-$30$-$15 provides one example (Otani et al.\markcite{105}
1996), and discrepancies between absorption profiles in high optical
depth warm absorbers and detailed single zone models was also
interpreted as evidence for two absorbers (George et al.\markcite{40}
1998).  Evidence of an association of the outer warm absorber with
dust via reddening (Reynolds\markcite{116} 1997; Reynolds et
al.\markcite{118} 1997) and polarization (Leighly et al.\markcite{77}
1997b) placed the outer warm absorber coincident with or exterior to
the broad line region.  Therefore, one possibility is that the inner
warm absorber in NLS1s may be too highly ionized to be observed.
Alternatively, it may not exist.  For example, it may have been blown
from the nucleus by radiation pressure.  If only the outer warm
absorber is detected, the cumulative ionization state will appear to
be low.  This somewhat fits our expectations because a fairly good
correlation between polarization and presence of a warm absorber is
observed in this sample; the dusty warm absorber should be the outer
one, as dust is not expected to survive the intense heating interior
to the broad-line region.

We may be able to test whether or not there is warm absorber gas
present in NLS1s, and in similar quantities as in broad-line Seyfert 1
galaxies, by measuring emission lines, such as coronal lines, that are
thought to associated with warm absorber gas (e.g.\ Porquet et
al.\markcite{111} 1998; Erkens, Appenzeller \& Wagner\markcite{30}
1997).  A complicating factor may be that the ionization potentials
for the common coronal line ions are quite high, around 0.2~keV.
NLS1s generally have steeper soft X-ray spectra than Seyfert 1s with
broader optical lines, and there is evidence that NLS1s are stronger
on average at 0.2~keV than broad-line Seyfert 1s compared with the
rest of the spectrum (Grupe et al.\markcite{48} 1998a).  Therefore,
stronger coronal line emission might be expected from NLS1s,
independent of whether warm absorber gas is present.  This expectation
appears to be roughly borne out in simulations by Porquet et al.  Of
the two continuum models used, the one that is stronger in soft X-rays
requires a lower column density to produce the observed coronal line
equivalent widths, at least for the lower densities.  Interestingly,
little dependence of the oxygen edges on the continuum shape is found.

\subsection{Models for NLS1s}

\subsubsection{Models Based on Orientation}

NLS1s have narrower optical emission lines than Seyfert 1 galaxies
with broad lines.  It has been suggested that the origin of this
behavior is that NLS1s are viewed with a face-on (pole-on)
orientation.  Narrower emission lines are seen because the optical
line emitting clouds are confined to a plane and therefore a smaller
degree of Doppler broadening is seen (e.g.\ Osterbrock \&
Pogge\markcite{103} 1985; Puchnarewicz et al.\markcite{114} 1992;
Boller, Brandt \& Fink\markcite{7} 1996).  Stronger soft X-ray
emission is expected for face-on orientations from some geometrically
thick accretion disk models (Madau\markcite{83} 1988).  Models in
which observed line width is based on orientation have recently been
supported by evidence for a correlation between radio power and the
monochromatic optical luminosity (an indication of the orientation
angle) and the width of the H$\beta$ lines (Wills \&
Brotherton\markcite{136} 1995).  It has been postulated that the
\ion{Fe}{II} emission may arise in the accretion disk
(Collin-Souffrin, Hameury \& Joly\markcite{16} 1988; Kwan et
al.\markcite{70} 1995), so a face-on viewing angle would enhance the
amount of \ion{Fe}{II} observed.
 
To be consistent with the {\it ASCA} results, the observed hard X-ray
photon indices should be orientation dependent.  The slab thermal
Comptonization model as implemented by Haardt \& Maraschi\markcite{57}
(1993) predicts softer spectra as the inclination angle increases.
This effect originates in the anisotropy of the first order
scattering, resulting in more photons directed downward toward the
slab.  A similar result was obtained by Dove, Wilms \&
Begelman\markcite{27} (1997) using a similar but improved slab model.
This trend is opposite of what the observations require.  When this
model is generalized, and the hot phase is assumed to be in localized
regions on the disk, the dependence on inclination decreases, and the
index depends rather on the height of the hot phase blobs above the
disk.

A face-on geometry might explain the low frequency of warm absorbers
in NLS1s inferred from this sample. It has been
suggested that intermediate Seyfert galaxy spectra are produced when
the line of sight to the broad line region is partially blocked by
absorbing material (e.g.\ Lawrence \& Elvis\markcite{74} 1982).  The
absorbing material may be the molecular torus thought to lie at a
distance intermediate to the broad and narrow-line regions; the warm
absorbing material and associated dust may be ablated from the
molecular torus.  Thus, warm absorbers may be more likely to be
observed when the inclination is moderate rather than face-on.

The enhanced variability observe in NLS1s should also be orientation
dependent.  These aspects are discussed in Part 1, and both face-on
and edge-on scenarios may be able to explain the enhanced variability.
The edge-on model may naturally explain the fact that the variability
parameters are correlated with the strength of the soft excess, since
in some disk models, enhanced soft X-ray emission is expected when the
viewing angle is edge-on (Laor \& Netzer\markcite{73} 1989). 

\subsubsection{Models Based on Accretion Rate and Black Hole Mass}

It has been suggested that a higher accretion rate relative to
Eddington in NLS1s can explain many of their characteristic
properties.  If so, assuming that the efficiency of conversion of
accretion energy to radiation is the same in both types of objects,
the black hole mass should be smaller in NLS1s.  The optical lines
then will be narrower, assuming that the broad-line region emission is
the same and that the motions of the broad line emitting clouds are
dominated by Keplerian velocities.  However, it may be more important
that a high accretion rate implies an enhanced photoionizing continuum
(Wandel \& Boller\markcite{134} 1998).

Several models of enhanced accretion predict a strong and hot soft
excess, as discussed in Section 4.2.3.  That an enhanced accretion
rate should be associated with a steeper hard X-ray spectral index was
first introduced first by Pounds, Done \& Osborne\markcite{112} 1995.
If the geometry and high energy cutoff are assumed to be the same in
all Seyfert 1 galaxies, then we can determine how much larger
intrinsic luminosity of the soft component would have to be.  It was
shown in Section 4.1.1 that $l_s/l_h$ should be a factor of 3.8 larger
in NLS1s to explain the difference in slope.  Assuming a covering
fraction of 0.5 and using Eq. 29 from Pietrini \& Krolik\markcite{109}
(1995), the intrinsic soft X-ray luminosity would be a factor of 5.5
larger in NLS1s.  However, the steeper hard X-ray spectrum may also
mean that the hot phase luminosity is smaller in NLS1s.  Assuming that
the normalization at 1~keV is the same in both types of objects (Grupe
et al.\markcite{48} 1998a), and the spectra show an exponential cutoff
at 100~keV, the hot phase luminosity in NLS1s may be 70\% that of
broad-line objects.  Therefore, the intrinsic luminosity in the soft
component would have to be about 7.5 times greater in NLS1s, assuming
that everything else remains the same.  If the efficiency of
conversion of gravitational potential energy to radiation is the same
in both types of objects, this would mean that the accretion rate
should be about 7.5 times greater in NLS1s compared with broad-line
objects.

As discussed in Part 1, the excess variance versus luminosity plot can
also be explained by a higher accretion rate and correspondingly
smaller black hole mass, assuming that the variability has the same
structure in NLS1s as in Seyfert galaxies with broad optical lines.
The inferred difference is a factor of 10, interestingly near the
enhancement required to produce the steep hard X-ray spectrum
discussed above.

The high accretion rate may also affect the properties of the warm
absorber.  Reynolds \& Fabian\markcite{117} 1995 hypothesize a
dynamical model for the warm absorber in which neutral gas is
accelerated by radiation pressure on the resonance lines.  In the
process, the gas is partially ionized.  When the gas becomes ionized,
the acceleration ceases, the inward gravitational acceleration takes
over until it becomes compressed enough to recombine.  The critical
luminosity required to balance these two effects is estimated by
Reynolds \& Fabian\markcite{117} (1995) to be about $0.05 L_{Edd}$.
They speculate that if the luminosity exceeds this critical
luminosity, an outflow of highly ionized material may be formed, and
this may explain the lower frequency of warm absorbers in quasars, as they are
thought to be radiating at a higher Eddington fraction than Seyfert
galaxies.

\subsubsection{Other Comments on Physical Models}

The two phase model has been a very successful scenario to describe
the behavior of broad-line Seyfert galaxies.  In this model, most of
the accretion energy, rather than being released in the disk, is
dissipated instead in the corona above the disk (Haardt \&
Maraschi\markcite{56} 1991,\markcite{57} 1993; Stern et
al.\markcite{128} 1995; Svensson \& Zdziarski\markcite{129} 1994).  An
important feature of this model is that if a sufficient amount of
energy is released in the corona, the thin disk solution can exist
close to the black hole (Svensson \& Zdziarski\markcite{129} 1994);
otherwise, a transition to another type of accretion is expected.

This model assumes that the source of power for the disk emission is
reprocessing of the coronal emission by the optically thick material.
However, in NLS1s, the soft excess is sufficiently strong and the hard
X-ray photon index is sufficiently steep that the soft excess cannot
be powered by reprocessing (e.g. Pounds, Done \& Osborne\markcite{112}
1996).  Although it has been shown that a thin disk with intrinsic
emission can exist close to the black hole in Galactic black hole
candidates (Gierli\'nski et al.\markcite{42} 1999), because of the
dependence of the stability criterion on black hole mass as well as
accretion rate, there can be no stable AGN thin disk solution in this
situation.  This fact marks a limitation of the BHC/AGN analogy as an
explanation for NLS1 behavior.  However, it may alleviate one of the
more uncomfortable problems with this analogy.  As has been widely
demonstrated, NLS1s are more variable than broad-line Seyfert
galaxies; in contrast, BHC in the soft state are less variable than in
the hard state.

What kind of accretion flow might be expected at small radii when the
radiation pressure becomes too large for the thin disk solution?  One
possibility that has been proposed for hard state Galactic black hole
candidates and broad-line radio galaxies is that the inner regions
form a hot, optically thin, geometrically thick torus, either the
two-temperature torus (SLE; e.g.\ Shapiro, Lightman \&
Eardley\markcite{123} 1976) or an advection dominated flow (ADAF; for
a review, Narayan, Mahadevan \& Quataert\markcite{97} 1999).  These
flows are typified by rather hard X-ray power-law emission and no soft
excess; therefore they are likely not to be applicable to NLS1s.

It was originally thought that when the accretion rate is high, there
would be a radiation pressure dominated disk (e.g.\ Shakura \&
Sunyaev\markcite{122} 1973).  However, this solution is shown to be
unstable if the viscosity is proportional to the radiation pressure,
although it may be stable if the viscosity is proportional to the gas
pressure only (Lightman \& Eardley\markcite{81} 1974).  At higher
accretion rates lies the slim disk solution, in which part of the
energy is advected into the black hole (e.g.\ Abramowicz et
al.\markcite{1} 1988).  In NLS1s, it is possible that the accretion
rate is in the radiation dominated region and it instead develops an
unsteady flow.  Such a scenario has recently been proposed for the
superluminal black hole binary GRS~1915+105 by Belloni et
al.\markcite{2} 1997.  This object may undergo unsteady accretion in a
similar way as dwarf novae, except the Lightman-Eardley instability
(which operates between the thin disk and slim disk solutions, with
the unstable radiation pressure dominated disk in between) operates
rather than the thermal-viscous instability.  The time scales for the
limit cycle are much longer for AGN (on the order of years); however,
on shorter time scales, during the accreting portion of the cycle, the
accretion happens unsteadily, and rapid variability is observed as
heating fronts move through the disk at the sound speed, and material
is removed on the infall time scale (Belloni et al.\markcite{2} 1997).

Such a scenario may be attractive for NLS1s.  Slim disks can produce
soft X-rays (Szuszkiewicz, Malkan \& Abramowicz\markcite{130} 1996).
The sound crossing speed and infall time scales in the inner region
may be appropriate for the rapid variability observed in NLS1s.  The
heating waves could conceivably be coherent, and therefore produce
large amplitude flares.  The longer time scale variability due to the
limit cycle behavior may provide a way to explain the very large
amplitude variability observed from some NLS1s (IC~3599: Grupe et
al.\markcite{46} 1995a; WPVS~007: Grupe et al.\markcite{47} 1995b;
RX~J0134$-$42: Grupe et al.\markcite{51} 1999b).

The correlation between the excess variance and strength of the soft
excess might be explained as a consequence of a variable transition
radius $r_t$ from the usual AGN geometry, that is, a thin disk plus
corona from magnetic flares existing at $r>r_t$, to the proposed highly
variable unsteady inner disk at $r<r_t$.  The location of this
transition radius may be a function of specific accretion rate.  When
$r_t$ is small, only a relatively small region is characterized by the
highly variable unstable disk, while there is a relatively large
region containing the usual thermal corona present in broad-line
objects.  Therefore, the soft excess is moderately strong, and the
hard power law is relatively strong, and, while the excess variance is
higher than that of broad-line Seyfert 1 galaxies, the variability is
not detectably non-Gaussian because the coherent variability from the
inner region is diluted by stochastic variability from the flaring
outer region.  When $r_t$ is large, the highly variable unstable inner
disk dominates the central parts of the AGN, leading to a very strong
soft excess. The hard tail is weaker, because less of the accretion
energy goes into the usual thermal corona.  Because the variability is
dominated by the very variable inner region, the excess variance is
again large, but since the coherent variability is minimally diluted,
non-Gaussianity is detected. Note that because the region sizes change
approximately in concordance with the emission from each region, the
compactnesses may stay approximately the same.

\section{Summary and Future Observations}

I present a comprehensive and uniform analysis of 25 {\it ASCA}
observations from 23 Narrow-line Seyfert 1 galaxies.  The results of
spectral analysis and correlations are reported here; the results of
the time series analysis are reported in Part 1.  The primary results
of this paper are the following:

\begin{itemize}
\item A maximum likelihood analysis confirms that the hard X-ray photon
index is significantly steeper at $>$90\% confidence in this sample of
NLS1s than in a random sample of Seyfert galaxies with broad optical
lines.  As previously noted, this can be explained by
two-phase thermal Comptonization models if the soft photon input is
greater in NLS1s than in  broad-line objects.  This view may be
supported by the observed correlation between the photon index and the
H$\beta$ FWHM among the objects in this sample.  Alternatively, the
steep photon indices may originate in a single scattering from a power
law distribution of electrons.
\item Soft excess emission was detected in 17 of the 19 objects which
had no significant absorption.  While the luminosities span two orders of
magnitude, the temperatures are roughly consistent, a result that
contradicts predictions of the simplest thin disk model.  The
soft excess strength above the power law spans a large range (a factor
of 50), and the strength of the soft excess was found to be correlated
with the variability parameters as well as the {\it ROSAT} photon
index.  Possible origins for the soft excess include primary emission
from an accretion disk characterized by a high accretion rate,
reprocessing in an ionized disk, and the signature of Comptonization
of black body emission.
\item Evidence was found that the warm absorber is less likely to be
observed in NLS1s and when present has typically lower ionization than
in broad-line Seyfert 1 galaxies.  However, this result was obtained
using a two-edge parameterization which would not necessarily be
sensitive to warm absorbers with very high ionization.  This result
may imply that the inner warm absorber is either absent or too highly
ionized to be detected in these objects, but the outer warm absorber,
typified by a lower ionization parameter, remains.
\item The iron line equivalent width average and distribution was
found to be the same in the sample of NLS1s as in a random sample of
Seyfert 1 galaxies with broad optical lines, a result which may imply
that the amount of reprocessing is the same in both types of objects.
However, this result is qualified by the generally poorer statistics
at the iron line in the NLS1 sample, and the fact that emission from
ionized iron, which may have either enhanced or reduced fluorescence
yield, was found in a few objects.
\end{itemize}

Models based on orientation and high accretion rate were examined in
the light of these results.  A model based a face-on orientation
possibly explains the narrow emission lines, stronger soft excess and
less frequent warm absorbers, but cannot explain the steep power law,
the variability results or the low ionization of the warm absorber.  A
model based on an edge on viewing angle may be able to explain the
variability results, the steep soft excess and perhaps also the steep
hard X-ray spectrum but requires NLS1 to consistently be inclined to
their host galaxies more than Seyfert 1 galaxies with broad optical
lines, and the dominant motion of broad line region clouds would be
required to be along the symmetry axis.

A model based on a higher accretion rate relative to Eddington
generally fairs the best.  It can explain the steep soft excess due to
a shift of the accretion disk spectrum to high energies; this also
explains the narrow optical emission lines through stronger ionization
and also the steep hard X-ray photon index through stronger
Comptonization cooling.  The rapid variability is naturally explained
by the requirement of a smaller mass black hole.  The nearly constant
soft excess temperature is predicted at high accretion rates in some
accretion disk models.  The less frequent warm absorber may be
explained if it is too highly ionized to be seen or has been blown out
of the system.

The {\it ASCA} observations and results presented here lead to further
questions which may be answerable in the future.

\begin{itemize}
\item Observations at higher energies are necessary to make significant
progress on understanding the origin of the steep hard X-ray spectrum
in NLS1s.  Observations up to 20 keV will allow us to search for
the Compton reflection component and obtain information on the amount
of reprocessing.  Observations to higher energies will allow us to
determine whether the hard X-ray emission process is thermal or
nonthermal.  In addition, studies of the photon index spectral
variability may yield some constraints on these processes.
\item The nature of the soft excess in the {\it ASCA} band remains a
mystery.  Observations with a broader band pass down to 0.1~keV and
better statistics would allow us to better constrain the shape.
Observations with better energy resolution may reveal line emission
characteristic of reprocessing in a ionized disk.  Observations of
variations in the shape of this component as a function of time also
may be valuable.
\item Better statistics in the iron line region are needed to
determine the importance of reprocessing and also to constrain the
inclination and properties of the reprocessing material including
ionization and distance from the black hole.
\item Observations with good statistics and energy resolution will
allow us to determine the properties of the warm
absorber in NLS1s with less ambiguity.
\end{itemize}

\acknowledgements

KML gratefully thanks all those people who built and operate {\it
ASCA}.  This project could have never attained this form without the
help and support of Jules Halpern.  Many thanks go to Dirk Grupe for
many various kinds of help and advice.  Useful discussions with Karl
Forster, Julian Krolik, Herman Marshall, Tahir Yaqoob and Andrzej
Zdziarski are acknowledged.  The following are thanked for a critical
reading of a draft: Joachim Siebert, Jules Halpern \& Tahir Yaqoob.
This research has made use of the NASA/IPAC extragalactic database
(NED) which is operated by the Jet Propulsion Laboratory, Caltech,
under contract with the National Aeronautics and Space Administration.
This research has made use of data obtained through the High Energy
Astrophysics Science Archive Research Center Online Service, provided
by the NASA/Goddard Space Flight Center.  KML gratefully acknowledges
support through NAG5-3307 and NAG5-7261 ({\it ASCA}) and NAG5-7971
(LTSA) .

\appendix

\section{Appendix -- Notes on Individual Objects}

\subsubsection{Ton S180:} 

GIS3 5.5--7.5 keV was excluded from spectral fitting.

Turner, George \& Nandra\markcite{132} (1998) report evidence for soft X-ray line
emission in this {\it ASCA} observation of Ton~S180.  The fact that
Turner et al.\ 	 model the spectra from both CCD detectors and also do
not model the soft excess component leads to some doubt about this
result.  As discussed in Section 2.1, data taken after the beginning
of 1996 tended to show large discrepancies between the SIS0 and SIS1
detectors.  The Ton~S180 observation was made in June 1996 and
displays one of the worst examples of this discrepancy found in this
sample of data.  The SIS1 residuals fall markedly below the SIS0
residuals for energies less than 0.8~keV (Leighly\markcite{76} 1998).
Thus, spectral complexity appears to be present in the SIS1 detector
but not in SIS0.  Spectral fitting confirms this conjecture; a soft
X-ray line appears significant in the SIS1 spectrum, but there is no
evidence for it in SIS0.  When a soft X-ray line is added to the best
fit model presented here, the improvement in $\chi^2$ was 6.2 for 2
d.o.f. and the equivalent width is 7.4~eV; therefore this feature is
not significant.

\subsubsection{PKS~0558$-$504:} GIS3 5.0--7.0 keV was excluded from spectral
fitting.

\subsubsection{Mrk~766:}  The spectral classification of Mrk~766 has
been debated.  Osterbrock \& Pogge\markcite{103} (1985) classify this
object as a Seyfert 1.5; later, however, Goodrich\markcite{44} (1989)
includes it in his NLS1 spectropolarimetry sample.
Goodrich\markcite{44} (1989) published a spectrum near H$\beta$ from
this object.  The typical Lorentzian NLS1 line profile can clearly be
seen, and measured FWHM from the plot is 1700~km/s.  I speculate that
the Seyfert 1.5 classification arose from a failure of a single
Gaussian fit to the Lorentzian profile; this is a common situation in
NLS1 optical spectra NLS1s (e.g.\ Goncalves, Veron \&
Veron-Cetty\markcite{43} 1998).  Furthermore, the H$\beta$ appears
weak compared with [\ion{O}{III}] in Goodrich\markcite{44} 1989;
however, this could be a consequence of a Balmer decrement arising
from a dusty warm absorber responsible for the ionized absorption in
the X-ray band (e.g.\ Leighly et al.\markcite{77} 1997b) and the red
optical spectrum (Molendi \& Maccacaro\markcite{88} 1994) and not
intrinsic to the emitted spectrum.

\subsubsection{IRAS~13349+2438:} The FWHM H$\beta$ of IRAS~13349+2438
is $2200\,\rm km\,s^{-1}$ and therefore it is not an NLS1, technically
speaking.  However, as noted by Brandt, Mathur \& Elvis\markcite{11}
1997, the steep spectrum and variability make it more similar to NLS1s
than to Seyfert 1s or quasars with broad optical lines.  Therefore it
is also included in this sample.

\subsubsection{PG 1244+226:} The soft X-ray line feature reported here
is qualitatively the same as that reported by Fiore et al.\markcite{34} 1998;
however, a smaller equivalent width is reported here (35 compared with
64 eV).  The origin of this difference is that here the spectra over
the full {\it ASCA} band pass is modeled and a soft excess component
is included whereas in Fiore et al., only 0.4--4.0 keV region is
modeled and the continuum consists only of a power law.  The soft
excess component models part of the positive residuals resulting in
smaller equivalent width.

The feature in the PG~1244+226 spectrum was alternatively modeled as
an edge at 1.17keV by Fiore et al.\markcite{34} 1998.  I found that an
edge at 1.17~keV also modeled the spectrum adequately
($\chi^2=487$/474 d.o.f.). The unresolved line model yielded a $\chi^2$
of 489 for the same number of degrees of freedom, indicating a
difference in $\chi^2$ of 2.  The likelihood ratio then is just 2.71
and therefore the difference is not significant.

\subsubsection{IRAS~17020+4544:} GIS3 6.0--8.0 keV was excluded from spectral
fitting.

\subsubsection{Kaz 163:} The ratio of data to model indicates that
there may be a weak soft excess in this object.  However, addition of
this component leads to only a small reduction in $\chi^2$
($\Delta\chi^2=8.7$) and therefore it is required with less than 99\%
confidence.

\subsubsection{IRAS~20181$-$2244:} Marginal evidence for a warm
absorber was found during this analysis which was not reported by
Halpern \& Moran\markcite{59} 1998.

\subsubsection{Ark~564:} Analysis of this object was complicated by
the fact that the observation was made partly with a level
discriminator and partly without, and the level discriminator was set
at different values during the time that it was on.  This required
different response matrices for the different level discriminator
states.  Spectra were accumulated excluding the data with the high
level discriminator.  The spectra were fit simultaneously with spectra
from a coordinate {\it RXTE} observation.  The details will be
reported in Leighly et al. in prep.

\newpage

\begin{deluxetable}{lllllllll}
\scriptsize
\tablewidth{0pc}
\tablenum{1}
\tablecaption{Target Properties and Optical Emission Line Parameters}
\tablehead{
\colhead{Target} & \colhead{z} 
& \colhead{H$\beta$ FWHM} &
\colhead{H$\beta$ Eq. W.} & \colhead{\ion{Fe}{II}/H$\beta$} & \colhead{[\ion{O}{III}]
FWHM} & \colhead{[\ion{O}{III}]/H$\beta$}  & 
 \colhead{Reference}\\ 
& & (km/s) & (\AA\ ) & & (km/s)\\}
\startdata

Mrk 335 & 0.025 & 1640 & 95 & 0.62 & & 0.23 & BG92 \\
I Zw 1 & 0.061 & 1240 & 51 & 1.47 & & 0.43 & BG92 \\
Ton S180 & 0.062 & 980 & 45 & 0.8 & 640 & 0.12 & G96 \\
PHL 1092 & 0.396 & 1790 & 63 $^1$ & 1.8 &  & 0.91 & BK80 LEWMB97 this paper \\
RX J0439$-$45 & 0.224 & 1010 & 65 & 0.55 & 1020 & 0.17 & G96 \\
NAB 0205+024 & 0.155 & 1050 & 58 & 0.62 &  & 0.36 & ZO90 GW94 BK84 K91 \\
PKS 0558$-$504 & 0.137 & 1250 & 45 & 1.56 $^2$ & & 0.04 & C97 R86 \\
1H 0707$-$495 & 0.0411 & 1050 & 32 & 1.36 & 1516 & 0.19 & this paper \\
Mrk 142 & 0.04494 & 1470 & 67 & 1.11 & 404 & 0.16 &  this paper; G96 \\
RE 1034+39 & 0.04244 & 840 & 20 & 0.32 & 543 & 1.25 & this paper MPJ96
GVV98 \\
NGC 4051 & 0.00242 & 1150 & 28 & 0.77 & 325 & 0.92 & this paper \\
PG 1211+143 & 0.0809 & 1860 & 84 & 0.52 & & 0.14  & BG92 \\ 
Mrk 766 & 0.01293 & 850 $^3$ & 71 & 0.52 & 360 & 1.85 & OP85 G89 GWWB98 \\
PG 1244+026 & 0.048 & 830 & 41 & 1.20 & & 0.41  & BG92 \\
IRAS 13224$-$3809 & 0.0667 & 650 & 23 & 2.42 & 810 & 0.60 & this paper; K98 \\
IRAS 13349+2438 & 0.108 & 2200 & 72 & 6.5 $^4$ & & 0.13 & GWWB98 \\
PG 1404+226 & 0.098 &  880 & 54 & 1.01 & & 0.12 &  BG92 \\
Mrk 478 & 0.079 & 1450 & 64 & 1.19 & & 0.15 & BG92 \\
IRAS 17020+4544 & 0.0604 & 1040 & 23 & 1.86 & 1021 & 2.48 & this paper; LKWWG97 \\
Mrk 507 & 0.055 & 1150 & 19 & 1.45 & 454 & 0.43 & this paper \\
KAZ 163 & 0.063 & 1620 & 76 & 0.57 & 534 & 0.86  & this paper \\
IRAS 20181$-$2244 & 0.185 & 370 & 22 & 0.89 & 537 & 6.12 & this paper; K98; HM98 \\
Ark 564 & 0.024 & 950 & 43 & 0.95 & 350 & 0.96 & this paper \\
\enddata
\tablenotetext{1}{This number is very uncertain and likely to be an
overestimate because of possible \ion{Fe}{II} emission in the vicinity of H$\beta$.}
\tablenotetext{2}{Total red and blue complexes, taken from Remillard
et al.\markcite{115} 1986, apparently measured between 4434\AA\ and 4684\AA\ (the
usual for the blue component) and 5147\AA\ and 5350\AA\ for the red
component.  For correlations I used 0.94, as 60\% of the \ion{Fe}{II} in the
above ranges is in the blue component in the  I~Zw~1 template.}
\tablenotetext{3}{The width of the so-called narrow component of
H$\beta$ listed by Osterbrock \& Pogge\markcite{103} 1985.  This corresponds most
closely to the FWHM measured from Figure 4 of Goodrich\markcite{44} 1989.}
\tablenotetext{4}{Grupe et al.\markcite{49} 1998b report the equivalent width of the
whole \ion{Fe}{II} complex from 4250\AA\ to 5880\AA\ found using the I~Zw~1
template.  I use 30\% of this value for the correlations.}
\tablerefs{BG92: Boroson \& Green\markcite{9} 1992; 
G96: Grupe\markcite{45} 1996; 
BK80: Bergeron \& Kunth\markcite{3} 1980; 
LEWMB97: Lawrence et al.\markcite{75} 1997; 
ZO90: Zheng \& O'Brien\markcite{141} 1990; 
GW94: Gelderman \& Whittle\markcite{38} 1994; 
BK84: Bergeron \& Kunth\markcite{4} 1984; 
K91: Korista\markcite{65} 1991; 
C97: Corbin\markcite{21} 1997; 
R86: Remillard et al.\markcite{115} 1986; 
MPG96: Mason, Puchnarewicz \& Jones\markcite{85} 1996; 
GVV98: Goncalves, Veron \& Veron-Cetty\markcite{43} 1998;  
OP85: Osterbrock \& Pogge\markcite{103} 1985;
G89: Goodrich\markcite{44} 1989; 
GWWB98: Grupe et al.\markcite{49} 1998b; 
K98: Kay et al.\markcite{64}  1998; 
LKWWG97: Leighly et al.\markcite{77} 1997b; 
HM98: Halpern \& Moran\markcite{59} 1998.}
\end{deluxetable}

\clearpage

\singlespace
\begin{deluxetable}{lccccc}
\scriptsize
\tablewidth{0pc}
\tablenum{2}
\tablecaption{Preliminary Spectral Fitting Results}
\tablehead{
\colhead{Target} & \colhead{d.o.f. $^1$} &
\colhead{Power Law $^2$} & 
 \colhead{Soft Excess $^2$} & \colhead{Two-Edge $^2$} &
\colhead{{\it Absori} $^2$} \\
& &  \colhead{$\chi^2_\nu$}  &
\colhead{$\chi^2_\nu$}  & \colhead{$\chi^2_\nu$}  & \colhead{$\chi^2_\nu$}   \\}
\startdata

Mrk 335 & 739 & 969  & 749 & 939 & 786 \nl
I Zw 1 & 549 & 627 & 627 & 626 & 623 \nl
Ton S180 & 856 & 1109 & 888 & 1109 & 1011 \nl
PHL 1092 & 246 & 335 & 239 & 335 & 265 \nl
RX J0439$-$45 & 330 & 475 & 320 & 475 & 341 \nl
NAB 0205+024 & 599 & 638 & 602 & 638 & 612  \nl
PKS 0558$-$504 & 1093 & 1126 & 1087 & 1126 & 1113 \nl
1H 0707$-$495 & 368 & 955 & 531 & 897 & 623 \nl
Mrk 142 & 365 & 364 & 336 & 364 & 345 \nl
RE 1034+39 & 324 & 415 & 315 & 400 & 330 \nl
NGC 4051(1) & 1177 & 2759 & 1313 & 2329 & 1698 \nl
NGC 4051(2) & 1850 & 4191 & 2288 & 3716 & 2646 \nl
PG 1211+143 & 582 & 1047 & 544 & 903 & 625 \nl
Mrk 766 & 1385 & 2102 & 1506 & 1535 & 1425  \nl
PG 1244+026 & 480 & 560 & 512 & 560 & 552  \nl
IRAS 13224$-$3809 & 404 & 1207 & 509 & 1207 & 1207 \nl
IRAS 13349+2438 & 568 & 583 & 569 & 541 & 549  \nl
PG 1404+226 & 230 & 565 & 276 & 565 & 371 \nl
Mrk 478 & 530 & 770 & 590 & 677 & 620 \nl
IRAS 17020+4544 & 939 & 1347 & 1347 & 1327 & 1080 \nl
Mrk 507 & 152 & 150 & 150 & 142 & 137  \nl
KAZ 163 & 249 & 221 & 212 & 220 & 218  \nl
IRAS 20181$-$2244 & 400 & 663 & 663 & 656 & 456 \nl
Ark 564 & 1402 & 2185 & 1590 & 2185 & 1819  \nl
\enddata
\tablenotetext{1}{Number of degrees of freedom for the power law plus
Galactic absorption model.  The number of degrees of freedom for the
other continuum models are these numbers less 2.}
\tablenotetext{2}{See text for description of the models.}
\end{deluxetable}

\clearpage

\renewcommand{\thepage}{\arabic{page}}
\setcounter{page}{53}

\begin{deluxetable}{lllllllllllllllll}
\scriptsize
\tablewidth{0pc}
\tablenum{4}
\tablecaption{{\it Absori} Warm Absorber Spectral Fitting Results}
\tablehead{
\colhead{Target} & \colhead{Photon Index} & \colhead{$N_W$} &
\colhead{$\xi$} & \colhead{$\rm T$ $^1$} & \colhead{$\Delta \chi^2$ $^2$} & 
\colhead{$\chi^2$/dof} \\
& & \colhead{($\times 10^{22}\rm \,cm^{-2}$)} & & \colhead{(K)} }
\startdata

Mrk 335 & $2.08 \pm 0.03$ & $2.6^{+2.3}_{-1.2}$ & $570^{+440}_{-250}$
& $1.0 \times 10^6$ & 19 & 705/732 \nl
I Zw 1 & $2.42^{+0.07}_{-0.08}$ & $0.06 \pm 0.04$ &
$0.03^{+0.24}_{-0.03}$ & $3.0 \times 10^4$ & 10 & 598/544 \nl 
NAB 0205+024 & $2.24\pm 0.07$ & $0.28^{+0.32}_{-0.18}$ &
$33^{+120}_{-33}$ & $3.0\times 10^4$ & 8(m) & 594/595 \nl
NGC 4051(1) & $2.05^{+0.04}_{-0.03}$ & $0.11^{+0.11}_{-0.05}$ &
$1.9^{+16}_{-0.9}$ & $1.0 \times 10^6$ & 15 & 1243/1167  & \nl
NGC 4051(2) & $2.00 \pm 0.02$ & $0.21 \pm 0.04$ & $1.7^{+0.6}_{-0.4}$
& $1.0 \times 10^6$ & 90 & 2058/1844 \nl
Mrk 766 & $1.98 \pm 0.03$ & $0.30^{+0.08}_{-0.06}$ &
$9.8^{+7.9}_{-4.0}$ & $3.0 \times 10^4$ & 154 & 1338/1378 \nl
PG 1244+026 & $2.40^{+0.12}_{-0.13}$ & $0.82^{+0.88}_{-0.56}$ &
$260^{+680}_{-210}$ & $3.0 \times 10^4$ & 6(m) & 483/471 \nl
IRAS 13224$-$3809 & $2.07^{+0.27}_{-0.30}$ & $1.12^{+0.48}_{-0.51}$ &
$86^{+104}_{-35}$ & $3.0 \times 10^4$ & 6(m) & 401/396 \nl
IRAS 13349+2438 & $2.39 \pm 0.05$ & $0.18^{+0.07}_{-0.05}$ &
$2.1^{+2.7}_{-1.1}$ & $1.0 \times 10^6$ & 38 & 540/563 \nl
Mrk 478 & $2.22^{+0.16}_{-0.13}$ & $0.54^{+0.38}_{-0.28}$ &
$21^{+27}_{-16}$ & $1.0\times 10^6$ & 14 & 567/523 \nl
IRAS 17020+4544 & $2.45\pm 0.05$ & $0.22^{+0.06}_{-0.05}$ &
$24^{+24}_{-15}$ & $3.0 \times 10^4$ & 80 & 1043/935 \nl
KAZ 163 & $1.50^{+0.28}_{-0.30}$ & $1.2^{+0.8}_{-0.7}$ &
$60^{+130}_{-40}$ & $3.0 \times 10^4$ & 7(m) & 205/245 \nl
IRAS 20181$-$2244 & $2.55^{+0.12}_{-0.11}$ & $0.24^{+0.21}_{-0.16}$ &
$60^{+290}_{-60}$ & $3.0 \times 10^4$ & 7(m) & 449/395 \nl
Ark 564 & $2.59^{+0.01}_{-0.02}$ & $1.0 \pm 0.4$ & $750^{+280}_{-160}$
& $1.0 \times 10^6$ & 10 & 1439/1393 \nl
\enddata
\tablenotetext{1}{The temperature of the ionized gas was fixed at two
different values for the spectral fitting; the one that gave the
better fit is listed.  }
\tablenotetext{2}{Increase in $\chi^2$ when the warm absorber was
removed from the model. }
\end{deluxetable}

\clearpage

\renewcommand{\thepage}{\arabic{page}}
\setcounter{page}{55}

\begin{figure}[t]
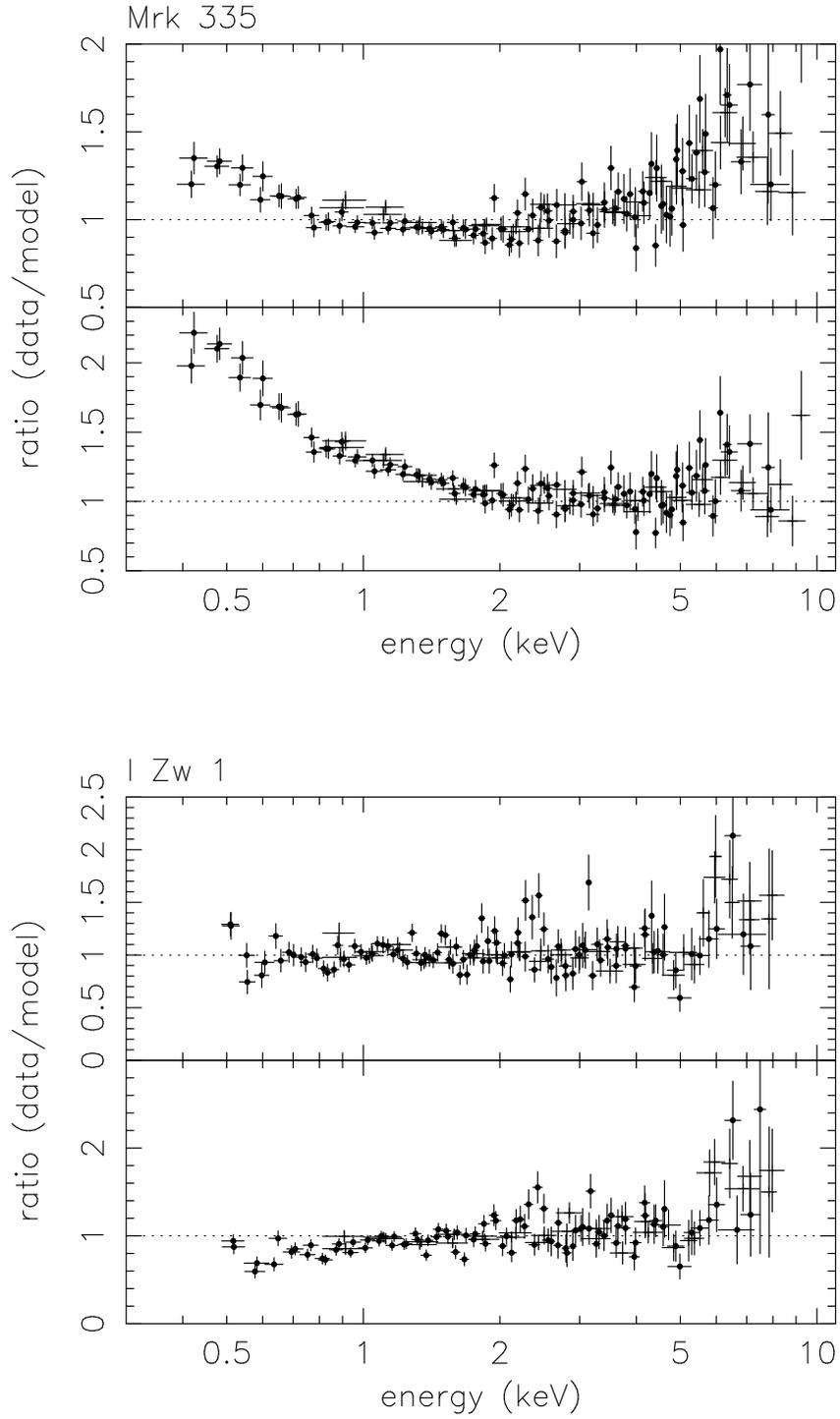

\vbox to4.0in{\rule{0pt}{4.0in}}
\includegraphics{fig1a.ps}
\vbox to4.0in{\rule{0pt}{4.0in}}
\includegraphics{fig1b.ps}
\caption{Top: Ratio of data to power law plus Galactic absorption
model.  Bottom: Ratio of data to best fitting power law model above
2~keV.  }
\end{figure}

\newpage

\renewcommand{\thefigure}{\arabic{figure}}
\setcounter{figure}{0}

\begin{figure}[t]
\vbox to4.0in{\rule{0pt}{4.0in}}
\includegraphics{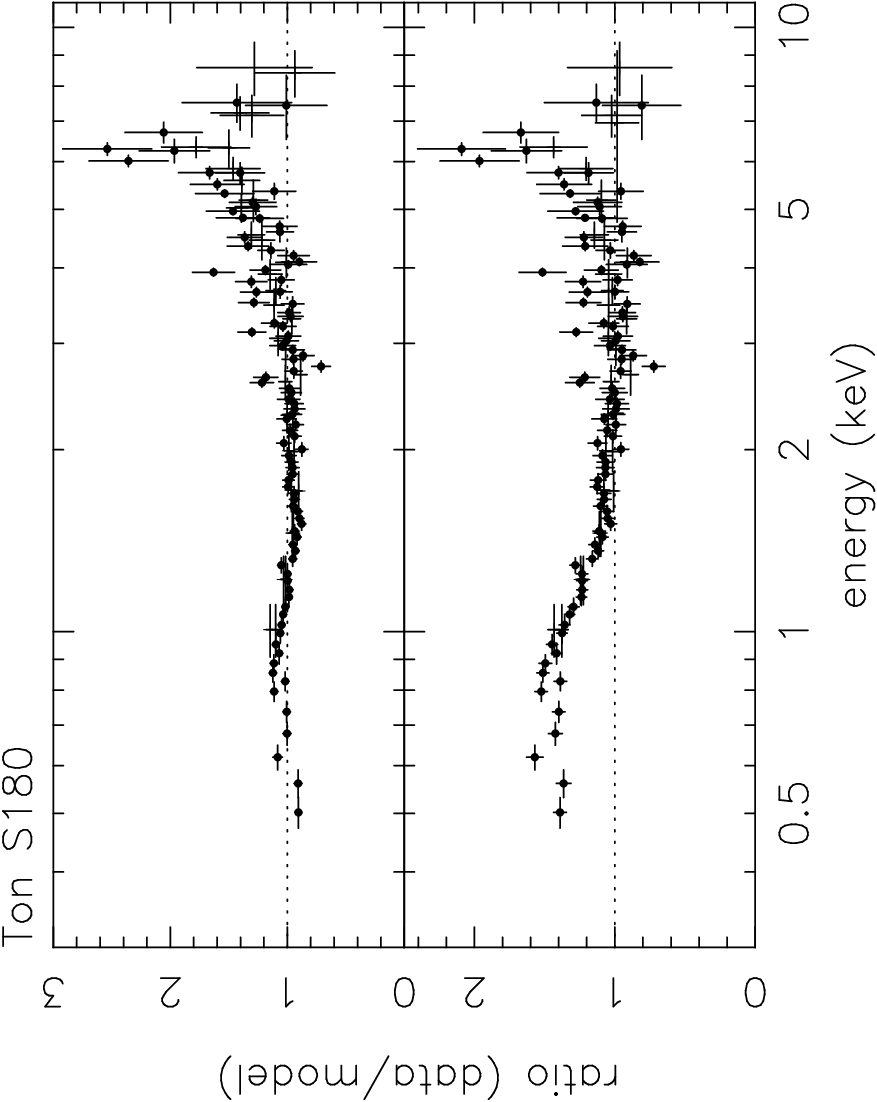}
\vbox to4.0in{\rule{0pt}{4.0in}}
\includegraphics{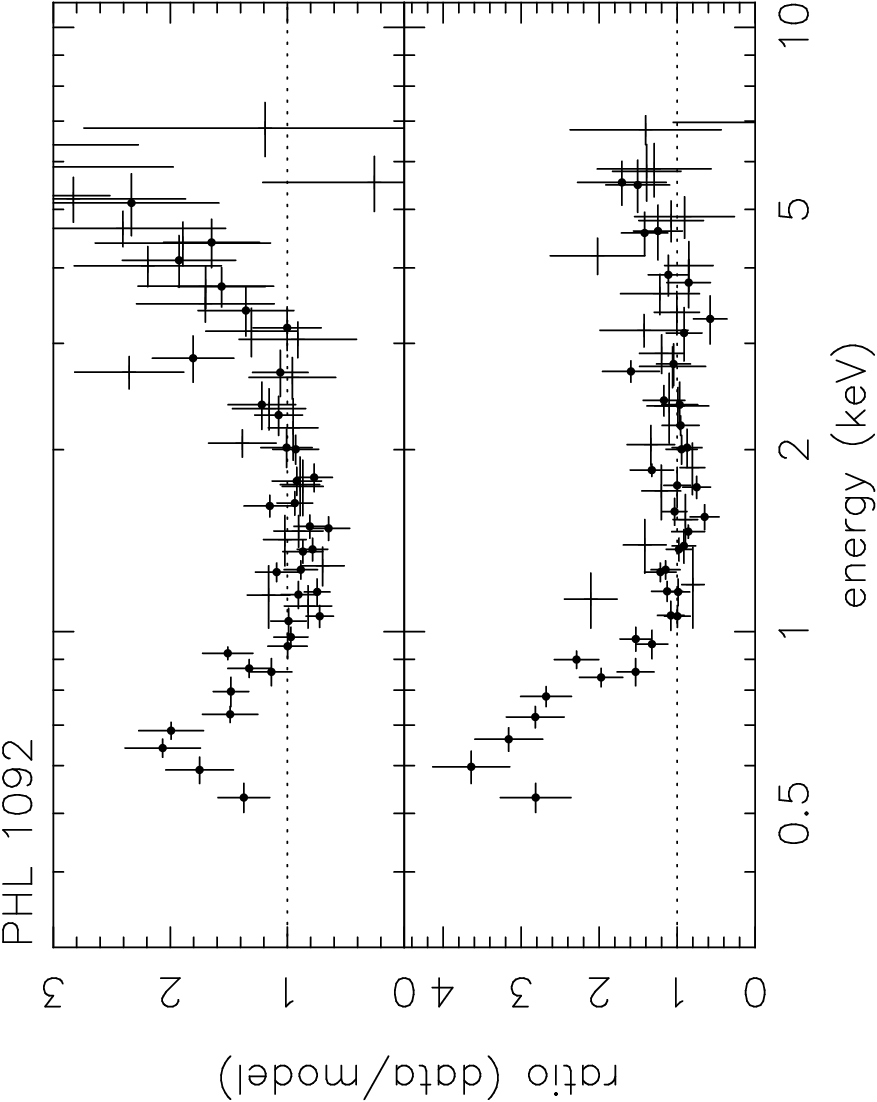}
\caption{continued.}
\end{figure}

\newpage

\renewcommand{\thefigure}{\arabic{figure}}
\setcounter{figure}{0}

\begin{figure}[t]
\vbox to4.0in{\rule{0pt}{4.0in}}
\includegraphics{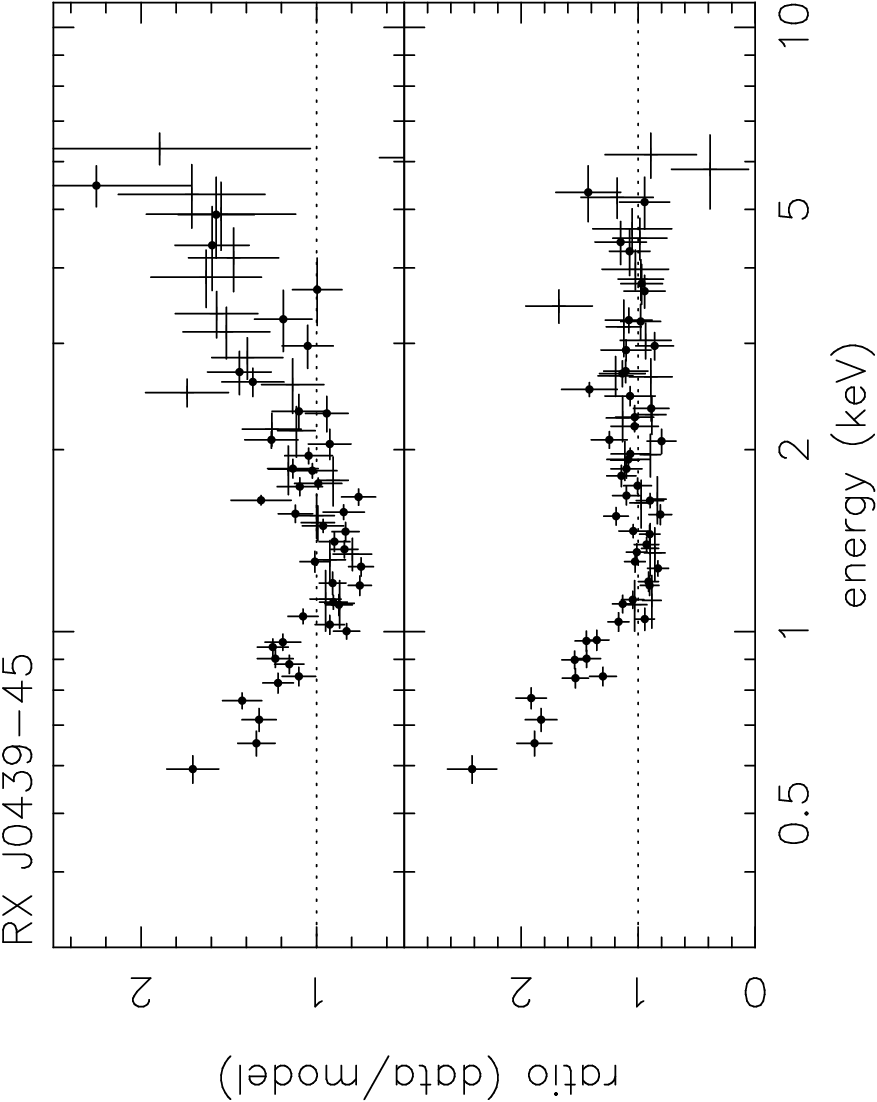}
\vbox to4.0in{\rule{0pt}{4.0in}}
\includegraphics{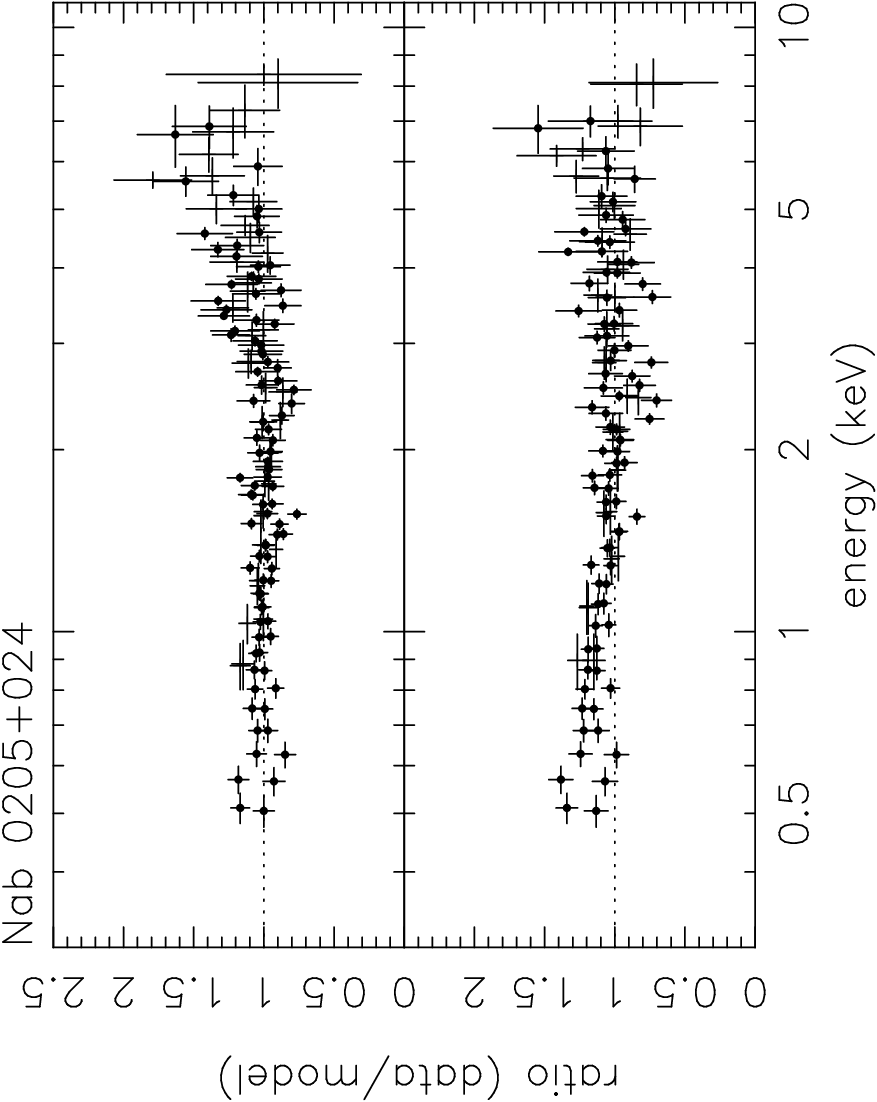}
\caption{continued.}
\end{figure}

\newpage

\renewcommand{\thefigure}{\arabic{figure}}
\setcounter{figure}{0}

\begin{figure}[t]
\vbox to4.0in{\rule{0pt}{4.0in}}
\includegraphics{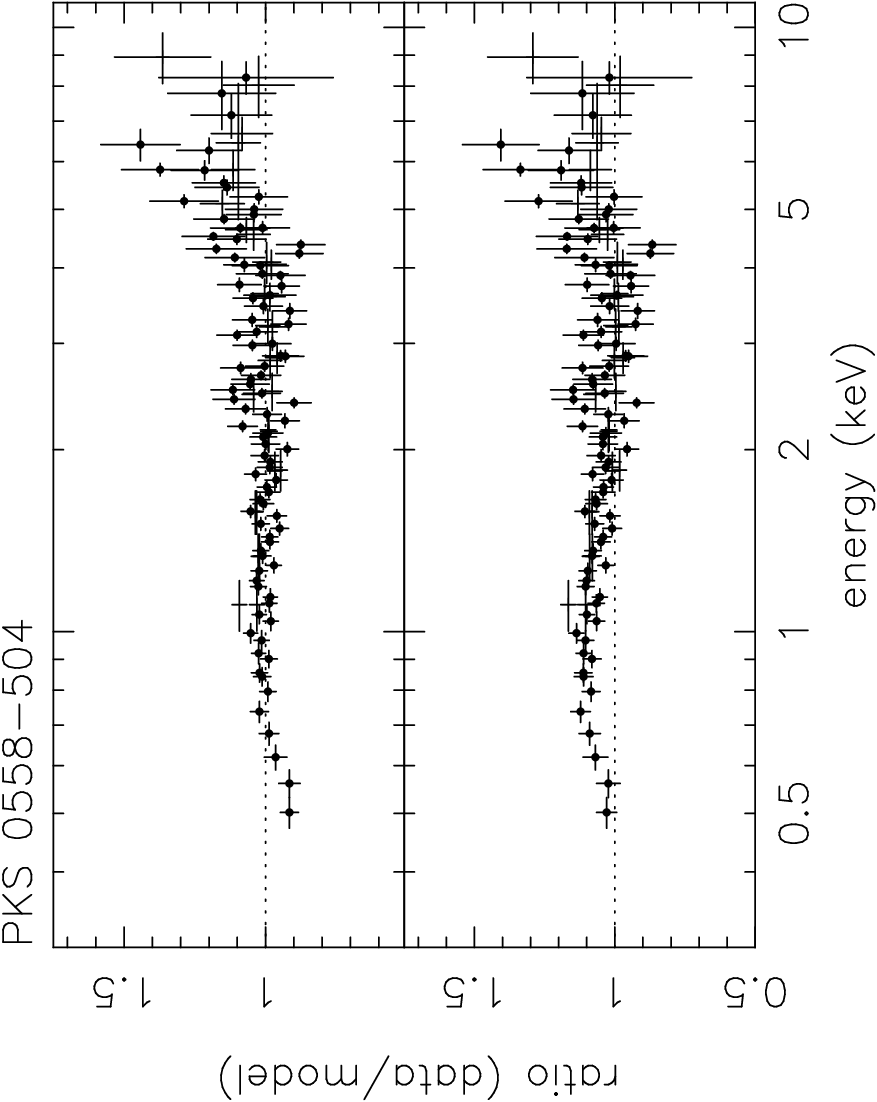}
\vbox to4.0in{\rule{0pt}{4.0in}}
\includegraphics{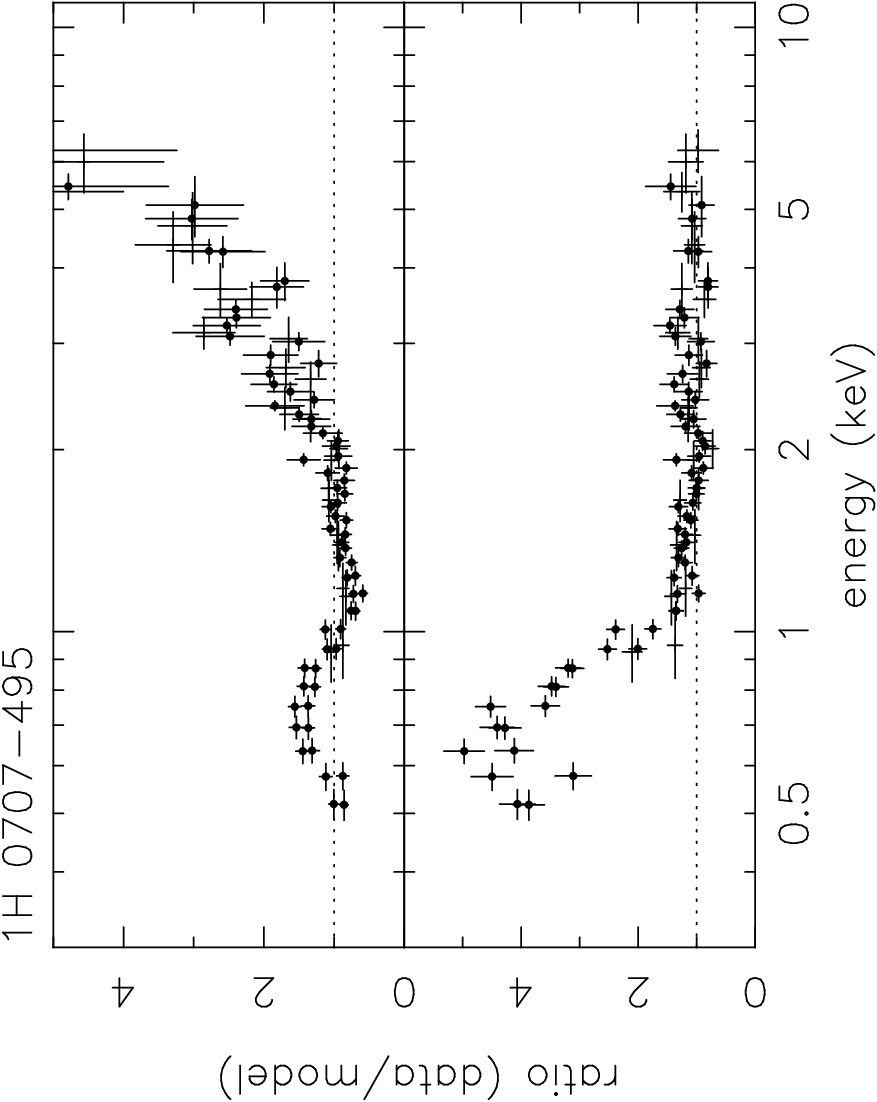}
\caption{continued.}
\end{figure}

\newpage

\renewcommand{\thefigure}{\arabic{figure}}
\setcounter{figure}{0}

\begin{figure}[t]
\vbox to4.0in{\rule{0pt}{4.0in}}
\includegraphics{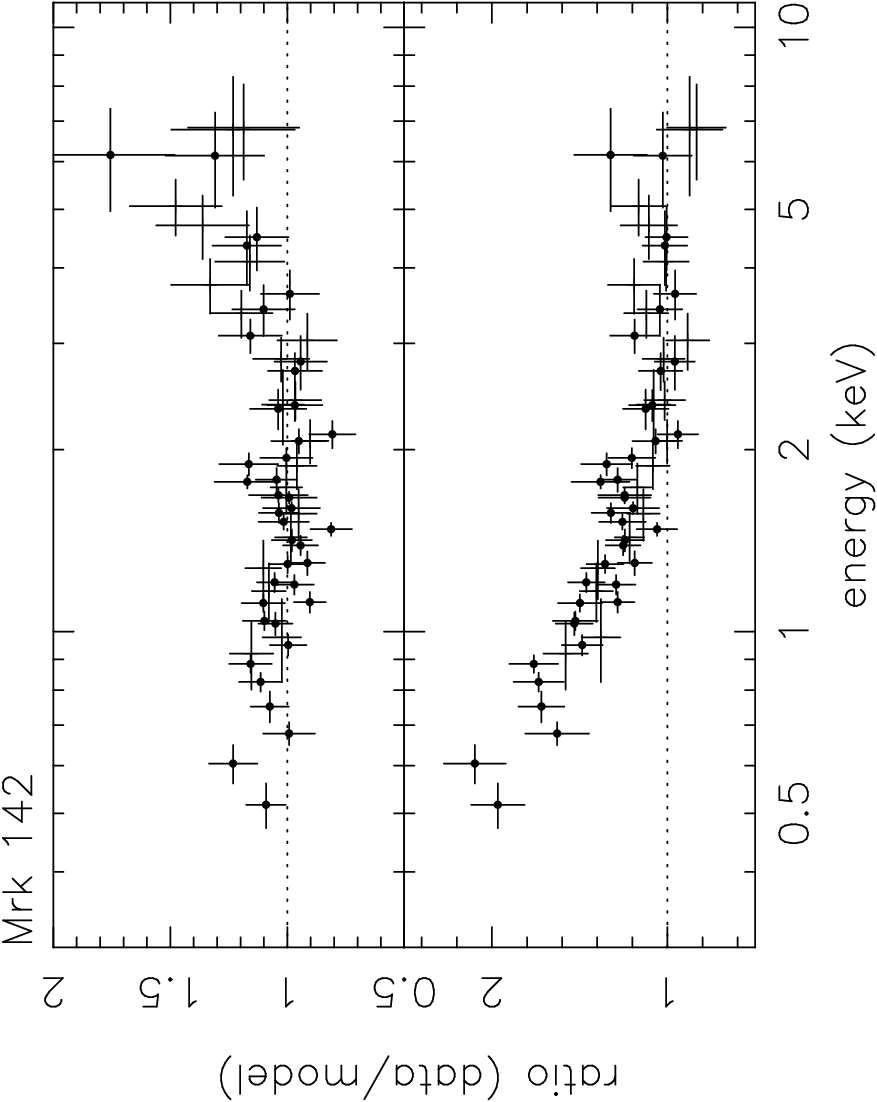}
\vbox to4.0in{\rule{0pt}{4.0in}}
\includegraphics{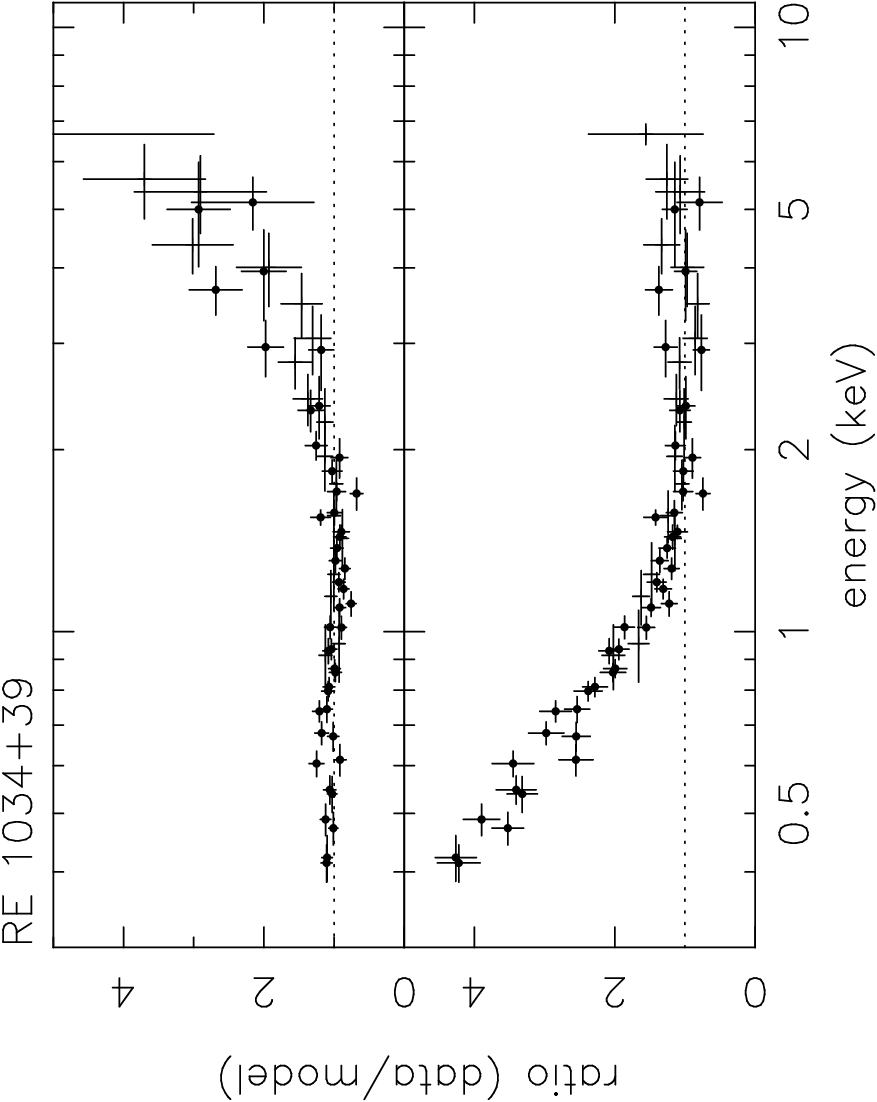}
\caption{continued.}
\end{figure}

\newpage

\renewcommand{\thefigure}{\arabic{figure}}
\setcounter{figure}{0}

\begin{figure}[t]
\vbox to4.0in{\rule{0pt}{4.0in}}
\includegraphics{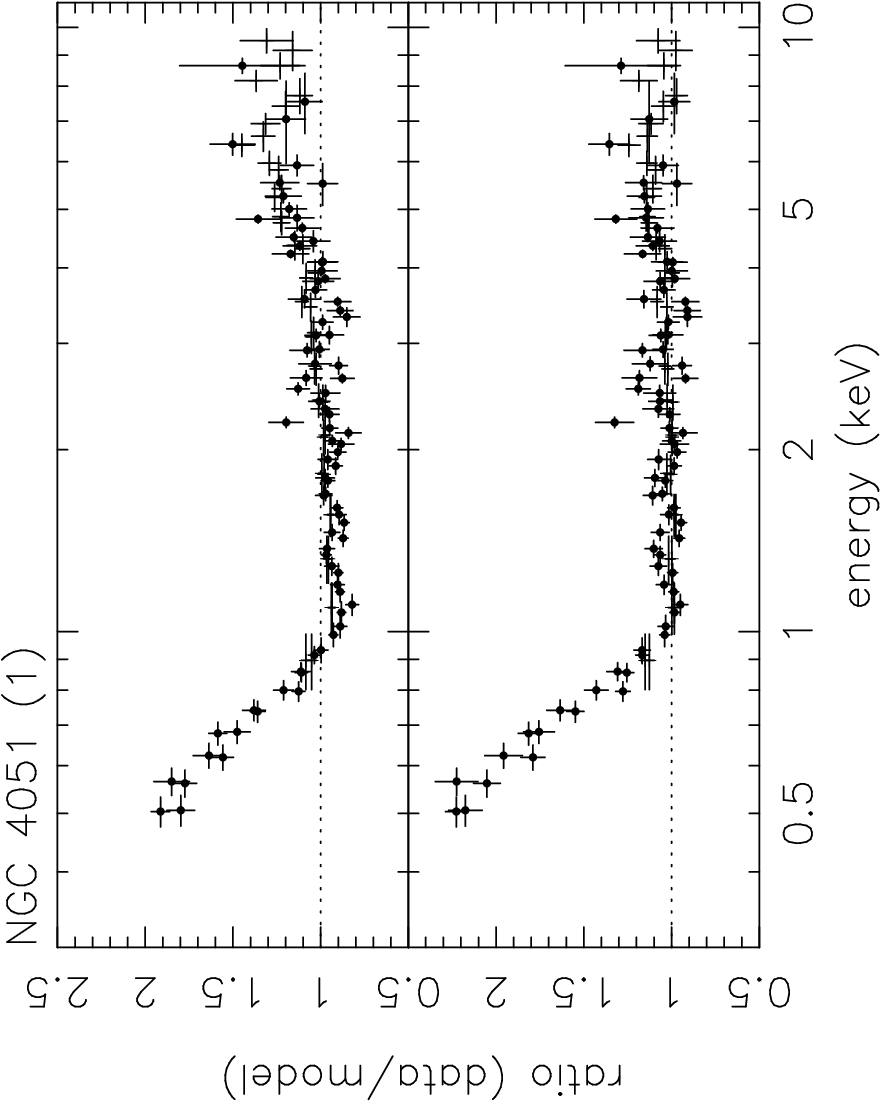}
\vbox to4.0in{\rule{0pt}{4.0in}}
\includegraphics{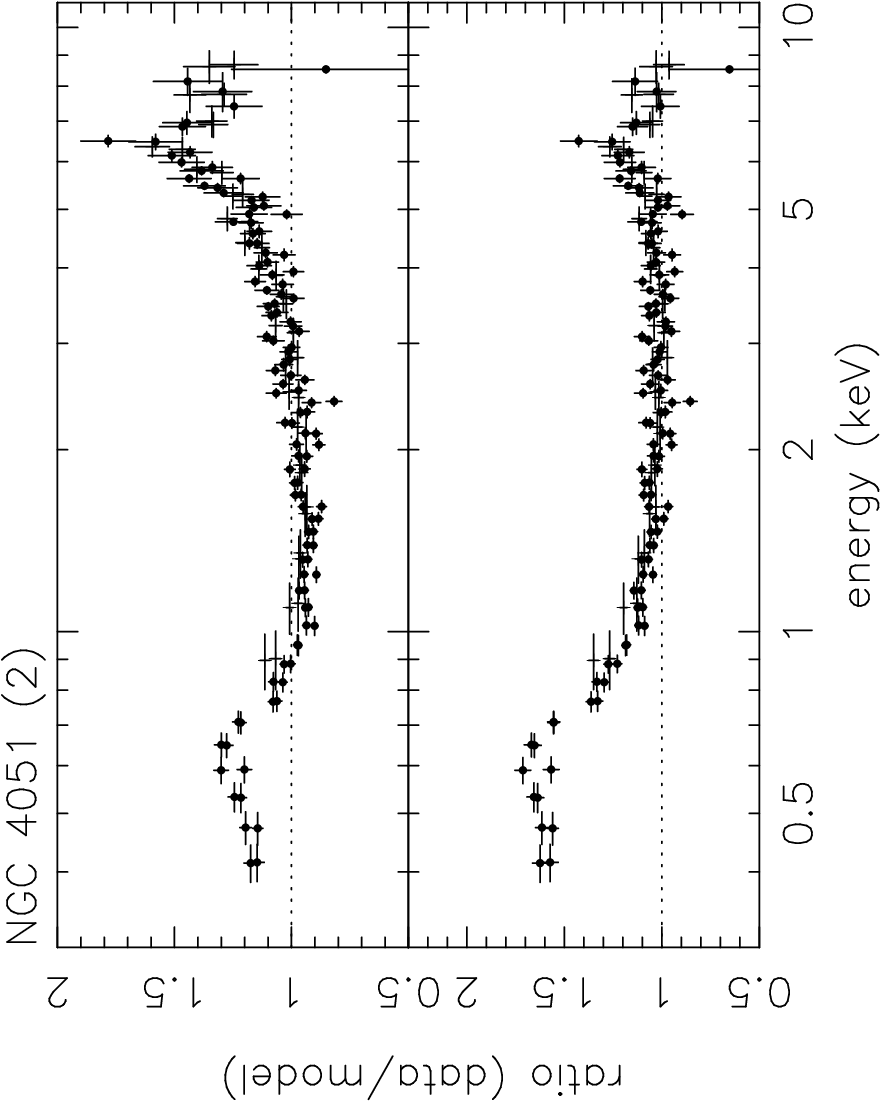}
\caption{continued.}
\end{figure}

\newpage

\renewcommand{\thefigure}{\arabic{figure}}
\setcounter{figure}{0}

\begin{figure}[t]
\vbox to4.0in{\rule{0pt}{4.0in}}
\includegraphics{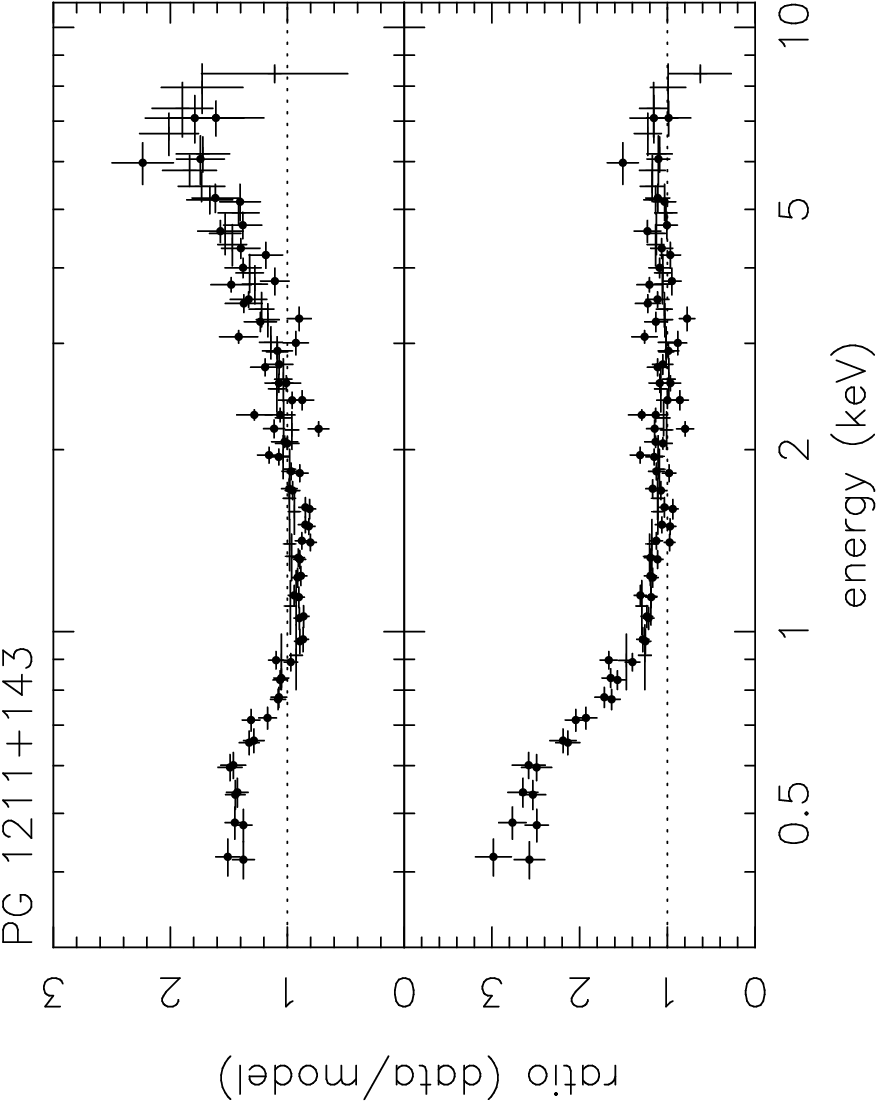}
\vbox to4.0in{\rule{0pt}{4.0in}}
\includegraphics{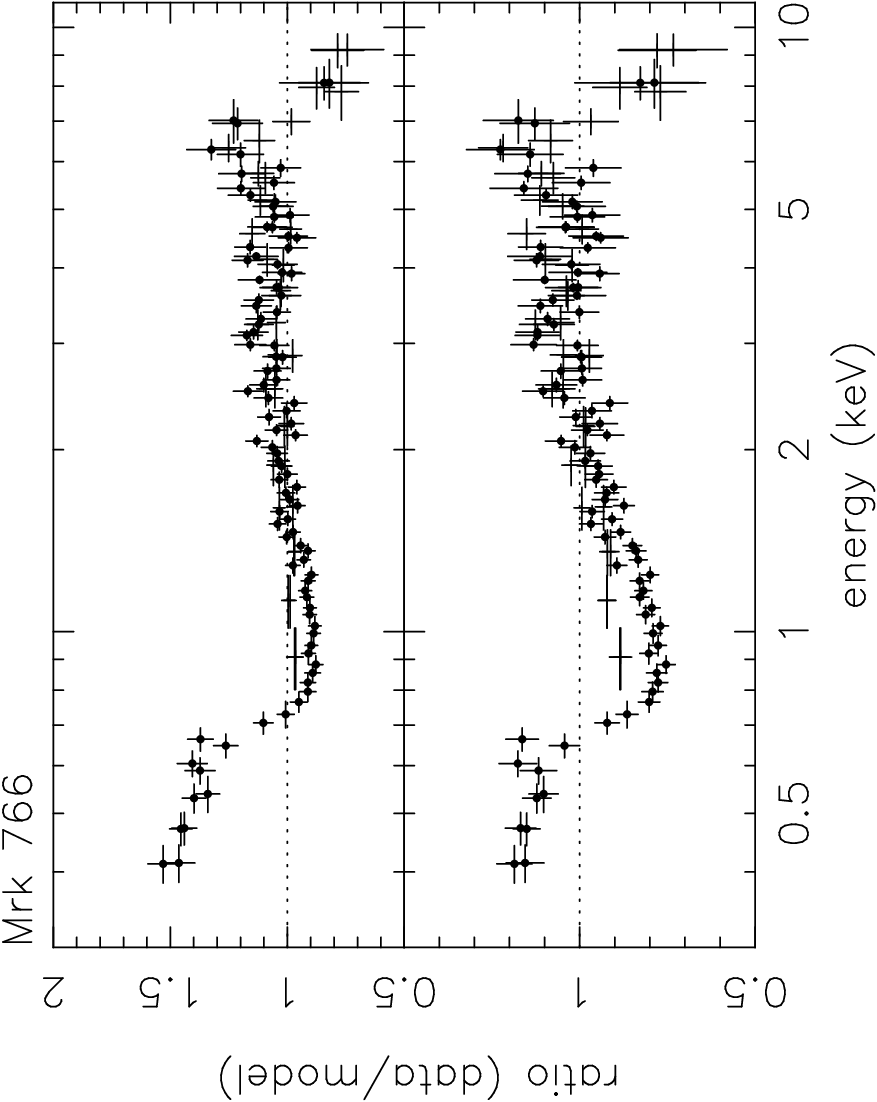}
\caption{continued.}
\end{figure}

\newpage

\renewcommand{\thefigure}{\arabic{figure}}
\setcounter{figure}{0}

\begin{figure}[t]
\vbox to4.0in{\rule{0pt}{4.0in}}
\includegraphics{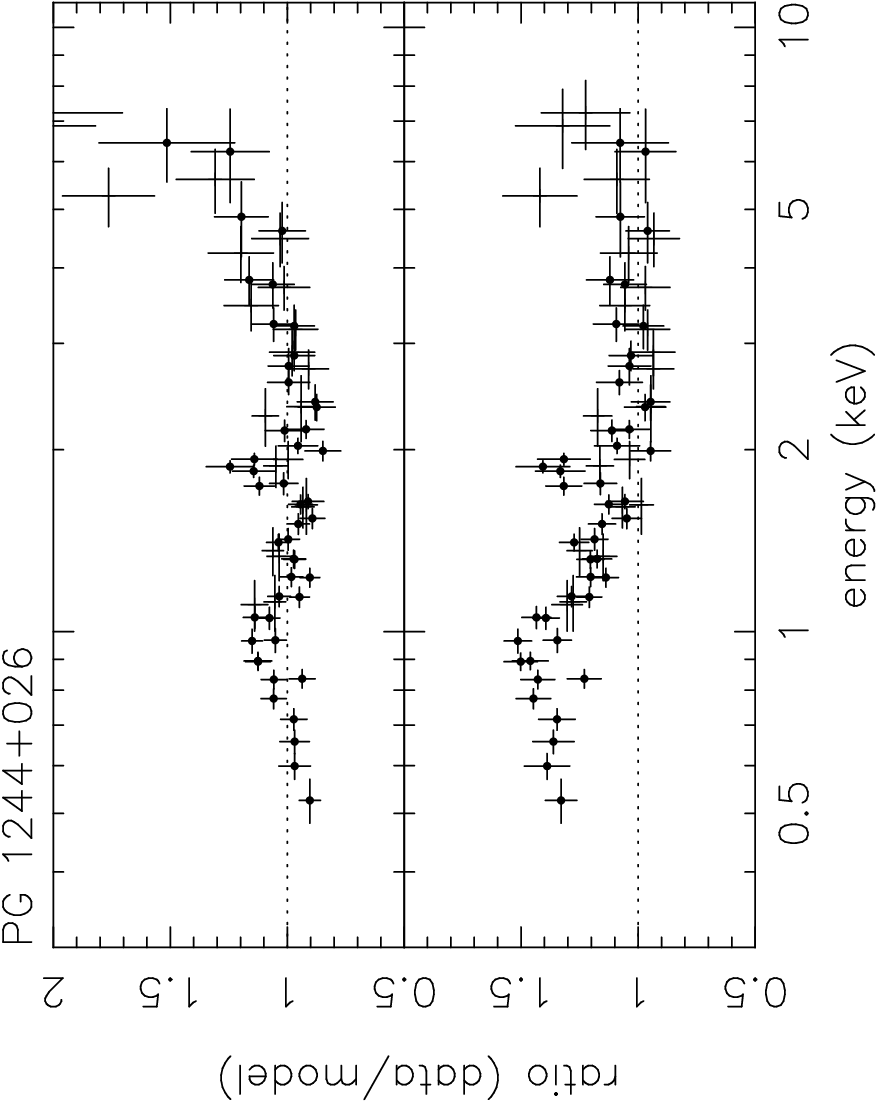}
\vbox to4.0in{\rule{0pt}{4.0in}}
\includegraphics{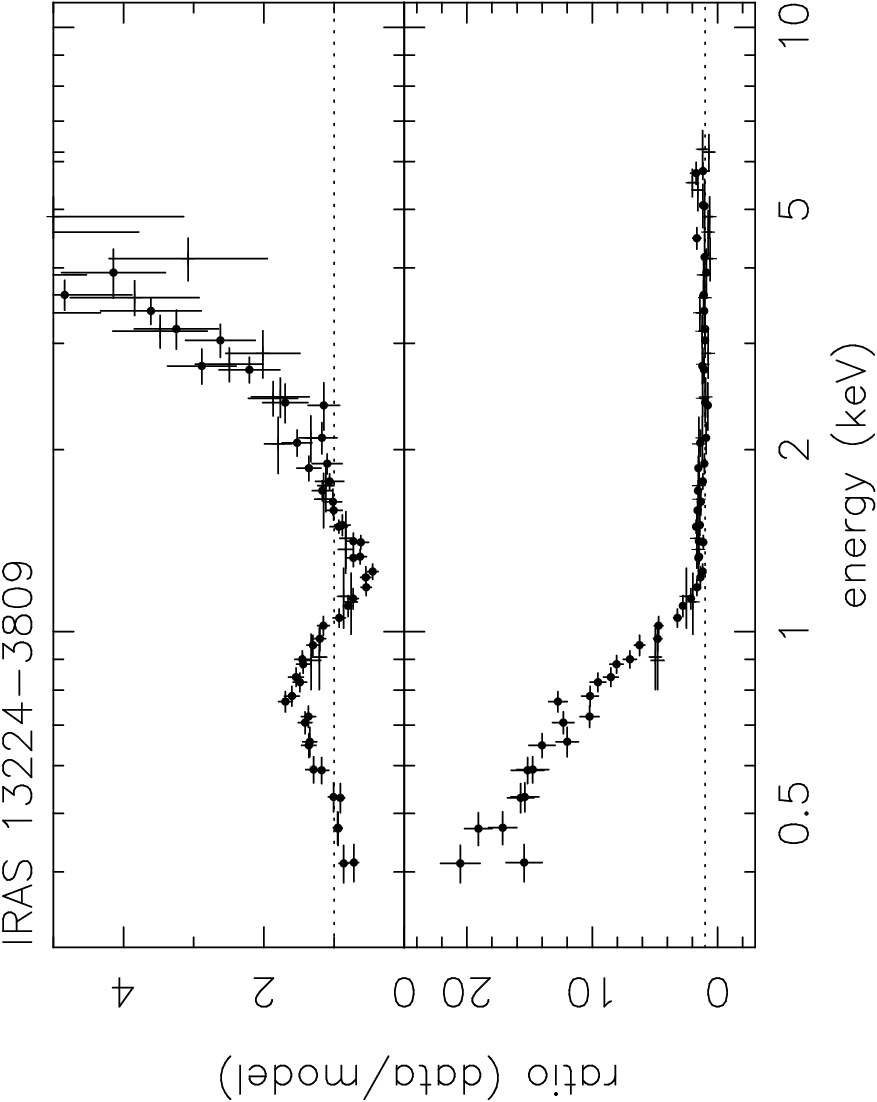}
\caption{continued.}
\end{figure}

\newpage

\renewcommand{\thefigure}{\arabic{figure}}
\setcounter{figure}{0}

\begin{figure}[t]
\vbox to4.0in{\rule{0pt}{4.0in}}
\includegraphics{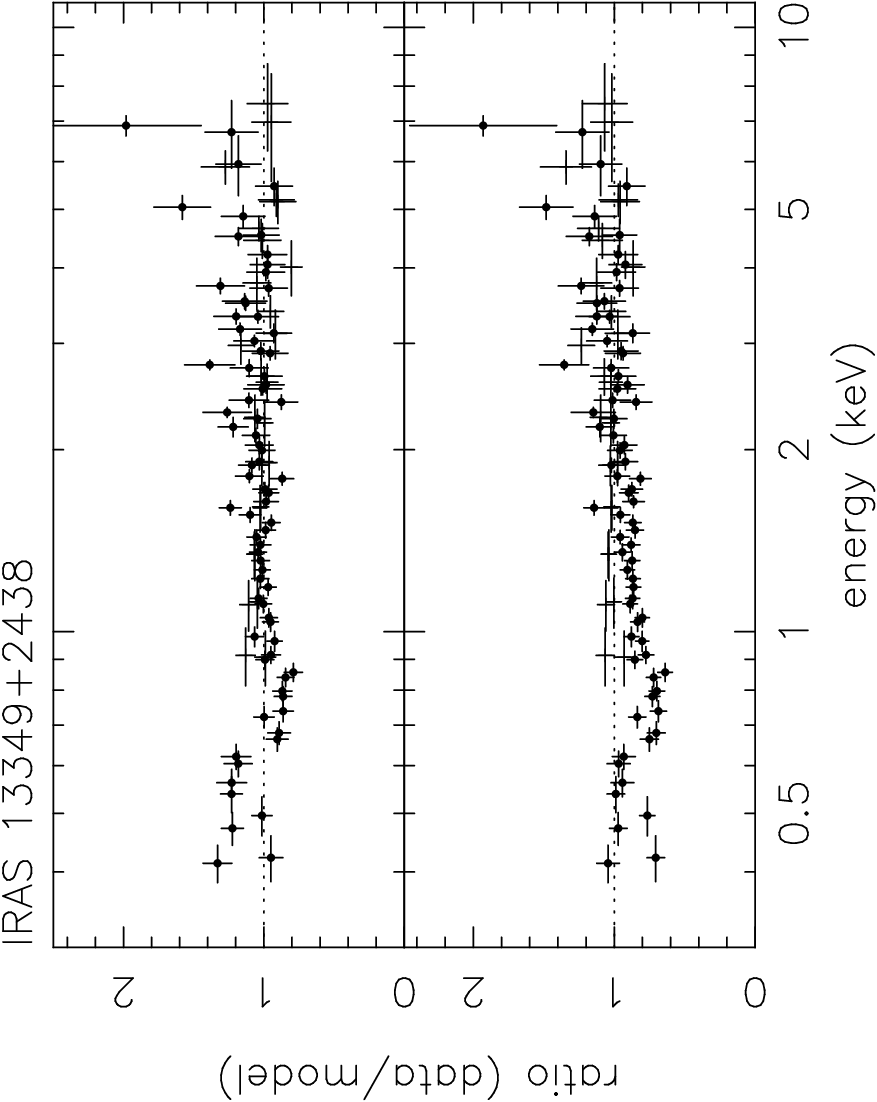}
\vbox to4.0in{\rule{0pt}{4.0in}}
\includegraphics{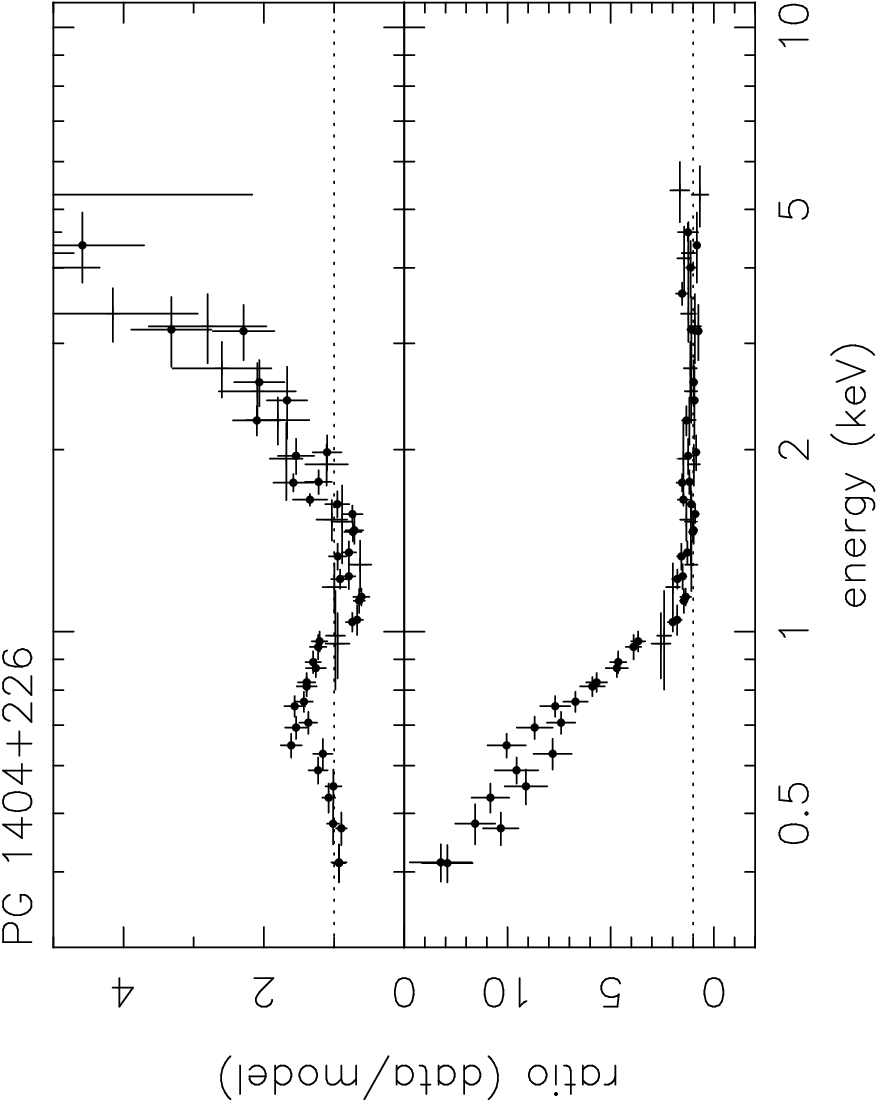}
\caption{continued.}
\end{figure}

\newpage

\renewcommand{\thefigure}{\arabic{figure}}
\setcounter{figure}{0}

\begin{figure}[t]
\vbox to4.0in{\rule{0pt}{4.0in}}
\includegraphics{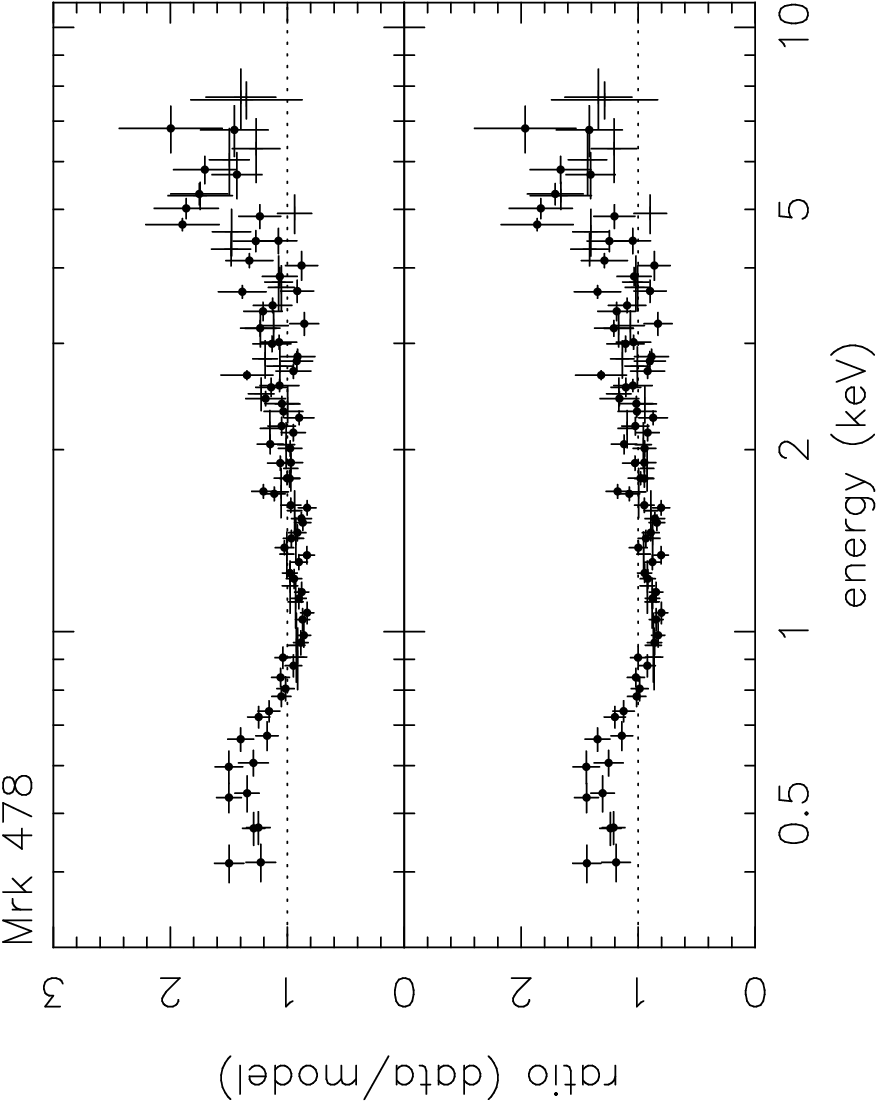}
\vbox to4.0in{\rule{0pt}{4.0in}}
\includegraphics{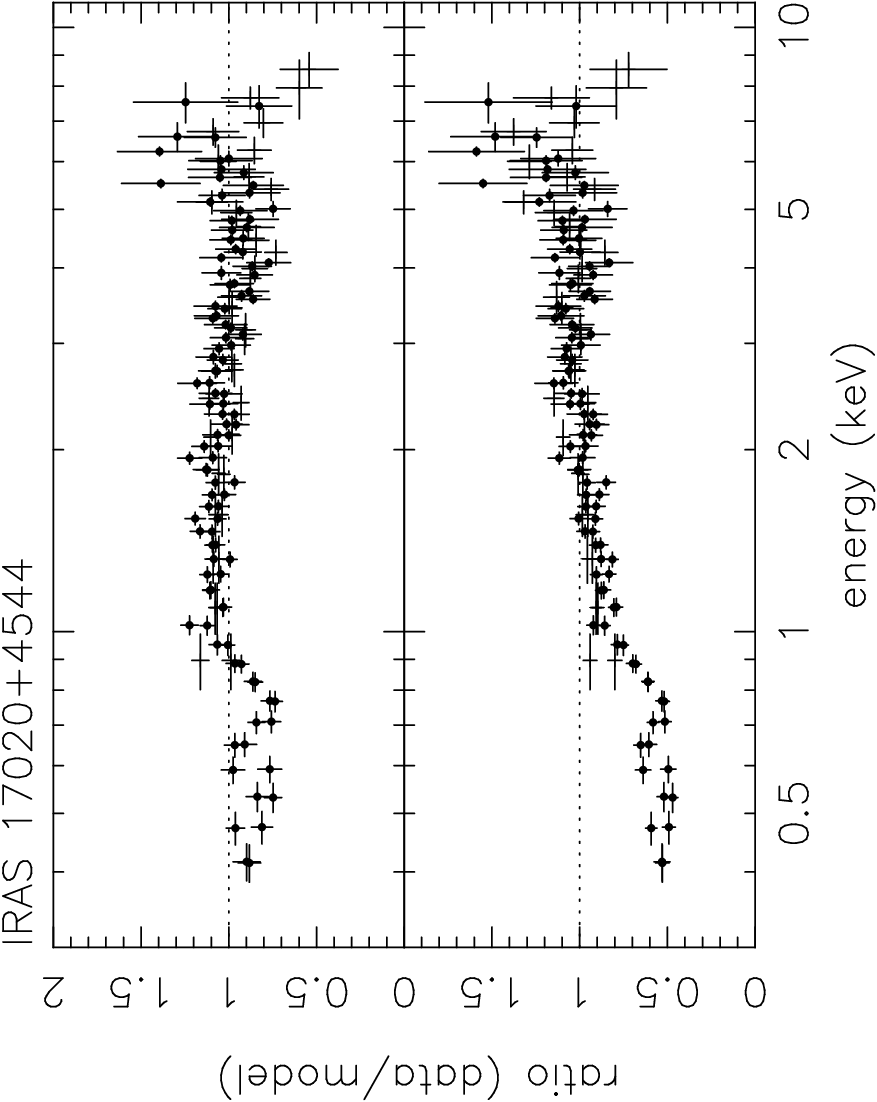}
\caption{continued.}
\end{figure}

\newpage

\renewcommand{\thefigure}{\arabic{figure}}
\setcounter{figure}{0}

\begin{figure}[t]
\vbox to4.0in{\rule{0pt}{4.0in}}
\includegraphics{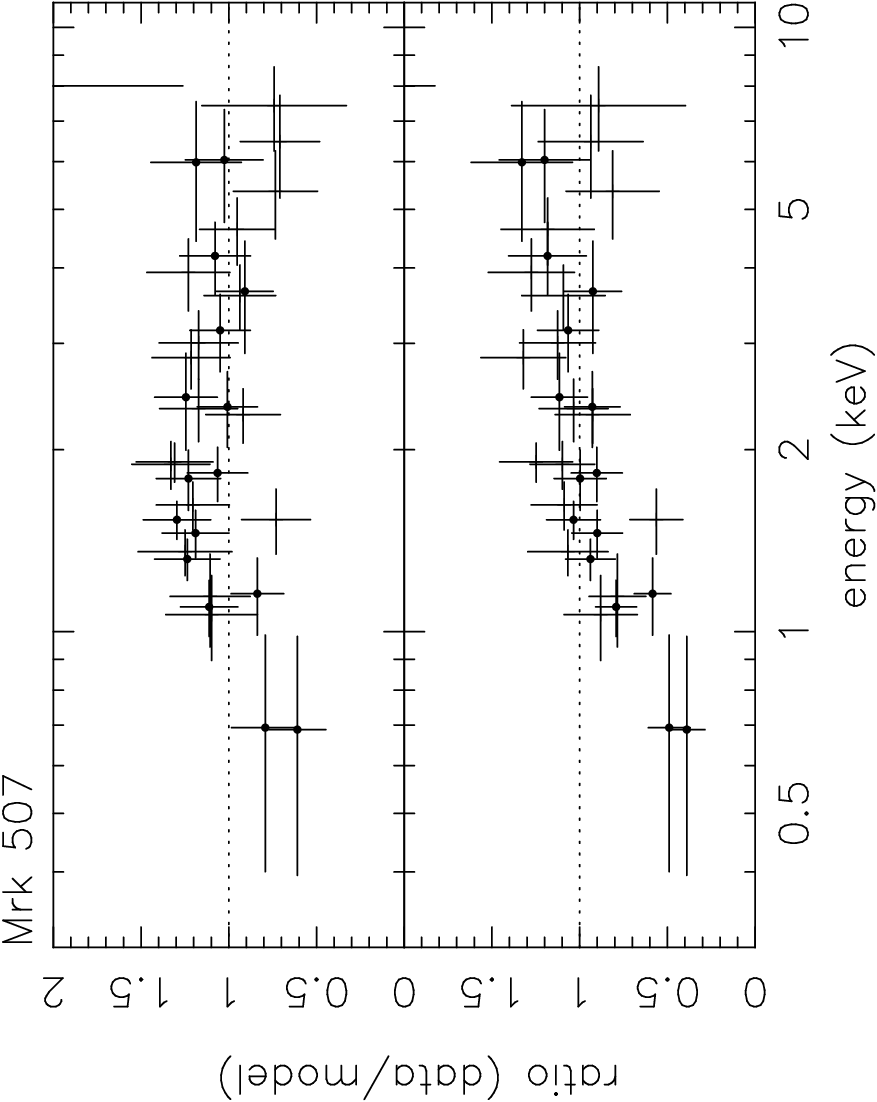}
\vbox to4.0in{\rule{0pt}{4.0in}}
\includegraphics{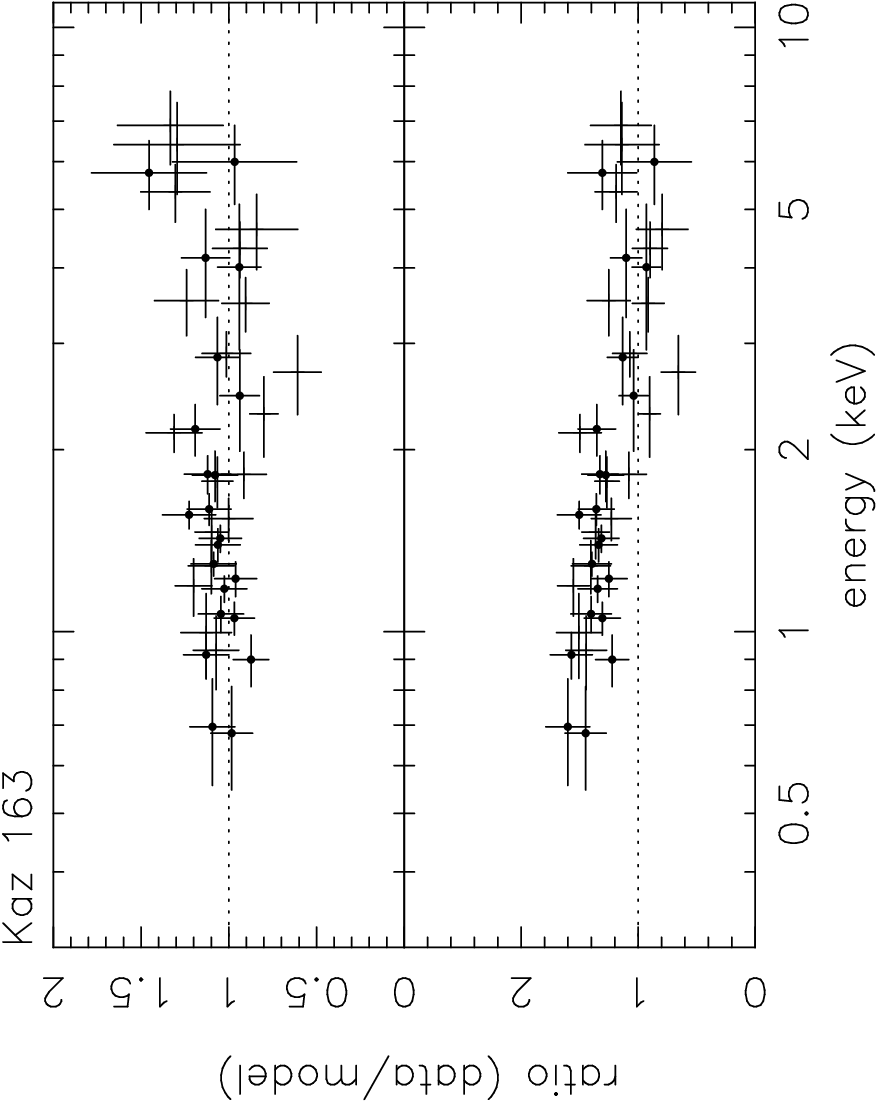}
\caption{continued.}
\end{figure}

\newpage

\renewcommand{\thefigure}{\arabic{figure}}
\setcounter{figure}{0}

\begin{figure}[t]
\vbox to4.0in{\rule{0pt}{4.0in}}
\includegraphics{fig1w.ps}
\vbox to4.0in{\rule{0pt}{4.0in}}
\includegraphics{fig1x.ps}
\caption{continued.}
\end{figure}

\newpage

\begin{figure}[t]
\vbox to3.5in{\rule{0pt}{3.5in}}
\includegraphics{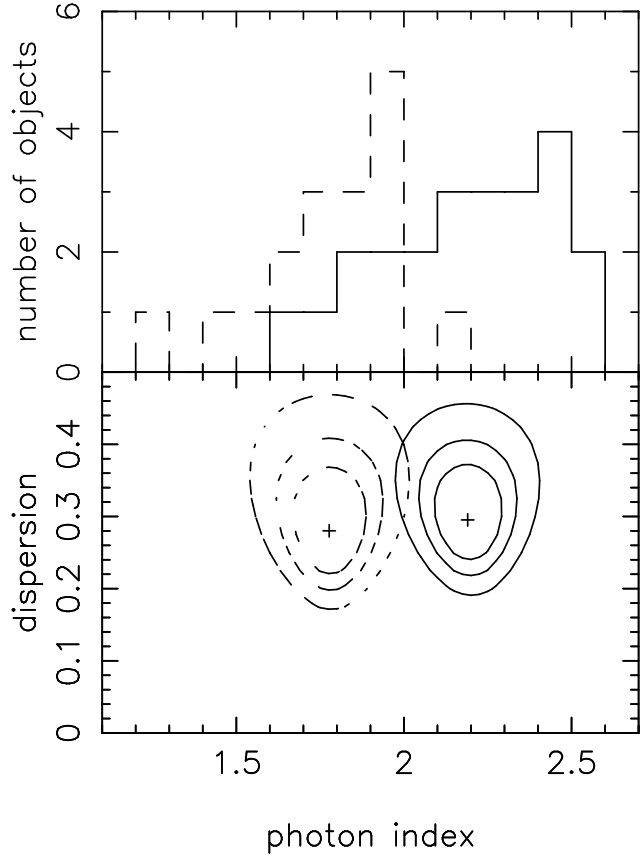}
\caption{The upper panel shows the histogram of the photon indices from the 23 narrow-line
Seyfert 1 galaxies analyzed in this paper (solid line) and from the 17
radio-quiet broad-line Seyfert 1 galaxies in Reynolds (1997) (dashed line).  The
lower panel shows the 68, 90 and 99\% $\chi^2$ contours for two
degrees of freedom from the maximum
likelihood determination of the average photon index and the
dispersion of the photon indices. For both broad and narrow-line
objects, the dispersion is significantly greater than zero, ruling out
a single photon index for each class of object, and the average photon
index for the two classes are significantly different at 90\%
confidence.}
\end{figure}

\newpage

\begin{figure}[t]
\vbox to4.0in{\rule{0pt}{4.0in}}
\includegraphics{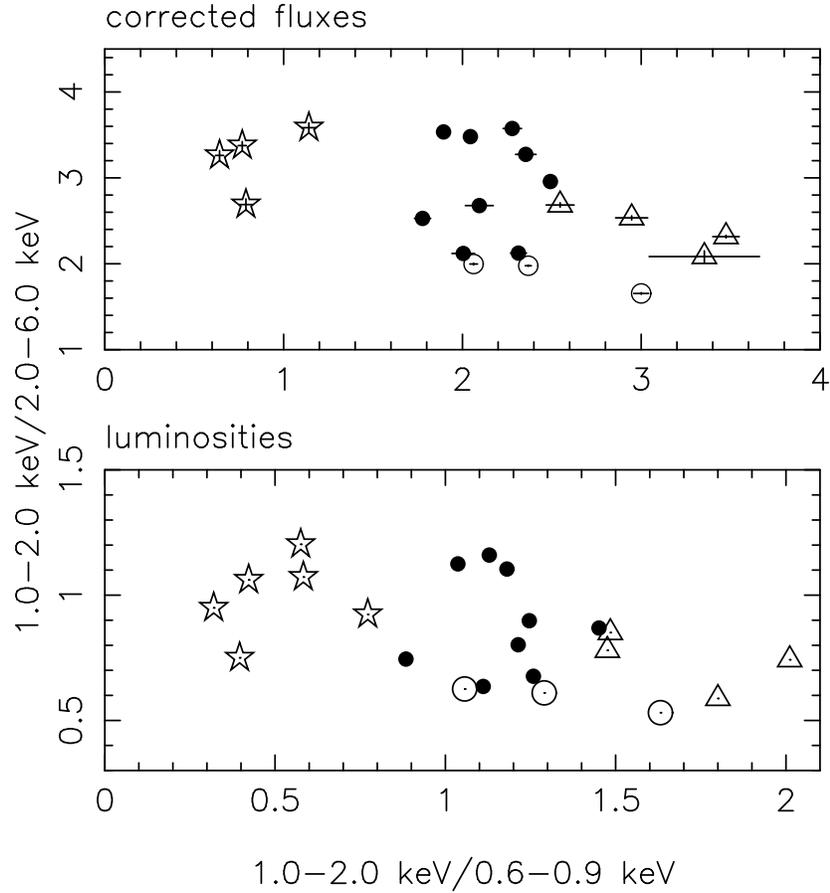}
\caption{Color ratios plots showing the medium (1.0-2.0~keV) to
soft (0.6-0.9~keV) ratio on the x axis, and the medium to hard ratio
(2.0--6.0~keV) on the y axis.  The objects separate according to
the x axis parameter, the strength of the soft excess.  Objects with
the strongest soft excesses are marked by stars and are found on the
left part of the graph, while objects marked by circles have weaker
but still statistically highly significant soft excesses.  Mrk~766 and
NGC~4051 are distinguished by open circles.  Objects marked by
triangles have no detected soft excess and are typically absorbed
sources.  The highly absorbed objects Mrk~507 and IRAS~20181$-$2244
are off the plot toward the right. Top: ratios from SIS0+SIS1 count
rates after correction for Galactic absorption and redshift.  The
ratios for the highest redshift objects cannot be correctly determined
this way, so PHL~1092 and RX~J0439$-$45 are not shown.  Bottom: ratios
from model luminosities.}
\end{figure}

\newpage

\begin{figure}[t]
\vbox to3.0in{\rule{0pt}{3.0in}}
\includegraphics{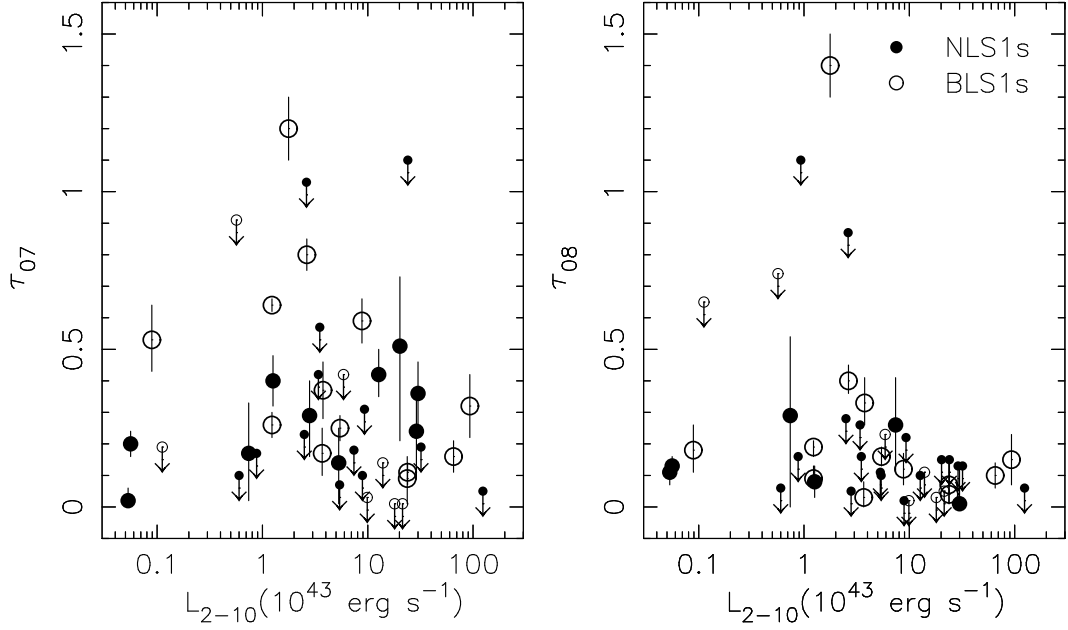}
\caption{The O~VII and O~VIII optical depths versus luminosity for the
narrow-line Seyfert 1 galaxies analyzed in this paper (solid points)
and 20 broad-line AGN presented in Reynolds (1997; open points).  Note
that 3C~273 ($L_x=1530 \times 10^{43}\rm\, ergs\,s^{-1}$) is not
plotted to preserve the scale.  Upper limits are plotted with a
smaller version of the same symbol and an arrow.  The distributions of
$\tau_{O\,VII}$ are the same for both NLS1s and broad-line AGN, but
the distributions of $\tau_{O\,VIII}$ are different at $>97$\%
confidence.  The difference is attributable to a lower optical depth
of O~VIII in the NLS1s.  The left panel indeed shows that the upper
limits of many of the NLS1s are smaller than the detections from the
broad-line Seyfert 1s.}
\end{figure}

\newpage

\begin{figure}[t]
\vbox to5.0in{\rule{0pt}{5.0in}}
\includegraphics{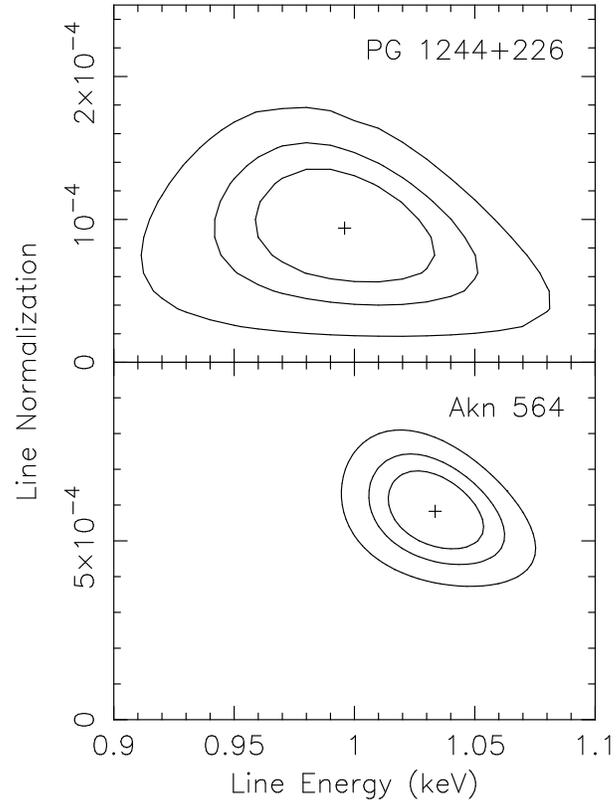}
\caption{$\chi^2$ contours (68\%, 90\% and 99\%) for the central energy in the
rest frame of the object and width $\sigma$ of a Gaussian line model
for the soft X-ray line in PG~1244+226 and Ark~564.  The width
$\sigma$ has been fixed at its best fit value for stability in
calculating the contour.}
\end{figure}

\newpage

\begin{figure}[t]
\vbox to5.0in{\rule{0pt}{5.0in}}
\includegraphics{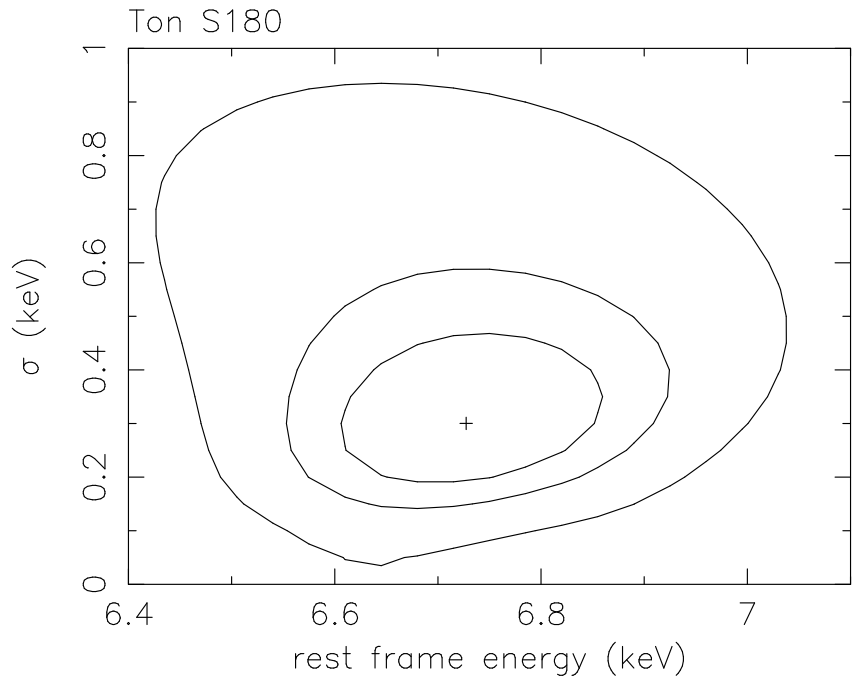}
\caption{$\chi^2$ contours (68\%, 90\% and 99\%) for the central energy in the
rest frame of the object and width $\sigma$ of a Gaussian line model
for the iron line in Ton~S180.  These show that the line is
significantly broad and the energy excludes the neutral iron emission
energy at $>99$\% confidence.}
\end{figure}

\newpage

\begin{figure}[t]
\vbox to8.0in{\rule{0pt}{8.0in}}
\includegraphics{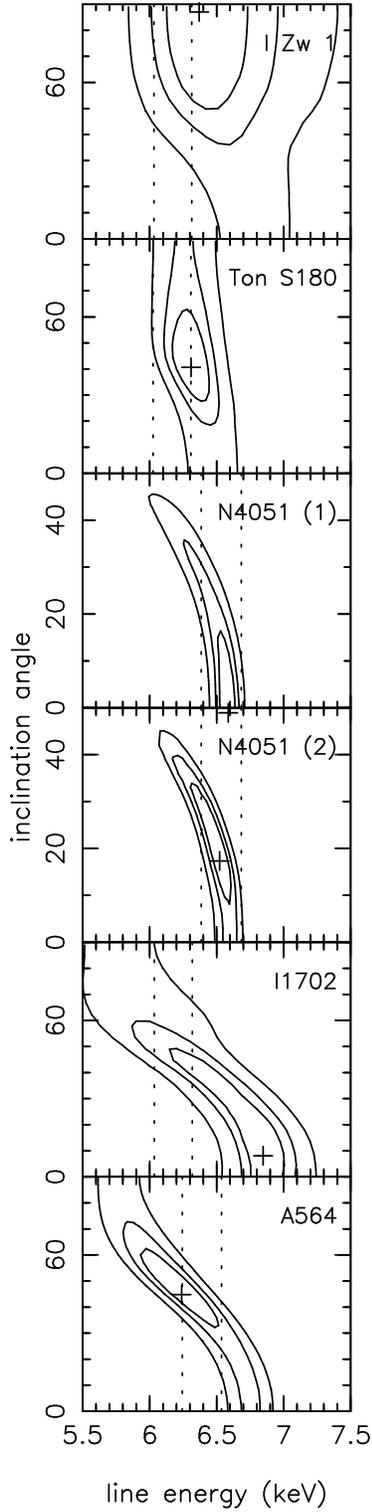}
\caption{Results from fitting a disk line model to 6 spectra from 5
objects in which a significantly broad iron line was detected.  The
inner and outer radii are fixed at $6$ and 1000 $R_g$ respectively,
and the emissivity is fixed at the best fit value between $-2$ and
$-3$.  The energy is in the observers frame and thus the positions of
rest 6.4 and 6.7~keV are shown by dotted lines.}
\end{figure}

\newpage


\clearpage

\begin{references}

\reference{1} Abramowicz, M. A., Czerny, B., Lasota, J. P., \&
Szuszkiewicz, E., 1988, \apj, 332, 646

\reference{2} Belloni, T., M\'endez, M., King, A. R., van der Klis, M.,
\& van Paradijis, J., 1997, ApJL, 145

\reference{3} Bergeron, J., \& Knuth, D., 1980, A\&A, 85, 11

\reference{4} Bergeron, J., \& Knuth, D., 1984, \mnras, 207, 263

\reference{5} Berriman, G., Schmidt, G. D., West, S. C. \& Stockman, H.
S. 1990, ApJS, 74, 869

\reference{6} Boller, Th., Bertoldi, F., Dennefeld, M., \& Voges, W.
1998, A\&AS, 129, 87

\reference{7} Boller, Th., Brandt, W. N., \& Fink, H. 1996, A\&A, 305,
53

\reference{8} Boller, Th., Meurs, E. J. A., Brinkmann, W., Fink, H.,
Zimmermann, U., \& Adorf, H.-M. 1992, A\&A, 261, 57

\reference{9} Boroson, T. A. \& Green, R. F. 1992, ApJS, 80, 109

\reference{10} Brandt, W. N., Fabian, A. C., \& Pounds, K. A., 1996,
\mnras, 278, 326

\reference{11} Brandt, W. N., Mathur, S., \& Elvis, M., 1997, \mnras,
285, 25p

\reference{12} Brandt, W. N., Mathur, S., Reynolds, C. S., \& Elvis, M.
1997, \mnras, 292, 407

\reference{13} Brinkmann, W., Kawai, N., Ogasaka, Y., \& Siebert, J.,
1996, A\&A, 316, 9

\reference{14} Chakrabarti, S., \& Titarchuk, L. G. 1995, \apj, 455, 623

\reference{15} Cohen, R. D. 1983, \apj, 273, 489

\reference{16} Collin-Souffrin, S., Hameury, J.-M., \& Joly, M., 1988,
A\&A, 205, 19

\reference{17} Colpi, M. 1988, \apj, 326, 223

\reference{18} Comastri, A., et al. 1998, A\&A, 333, 31

\reference{19} Comastri, A., Molendi, S., \& Ulrich, M. H., 1997,
``X-ray Imaging and Spectroscopy of Cosmic Hot Plasmas'', ed. F.
Makino \& K. Mitsuda (Tokyo: University Academy Press), 279

\reference{20} Coppi, P. S., 1999, in proc. ``High Energy Processes in
Accreting Black Holes'', eds. J. Poutanen \& R. Svensson, ASP Conf.
Series, Vol. 161, p. 375

\reference{21} Corbin, M. R., 1997, ApJS, 113, 245

\reference{22} Cui, W., Ebisawa, K., Dotani, T., \& Kubota, A., 1998,
ApJL, 493, 75

\reference{23} Czerny, B., \& Elvis, M., 1987, \apj, 321, 305

\reference{24} Czerny, B., \& \.Zycki, P. T., 1994, ApJL, 431, 5

\reference{25} Dickey, J. M., \& Lockman, F. J., 1990, Ann. Rev. Astron.
\& Astrophys. 28, 215

\reference{26} Dotani T. 1998, ASCA Calibration Report, http://www.astro.isas.ac.jp/~dotani/rdd.html

\reference{27} Dove, J. B., Wilms, J., \& Begelman, M. C., 1997, \apj,
487, 747

\reference{28} Ebisawa, K., Titarchuk, L., \& Chakrabarti, S. K., 1996,
PASJ, 48, 59

\reference{29} Elvis, M., Lockman, F. J., \& Wilkes, B. J., 1989, \aj, 97, 777

\reference{30} Erkens, U., Appenzeller, I., \& Wagner, S., 1997, A\&A,
323, 707

\reference{31} Fabian, A. C., Guilbert, P. W., Arnaud, K. A., Shafer,
R. A., Tennant, A. F., \& Ward, M. J., 1986, \mnras, 218, 457

\reference{32} Fabian, A. C., Rees, M. J., Stella, L., \& White, N. E.,
1989, \mnras, 238, 729

\reference{33} Feigelson, E. D., \& Nelson, P. I., 1985, \apj, 293, 192

\reference{34} Fiore, F., Matt, G., Cappi, M., Elvis, M., Leighly, K.
M., Nicastro, F., Piro, L., Siemiginowska, A., \& Wilkes, B. J., 1998,
\mnras, 298, 103

\reference{35} Fiore, F., Matt, G., \& Nicastro, F., 1997, \apj, 284, 731

\reference{36} Forster, K. 1998, PhD Thesis, Columbia University

\reference{37} Forster, K. \& Halpern, J. P., 1996, \apj, 468, 565 

\reference{38} Gelderman, R., \& Whittle, M., 1994, ApJS, 91, 491

\reference{39} George, I. M. \& Fabian, A. C., 1991, \mnras, 249, 352

\reference{40} George, I. M., Turner, T. J., Netzer, H., Nandra, K.,
Mushotzky, R. F., \& Yaqoob, T., 1998, ApJS, 114, 73

\reference{41} Ghisellini, G., \& Haardt, F., 1994, ApJL, 429, 53

\reference{42} Gierli\'nski, M., Zdziarski, A. A., Poutanen, J., Coppi,
P., Ebisawa, K., \& Johnson, W. N., 1999, MNRAS, in press

\reference{43} Goncalves, A. C., V\'eron, P., \& V\'eron-Cetty, M.-P.
1998, in Proc. ``Structure and Kinematics of Quasar Broad Line
Regions'', eds. C. M. Gaskell, W. N. Brandt, M. Dietrich, D.
Dultzin-Hacyan \& M. Eracleous, in press

\reference{44} Goodrich, R. W. 1989, \apj, 342, 224

\reference{45} Grupe, D. 1996, PhD Thesis, University of G\"ottingen

\reference{46} Grupe, D., Beuermann, K., Mannheim, K., Bade, N., Thomas,
H.-C., De~Martino, D., \& Schwope, A., 1995a, A\&A, 299, 5

\reference{47} Grupe, D., Beuermann, K., Mannheim, K., Thomas, H.-C.,
Fink, H. H., \& De~Martino, D., 1995b, A\&A, 300, 21

\reference{48} Grupe, D., Beuermann, K., Thomas, H.-C., Mannheim, K. \&
Fink, H. H. 1998a, A\&A, 330, 25

\reference{49} Grupe, D., Wills, B. J., Wills, D., \& Beuermann, K.,
1998b, A\&A, 333, 827 

\reference{50} Grupe, D., Beuermann, K., Mannheim, K., \& Thomas, H.-C.,
1999a, A\&A, in press

\reference{51} Grupe, D., Leighly, K. M., Thomas, H.-C.,
Laurent-Muehleisen, S. A., 1999b, A\&A, submitted
	
\reference{52} Guainazzi, M., Matsuoka, M., Piro, L., Mihara, T. \&
Yamauchi, M., 1994, ApJL, 436, 35

\reference{53} Guainazzi, M., Mihara, T., Otani, C., \& Matsuoka, M.,
1996, PASJ, 48, 781

\reference{54} Guainazzi, M., Piro, L., Capalbi, M., Parmar, A. N.,
Yamauchi, M., \& Matsuoka, M., 1998a, A\&A, 339, 337

\reference{55} Guainazzi, M., et al., 1998b, A\&A, 339, 327

\reference{56} Haardt, F., \& Maraschi, L., 1991, ApJL, 380, 51

\reference{57} Haardt, F., \& Maraschi, L., 1993, \apj, 413, 507

\reference{58} Haardt, F., Maraschi, L., \& Ghisellini, G., 1994, \apj,
432, 95

\reference{59} Halpern, J. P., \& Moran, E. C. 1998, \apj, 494, 194

\reference{60} Hayashida, K. 1996, Proc. ``Emission Lines in AGN: New
Methods and Techniques'', ed. B. M. Peterson, F.-Z. Cheng \& A. S.
Wilson, (San Fransisco: ASP), 40

\reference{61} Isobe, T., \& Feigelson, E. D., 1990, BAAS, 22, 917

\reference{62} Iwasawa, K., Brandt, W. N., \& Fabian, A. C., 1998,
\mnras, 293, 251

\reference{63} Joly, M., 1993, Ann.\ Phys.\ Fr., 18, 241

\reference{64} Kay, L. E., Magalh\~aes, A. M., Elizalde, F.,
Rodrigues, C., 1999, ApJ, 518, 219

\reference{65} Korista, K. T. 1991, \aj, 102, 41

\reference{66} Korista, K., Ferland, G., \& Baldwin, J., 1997, ApJ, 487, 555

\reference{67} Koski, A. T. 1978, \apj, 223, 56

\reference{68} Krolik, J. H., \& Kallman, T. R. 1988, \apj, 324, 714

\reference{69} Krolik, J. H., \& Kriss, G. A., 1995, \apj, 447, 512

\reference{70} Kwan, J., Cheng, F.-Z., Fang, L.-Z., Zheng, W., \& Ge,
J., 1995, \apj, 440, 628

\reference{71} Laor, A., Fiore, F., Elvis, M., Wilkes, B. J., \&
McDowell, J. C., 1997a, \apj, 477, 93

\reference{72} Laor, A., Jannuzzi, B. T., Green, R. F., \& Boroson, T.
A., 1997b, \apj, 489, 656

\reference{73} Laor, A., \& Netzer, H., 1989, \mnras, 238, 897

\reference{74} Lawrence, A., \& Elvis, M., 1982, \apj, 256, 410

\reference{75} Lawrence, A., Elvis, M., Wilkes, B. J., McHardy, I., \&
Brandt, N. 1997, \mnras, 286, 879

\reference{76} Leighly, K. M. 1998, in proc. ``Accretion Processes in
Astrophysical Systems: Some Like it Hot!'', eds. S. S. Holt, T. R.
Kallman, (AIP: Woodbury, New York), p. 199

\reference{77} Leighly, K. M., Kay, L. E., Wills, B. J., Wills, D., \&
Grupe, D. 1997b, ApJL, 489, 137

\reference{78} Leighly, K. M., Mushotzky, R. F., Yaqoob, T., Kunieda,
K., \& Edelson, R., 1996, \apj, 469, 14

\reference{79} Leighly, K. M., Mushotzky, R. F., Nandra, K., Forster,
K., 1997a, ApJL, 489, 25

\reference{80} Leighly, K. M., \& O'Brien, P. T., 1997, ApJL, 481, 15

\reference{81} Lightman, A. P. \& Eardley, D. M. 1974, ApJL, 187, 1

\reference{82} Maccacaro, T., Gioia, I. M., Wolter, A., Zamorani, G.,
\& Stocke, J. T., 1988, \apj, 326, 680

\reference{83} Madau, P. 1998, \apj, 327, 116

\reference{84} Makuch, R. W., Escobar, M., \& Merrill III, S., 1991,
Appl. Statist. 40, 19

\reference{85} Mason, K. O., Puchnarewicz, E. M., \& Jones, L. R., 1996,
\mnras, 283, 26

\reference{86} Matt, G., Fabian, A. C., \& Ross, R. R., 1993b, \mnras,
262, 179

\reference{87} Matt, G., Fabian, A. C., \& Ross, R. R. 1993a, \mnras, 264, 839

\reference{88} Molendi, S., \& Maccacaro, T., 1994, A\&A, 291, 420

\reference{89} Moran, E. C., Halpern, J. P., \& Helfand, D. J., 1996, ApJS, 106, 341

\reference{90} Morrison, R., \& McCammon, D., 1983, \apj, 270, 119

\reference{91} Murphy, E. M., Lockman, F. J., Laor, A., \& Elvis, M.
1996, ApJS, 105, 369

\reference{92} Mushotzky, R. F. 1982, \apj, 256, 92

\reference{93} Mushotzky, R. F., Fabian, A. C., Iwasawa, K., Kunieda,
H., Matsuoka, M., Nandra, K., \& Tanaka, Y. 1995, \mnras, 272, 9

\reference{94} Nandra, K., George, I. M., Mushotzky, R. F., Turner, T.
J., \& Yaqoob, T., 1997a, \apj, 477, 602

\reference{95} Nandra, K., George, I. M., Mushotzky, R. F., Turner, T.
J., \& Yaqoob, T., 1997b, ApJL, 488, 91

\reference{96} Nandra, K., \& Pounds, K. A., 1994, \mnras, 267, 974

\reference{97} Narayan, R., Mahadevan, R., \& Quataert, E., 1999, in
``Theory of Black Hole Accretion Disks'', eds. M. A. Abramowicz, G.
Bjornsson, \& J. E. Pringle, in press

\reference{98} Netzer, H., 1996, \apj, 473, 781

\reference{99} Nicastro, F., Fiore, F., Perola, G. C., \& Elvis, M.,
1999, \apj, 512, 184

\reference{100} Nicastro, F., Fiore, F., \& Matt, G., 1999, \apj, 517, 108

\reference{101} Nishimura, J., Mitsuda, K, \& Itoh, M., 1986, PASJ, 38, 819

\reference{102} Nowak, M. A. 1995, PASP, 107, 1207

\reference{103} Osterbrock, D. E., \& Pogge, R. W. 1985, \apj, 297, 166

\reference{104} Otani, C., Kii, T. \& Miya, K. 1996, Proc.
``R\"ontgenstrahlung from the Universe'', eds. Zimmermann, H. U., Tr\"umper, J.,
   and Yorke H.,  1996, MPE Report 263, 491

\reference{105} Otani, C. et al. 1996, PASJ, 48, 211

\reference{106} Phillips, M. M., 1976, \apj, 208, 37

\reference{107} Phlips, B.\ F.\, et al., 1996, ApJ, 465, 907

\reference{108} Piccinotti, G., Mushotzky, R. F., Boldt, E. A., Holt, S.
S., Marshall, F. E., Serlemitsos, P. J., \& Shafer, R. A., 1982, \apj,
253, 485

\reference{109} Pietrini, P., \& Krolik, J. H. 1995, \apj, 447, 526

\reference{110} Pogge, R. W., \& Owen, J. M. 1993, OSU Internal Report
93-01 

\reference{111} Porquet, D., Dumont, A.-M., Collin, S., \& Mouchet, M.,
1999, A\&A, 341, 58

\reference{112} Pounds, K. A., Done, C., \& Osborne, J., 1996, \mnras,
277, L5

\reference{113} Press, W. H., Teukolsky, S. A., Vettering, W. T., \&
Flannery, B. P. 1992, {\it Numerical Recipes}, (Cambridge University
Press: Cambridge)

\reference{114} Puchnarewicz, E. M. et al. 1992, \mnras, 256, 589

\reference{115} Remillard, R. A., Bradt, H. V., Buckley, D. A. H.,
Roberts, W., Schwartz, D. A., Tuohy, I. R., \& Wood, K. 1986, \apj,
301, 742

\reference{116} Reynolds, C. S., 1997, \mnras, 286, 513

\reference{117} Reynolds, C. S., \& Fabian, A. C., 1995, \mnras, 273, 1167

\reference{118} Reynolds, C. S., Ward, M. J., Fabian, A. C., \& Celotti,
A. 1997, \mnras, 291, 403

\reference{119} Ross, R. R., Fabian, A. C., \& Mineshige, S., 1992,
\mnras, 258, 189

\reference{120} Rybicki, G. B., \& Lightman, A. P., 1979, ``Radiative
Processes in Astrophysics'' (Wiley: New York)

\reference{121} Schartel, N., Walter, R., Fink, H. H., \& Tr\"umper, J.,
1996, A\&A, 307, 33

\reference{122} Shakura, N. I., \& Sunyaev, R. A., 1973, A\&A, 24, 337

\reference{123} Shapiro, S. L., Lightman, A. P., \& Eardley, D. M., 1976,
\apj, 204, 187

\reference{124} Shields, J. C., Ferland, G. J., \& Peterson, B. M., 1995,
\apj, 441, 507

\reference{125} Shimura, T., \& Takahara, F., 1993, \apj, 419, 78

\reference{126} Shrader, C., \& Titarchuk, L., 1998, ApJL, 499, 31

\reference{127} Smith, P. S., Schmidt, G. D., Allen,  R. G. \& Hines, D.
C. 1998, \apj, 448, 202

\reference{128} Stern, B. E., Poutanen, J., \& Svensson, R. 1995, ApJL,
449, 13

\reference{144} Swensson, R., 1994, ApJS, 92, 585

\reference{129} Svensson, R., \& Zdziarski, A. A., 1994, \apj, 436, 599

\reference{130} Szuszkiewicz, E., Malkan, M. A., \& Abramowicz, M. A.,
1996, \apj, 458, 474.

\reference{131} Tanaka, Y., et al. 1995, Nature, 375, 659

\reference{132} Turner, T. J., George, I. M., \& Nandra, K., 1998,
\apj, 508, 648

\reference{133} Turner, T. J., \& Pounds, K. A., 1989, \mnras, 240, 833

\reference{134} Wandel, A., \& Boller, T., 1998, A\&A, 331, 884

\reference{135} Wilkes, B. J., Elvis, M. \& McHardy, I. 1987, \apj,
321, 23

\reference{136} Wills, B. J., \& Brotherton, M. S., 1995, ApJL, 448, 81

\reference{137} Wills, B., J., Laor, A., Brotherton, M. S., Wills, D.,
Wilkes, B. J., Ferland, G. J., \& Shang, Z., 1999, ApJL, 515, 53

\reference{138} Wills, B. J., Wills, D., Evans, N. J., Natta, A.,
Thompson, K. L., Breger, M., \& Sitko, M. L. 1992, \apj, 400, 96

\reference{139} Yaqoob, T., 1998, ApJ, 500, 893

\reference{140} Zheng, W., Kriss, G. A., \& Davidsen, A. F. 1995, ApJ,
440, 606

\reference{141} Zheng, W., \& O'Brien, P. T. 1990, \apj, 353, 433

\reference{142} \.Zycki, P. T., Krolik, J. H., Zdziarski, A. A., \&
Kallman, T. R., 1994, \apj, 437, 597

\end{references}
\end{document}